\documentclass[twocolumn,showpacs,floatfix,pre,superscriptaddress]{revtex4-1}
\RequirePackage{lineno}

\usepackage{graphicx} 
\usepackage{dcolumn} 
\usepackage{epsf} 
\usepackage{amsmath}
\usepackage{bm} 
\usepackage{setspace}
\usepackage{color}

\newcommand{\ppdfs}{{\emph{p-pdf}}-s}
\newcommand{\rij}{$\bm{r}_{ij}$\;}
\newcommand{\rijF}{$\left\{\bm{r}_{ij}\right\}$\;}
\newcommand{\xax}{$\bm{\hat{x}}\;$}
\newcommand{\yax}{$\bm{\hat{y}}\;$}
\newcommand{\zax}{$\bm{\hat{z}}\;$}
\newcommand{\ril}{$RI\lambda\;$}
\newcommand{\ris}{$RI\sigma^{sph}\;$}
\newcommand{\pd}{\emph{PD\;\,}}
\newcommand{\pds}{\emph{PDs\;\,}}
\newcommand{\lto}{$(\lambda_2/\lambda_1)$\;\,}
\newcommand{\ltt}{$(\lambda_3/\lambda_2)$\;\,}
\newcommand{\ssc}{$\sigma_i^{sph}$s}

\begin{document}


\title{Analysis of structural correlations in a model binary 3D liquid through the eigenvalues
and eigenvectors of the atomic stress tensors}

 \author{V.A.~Levashov}

 \affiliation{Technological Design Institute of Scientific Instrument Engineering,
 Novosibirsk, 630058, Russia}


\begin{abstract}
It is possible to associate with every atom or molecule in a liquid its own
atomic stress tensor. These atomic stress tensors can be used to describe
liquids' structures and to investigate the connection between structural and   
dynamic properties. In particular, atomic stresses allow to 
address atomic scale correlations relevant to the Green-Kubo 
expression for viscosity.
Previously correlations between the atomic stresses of different atoms
were studied using the Cartesian representation of the stress tensors or 
the representation based on spherical harmonics.

In this paper we address structural correlations 
in a model 3D binary liquid using the eigenvalues and eigenvectors of the atomic stress tensors.
Thus correlations relevant to the Green-Kubo expression
for viscosity are interpreted in a simple geometric way.
On decrease of temperature the changes in the relevant stress correlation
function between different atoms are significantly more pronounced than the changes in the pair density function.  
We demonstrate that this behaviour originates from the orientational correlations between the eigenvectors of the 
atomic stress tensors.

We also found correlations between the eigenvalues of the same atomic stress tensor.
For the studied system, with purely repulsive interactions between the particles, the eigenvalues 
of every atomic stress tensor are positive and they can be ordered: 
 $\lambda_1 \geq \lambda_2 \geq \lambda_3 \geq 0$. 
We found that, for the particles of a given type, the probability distributions of the ratios \lto and \ltt
are essentially identical to each other in the liquids state.
We also found that $\lambda_2$ tends to be equal to the geometric average of $\lambda_1$ and $\lambda_3$.
In our view, correlations between the eigenvalues may represent 
``the Poisson ratio effect" at the atomic scale.  
\end{abstract}

\today

\maketitle


\section{Introduction}\label{s:intro}

Despite many years of investigations there is still no commonly 
accepted vision of the slowdown mechanism in supercooled 
liquids \cite{Ediger20121,Biroli20131}. 
It is natural to expect that in order to understand 
the behaviour of supercooled liquids and the phenomenon 
of the glass transition it is necessary to be able to describe 
structural changes in a way that would allow to make connection
to the dynamical properties \cite{EMa20111,ChenYQ2013,Tanaka20131,Wei20151}. 

Viscosity represents a standard parameter 
that is used to characterize dynamical slowdown in supercooled liquids. 
In molecular dynamics simulations liquids' viscosities are commonly 
calculated using the Green-Kubo expression 
\cite{Green1954,Kubo1957,Helfand1960,HansenJP20061,Boon19911,EvansDJ19901}:
\begin{eqnarray}
\eta = \frac{V}{k_{\scalebox{0.5}{B}} T}\int_{0}^{\infty}
<\Pi^{xy}(t_o)\Pi^{xy}(t_o + t)>_{t_o} dt\;\;,
\label{eq;Green-Kubo-01}
\end{eqnarray}
where $k_{\scalebox{0.5}{B}}$-is the Boltzmann constant, $T$ -is the temperature, $V$-is the volume of the system,
$\Pi^{xy}(t)$-is the value of the $xy$ component of the macroscopic stress tensor at time $t$.
Expression (\ref{eq;Green-Kubo-01}) is the limit for zero-frequency and zero-wavevector 
viscosity $\eta(\omega =0, q = 0)$ \cite{HansenJP20061,Boon19911,EvansDJ19901}.

For a given interaction potential, the macroscopic stress tensor $\Pi^{xy}(t)$ 
depends on particles' velocities and coordinates.
In the past there have been multiple investigations of the behaviour
of the integration kernel in (\ref{eq;Green-Kubo-01}) \cite{Hoheisel19881}. 
In particular, it has been found that
in supercooled liquids the value of the correlation 
function $<\Pi^{xy}(t_o)\Pi^{xy}(t_o + t)>_{t_o}$  
is almost completely determined by the liquids structure, i.e.,
by particles coordinates, while the contribution from the 
terms associated with the particles' velocities usually represents
less than $5\%$ of the correlation function value \cite{Hoheisel19881}.
For this reason we neglect the velocity dependent 
terms and provide a simple version of the definition of the macroscopic stress tensor.
Thus, as it was discussed before, the macroscopic stress tensor can be written as the sum of 
the atomic level stress elements \cite{Levashov20111,Levashov2013}: 
\begin{eqnarray}
&&\Pi^{xy}(t) = \frac{1}{V}\sum_{i=1}^{N} s_i^{xy},\label{eq;Green-Kubo-02}\\
&&s_i^{xy}=\sum_{j \neq i} 
\left(\frac{dU_{ab}(r_{ij})}{dr_{ij}}\right)\left(\frac{r_{ij}^a r_{ij}^b}{r_{ij}}\right),
\label{eq;Green-Kubo-03}
\end{eqnarray}
where $U_{ab}(r_{ij})$ is the interaction pair potential between the particles of type
$a$ and $b$, while $\bm{r_{ij}}=\bm{r_j-r_i}$ is the radius vector from particle $i$ to
particle $j$. The sum over $j$ in (\ref{eq;Green-Kubo-03}) is over all particles with 
which particle $i$ interacts.

In principle, expressions (\ref{eq;Green-Kubo-01},\ref{eq;Green-Kubo-02},\ref{eq;Green-Kubo-03}) 
establish the relationship between the structure and the dynamic quantity, i.e., viscosity.
However, the form of the expressions (\ref{eq;Green-Kubo-01},\ref{eq;Green-Kubo-02},\ref{eq;Green-Kubo-03})  
does not provide an explicit answer with respect to what kind of structural correlations
determine viscosity. 
There were studies that addressed how local structural perturbations affect 
the the stress fields in glasses and liquids \cite{Picard20041,Tanguy20061,Maloney20061,Lemaitre20071,Tsamados20081,Lemaitre20091,Chattorai20131,Puosi20141,Lemaitre20141}. 
Yet, the geometric meaning of the atomic scale correlations relevant to the Green-Kubo 
expression for viscosity has not been completely elucidated.
What is meant by the last statement will become more clear from the following.

In several previous publications behaviour of the 
correlation function  $<\Pi^{xy}(t_o)\Pi^{xy}(t_o + t)>_{t_o}$ 
has been studied from an atomic scale perspective 
\cite{Levashov20111,Levashov2013,Levashov20141,Levashov2014B,Bin20151}. 
In these studies macroscopic shear stress correlation function 
in (\ref{eq;Green-Kubo-01})
has been expanded into the correlation 
functions, $\langle s_i^{xy} s_j^{xy} \rangle$ between the
atomic level stress elements from (\ref{eq;Green-Kubo-02},\ref{eq;Green-Kubo-03}). 
In this way, in particular, a relationship between the propagation 
of shear stress waves and viscosity has been demonstrated on atomic scale \cite{Levashov20111,Levashov2013,Levashov20141,Levashov2014B}.

In order to understand the correlation 
function $\langle s_i^{xy} s_j^{xy} \rangle$, as
follows from this paper,
it is useful to realize that the form $\langle s_i^{xy} s_j^{xy} \rangle$ reflects not only 
the nature of the physical correlations in liquids, but also the properties of the
chosen Cartesian representation. 
On the other hand, it is a common practice in considerations of tensors 
to speak about their properties in terms of representation-invariant 
parameters \cite{Ruiz2005,Algebra}.
Surprisingly (to the best of our knowledge) the correlation function
$\langle s_i^{xy} s_j^{xy} \rangle$ has not been studied before 
in 3D in terms of representation-invariant variables. 
Our present paper is devoted to these kind of investigations. 

Our approach is based on the concept of atomic stress 
elements (or atomic level stresses) \cite{Egami19801,Egami19802,Egami19821}.
In the framework of this approach the atomic environment of every atom is described
by a symmetric and real atomic stress tensor.
In this paper we define the atomic stress tensor in a way which is somewhat different
from the previously used definition \cite{Egami19801,Egami19802,Egami19821,Chen19881,Levashov2008B}. 
We explain this difference in the following.
Thus, for an atom $i$ we define its atomic stress tensor as:
\begin{eqnarray}
\sigma_i^{\alpha \beta}=-\frac{1}{2\langle V_i \rangle}\sum_{j \ne i} 
\left[\frac{d U_{ab}(r_{ij})}{d r_{ij}}\right]\left(\frac{r_{ij}^{\alpha} r_{ij}^{\beta}}{r_{ij}}\right)\;,\;\;
\label{eq;stressdef1}
\end{eqnarray}
where $\langle V_i \rangle$ is the 
average atomic volume, $\langle V_i \rangle \equiv 1/\rho_o$, 
while $\rho_o$ is the average atomic number density.
Note that the definition without $\langle V_i \rangle$ and with the opposite sign
corresponds to the atomic stress element from (\ref{eq;Green-Kubo-02},\ref{eq;Green-Kubo-03}) 
\cite{Levashov2013,Levashov20111}. Also note that $\alpha$-component of the force 
acting on particle $i$ from particle $j$ is 
$f^{\alpha}_{ij}=
\left[ d U_{ab}(r_{ij})/d 
r_{ij}\right] \left(r_{ij}^{\alpha}/r_{ij}\right)$. 
Finally note that the atomic stress tensor 
(\ref{eq;stressdef1}) is symmetric with respect to the indexes $\alpha$ and 
$\beta$. Thus in 3D it has 6 independent components 
\cite{Egami19801,Egami19802,Egami19821,Chen19881,Levashov2008B}.

We introduced the  $\langle V_i \rangle$ in (\ref{eq;stressdef1}) in order to
use variables, $\sigma_i^{\alpha \beta}$, that have units of stress. 
Since, for a given density, 
$\langle V_i \rangle$ is just a constant its introduction does not affect
any conclusions. In comparison to the previous 
definition \cite{Egami19801,Egami19802,Egami19821,Chen19881,Levashov2008B}
we also introduced the minus sign in (\ref{eq;stressdef1}).
Our definition makes an atom under compression to have a positive pressure, while
under the previous definition atoms under compression had negative pressure. 
Thus, the minus sign in (\ref{eq;stressdef1}) is also only a matter of convenience, 
which makes the results look more intuitive.

In the previous definition of the atomic level stresses instead of a
constant $\langle V_i \rangle$ was used essentially the volume of
the Voronoi cell of an 
atom \cite{Egami19801,Egami19802,Egami19821,Chen19881,Levashov2008B}. 
Thus every atom, in the previously used definition,
had its own characteristic volume. 
We, in our considerations, avoid using the atom-dependent Voronoi volume
because it is not present in the expressions for viscosity 
(\ref{eq;Green-Kubo-01},\ref{eq;Green-Kubo-02},\ref{eq;Green-Kubo-03}).
Thus we would like to use variables which are directly related to viscosity,
but at the same time we want them to have convenient ``stress" units.
For this reason we introduced the constant multiplication factor $\langle V_i \rangle$.

The concept of atomic level stresses was introduced to describe 
structures of metallic glasses and their liquids 
\cite{Egami19801,Egami19802,Egami19821}. 
There are several important results associated
with this concept. 
One result is the equipartition of the atomic level stress energies in 
liquids \cite{Egami19821,Chen19881,Levashov2008B}.
Thus, the energies of the atomic level stress components can be defined and it was 
demonstrated, for the studied model liquids in 3D, that the energy of every 
stress component is equal to $k_{\scalebox{0.5}{B}}T/4$. 
Thus the total stress energy, which is the sum of the energies of all six stress components,
is equal to $6 \cdot k_{\scalebox{0.5}{B}}T/4 = (3/2)k_{\scalebox{0.5}{B}}T$, i.e., the potential energy of a 
classical 3D harmonic oscillator.
An explanation of this result has been suggested \cite{Egami19821,Chen19881,Levashov2008B}. 
Then there was an attempt to describe the glass transition and fragilities
of liquids using atomic level stresses \cite{Egami20071}.
Another result is related to the Green-Kubo expression for viscosity.
Thus the macroscopic stress correlation function that
enters into the Green-Kubo expression for viscosity was decomposed into
the correlation functions between the atomic stress elements. 
Considerations of the obtained atomic stress correlation functions 
demonstrated the relationship between the propagation and dissipation 
of shear waves and viscosity. 
This result, after all, is not surprising
in view of the existing generalized 
hydrodynamics and mode-coupling theories \cite{HansenJP20061,Boon19911}.
However, in Ref.\cite{Levashov2013,Levashov20111,Levashov20141,Levashov2014B} the 
issue has been addressed from a new perspective and the relationship
between viscosity and shear waves was demonstrated very explicitly
at the atomic level.

Since the atomic stress tensor is real and symmetric it can be diagonalized and its eigenvalues
and eigenvectors can be found \cite{Algebra}. 
In the Cartesian coordinate frame based on the eigenvectors
the atomic stress tensor is diagonal with the eigenvalues on the diagonal. 
Thus six stress elements of the symmetric atomic stress tensor (in our reference coordinate frame)
contain the information about the eigenvalues 
and orientations of the eigenvectors. 
It follows from this viewpoint that in the approach based on considerations 
of the atomic stresses it is possible in 3D to associate with every 
atom (and its environment) an ellipsoid
whose axes have the lengths of the eigenvalues and whose orientation is described by
its eigenvectors. 

Analysis in terms of the eigenvalues and eigenvectors
of the atomic stresses represents simple, geometric and representation-invariant 
approach that can be used to describe liquids' structures.
In this paper we express, in particular, the correlation function $\langle \sigma_i^{xy}\sigma_j^{xy} \rangle$ 
in terms of the correlation functions between the eigenvalues and the angles between the eigenvectors of atoms $i$ and $j$.
This result provides a new insight into the nature of the structural correlations
that determine the Green-Kubo correlation function. 
Our results show that on supercooling correlations in the orientations of the stress ellipsoids develop. 
These orientational correlations cause more significant changes in 
the $\langle \sigma_i^{xy}\sigma_j^{xy} \rangle$ correlation function than the changes 
associated with correlations between the eigenvalues.

Effectively this paper has four parts. 
In the first part we describe the formalism that allows to address structure
of liquids in terms of the eigenvalues and eigenvectors of the atomic stresses. 
In the second part, using MD and MC simulations we analyse correlations 
between the eigenvalues of the same atomic stress tensor and some other related issues. 
In the third part, we present the results
on correlations between the eigenvalues and eigenvectors of different atoms. 
The forth part contains appendices that provide additional analytical 
insights into the data obtained with computer simulations.

\section{Stress tensor ellipsoids \label{sec:idea}}

The atomic stress tensor, $\Sigma_i$, 
defined with equation 
(\ref{eq;stressdef1}) is real and symmetric.  Thus it can be 
diagonalised and, in 3D, three real eigenvalues $\lambda^1_i$, 
$\lambda^2_i$, $\lambda^3_i$ and three 
real orthogonal eigenvectors of the stress tensor can be found \cite{Algebra}.
For this reason we can associate with every atom an ellipsoid with
the axes of length $\lambda^1_i$, $\lambda^2_i$, $\lambda^3_i$.
These axes are parallel to the corresponding eigenvectors.
In the frame of the ellipsoid's axes the stress tensor is diagonal.
 
In the following we refer to the coordinate frame based on the eigenvectors
of atom $i$ as to the \emph{eigenframe} of atom $i$. 
In a different reference coordinate frame for the normalized 
eigenvector $\bm{\bm{\hat{v}}}_i^{1}$ we introduce the following notations:
\begin{eqnarray}
\bm{\bm{\hat{v}}}_i^{1} = \left(c_i^{11},c_i^{12},c_i^{13}\right)\;\;,\;
\label{eq;vi1}
\end{eqnarray}
where the vector's components are the directional cosines defined through the scalar products:
\begin{eqnarray}
c_i^{11}=\bm{\bm{\hat{v}}}_i^1\bm{\hat{x}}\;,\;\;\;c_i^{12}=\bm{\bm{\hat{v}}}_i^1\bm{\hat{y}}\;,
\;\;\;c_i^{13}=\bm{\bm{\hat{v}}}_i^1\bm{\hat{z}}\;\;.\;\;\;\;\;\;
\label{eq;vi12}
\end{eqnarray}
Similar notations are assumed for the other two eigenvectors. 
Further we define the matrix of the column-eigenvectors, $V_i$, 
and the matrix of the eigenvalues, $\Lambda_{i}$:
\begin{eqnarray}
V_i \equiv \begin{pmatrix} 
c_i^{11} & c_i^{21} & c_i^{31} \\ 
c_i^{12} & c_i^{22} & c_i^{32} \\ 
c_i^{13} & c_i^{23} & c_i^{33}  
\end{pmatrix}\;\;,\;\;\;\;
\Lambda_{i} \equiv \begin{pmatrix} 
\lambda_i^1 & 0 & 0 \\ 
0 & \lambda_i^2 & 0 \\ 
0 & 0 & \lambda_i^3  
\end{pmatrix}\;\;.\;\;\;\;
\label{eq;Vmi1}
\end{eqnarray}
If follows from the definitions of $V_i$, $\Lambda_i$, and the known relations from linear algebra that: 
\begin{eqnarray}
\sigma_i V_i = V_i \Lambda_{i}\;\;.\;\;
\label{eq;Veigen1}
\end{eqnarray}
From (\ref{eq;Veigen1}) we get:
\begin{eqnarray}
\Lambda_{i} = V_i^T \Sigma_i V_i\;\;,\;\;\;\;\;\;\;\Sigma_i= V_i \Lambda_{i} V_i^T\;\;,\;\;
\label{eq;Veigen2}
\end{eqnarray}
where matrix $V_i^T$ is the transposed and also the inverse of $V_i$.

In our further considerations we use some well known results \cite{Ruiz2005,Algebra} 
from linear algebra and tensor analysis about which we remind here.
Let us suppose that the stress tensors of atoms $i$ in a particular 
coordinate frame is:
\begin{eqnarray}
\Sigma_i \equiv \begin{pmatrix} 
\sigma_i^{x} & \tau_i^{xy} & \tau_i^{xz} \\ 
\tau_i^{xy} & \sigma_i^{y} & \tau_i^{yz} \\ 
\tau_i^{xz} & \tau_i^{yz} & \sigma_i^{z}  
\end{pmatrix}.\;\;\;\;
\label{eq;Sitensor}
\end{eqnarray}

The equation (the determinant) for the eigenvalues of a $3 \times 3$ real 
symmetric (stress) matrix can be written in terms 
of the rotational invariants, $I_1$, $I_2$, and $I_3$ as follows:
\begin{eqnarray}
\lambda^3 - I_1 \lambda^2 + I_2 \lambda - I_3 = 0 \;\;,\;\;
\label{eq;eigenequation01}
\end{eqnarray}
where (for briefness we omit index $i$):
\begin{eqnarray}
I_1 =&& +\sigma^{x} + \sigma^{y} + \sigma^{z}\;,\;\;\label{eq;eigenI01}\\
I_2 =&& +\sigma^{x}\sigma^{y} + \sigma^{x}\sigma^{z} + \sigma^{y}\sigma^{z}\label{eq;eigenI02}\\
&&-(\tau^{xy})^2 - (\tau^{xz})^2 - (\tau^{yz})^2 \nonumber\\ 
I_3 =&& + \sigma^{x}\sigma^{y}\sigma^{z} + 2\tau^{xy}\tau^{xz}\tau^{yz} \label{eq;eigenI03}\\
&& - (\tau^{xy})^2\sigma^{z} - (\tau^{xz})^2\sigma^{y} - (\tau^{yz})^2\sigma^{x} \;\;.\;\;\nonumber
\end{eqnarray}
If the eigenvalues are known then these invariants can be rewritten in terms of the eigenvalues:
\begin{eqnarray}
I_1 \;=\;\;&& \lambda_{1} + \lambda_{2} + \lambda_{3}\;,\;\;\label{eq;eigenI012}\\
I_2 \;=\;\;&& \lambda_{1}\lambda_{2} + \lambda_{1}\lambda_{3} + \lambda_{2}\lambda_{3}\;,\;\;\label{eq;eigenI022}\\
I_3 \;=\;\;&& \lambda_{1}\lambda_{2}\lambda_{3}\;.\;\; \label{eq;eigenI032}
\end{eqnarray}
For further convenience let us also notice that:
\begin{eqnarray}
(I_1)^2 - 3I_2 &&= \lambda_{1}^2 + \lambda_{2}^2 + \lambda_{3}^2 -\lambda_{1}\lambda_{2} - \lambda_{1}\lambda_{3} - \lambda_{2}\lambda_{3}\label{eq;eigenI013}\\
&&=\frac{1}{2}\left[(\lambda_1 - \lambda_2)^2 + (\lambda_1 - \lambda_3)^2 + (\lambda_2 - \lambda_3)^2\right]\;.\nonumber
\end{eqnarray}
It is demonstrated in the following subsection that $(I_1)^2 - 3I_2$ in (\ref{eq;eigenI013}) is 
essentially the square of the von Mises shear stress.

\subsection{Elements of the atomic stress tensors in the spherical representation}

Previously it has been argued that it is useful to assume that 
the nearest neighbour atomic environment of every atom
is approximately spherical \cite{Egami19821,Chen19881,Levashov2008B}. 
This assumption, in particular, allows to introduce and rationalise 
the concept of the atomic stress energies as excitations from some average atomic environment. 
The relevant derivation is based on the representation of the atomic stresses in terms of the spherical harmonics
\cite{Egami19821,Chen19881}. 
In the following sections we present the data that justify consideration
of this approach in our context. 

The atomic stress elements defined through (\ref{eq;stressdef1}) do not reflect the vision
that the nearest neighbour atomic environment of every atom is approximately spherical 
(however good or bad this approximation is).
The consideration of the atomic stresses in terms of the spherical harmonics leads
to the following linear combinations of the Cartesian stress components that reflect the vision that atomic environment 
of every atom is approximately spherical:
\begin{eqnarray}
p_i \equiv &&s_{0,i}\equiv \tfrac{1}{3}\left[ \sigma^{x}_i + \sigma^{y}_i + \sigma^{z}_i \right]\;,\;\;\;\;\label{eq:to01}\\
&&s_{1,i} \equiv \tau^{xy}_i\;,\;\;s_{2,i} \equiv \tau^{xz}_i\;,\;\;s_{3,i} \equiv \tau^{yz}_i\;,\;\label{eq:to02}\\
&&s_{4,i} \equiv \tfrac{1}{2}\left(\sigma^{x}_i - \sigma^{y}_i\right)\;,\;\label{eq:to03}\\
&&s_{5,i} \equiv \tfrac{1}{\sqrt{3}}\left[\sigma_i^{z}-\tfrac{1}{2}\left(s^{x}_i + s^{y}_i)\right)\right]\;,\;\label{eq:to04}
\end{eqnarray}
Formulas (\ref{eq:to01},\ref{eq:to02},\ref{eq:to03},\ref{eq:to04})
define one pressure component, $p_i \equiv s_{0,i}$, and 5 \emph{equivalent} to each other shear stress 
components that reflect sphericity of the atomic environment of atom $i$ \cite{Egami19821,Chen19881,Levashov2008B}:
The notations that we use are slightly different from those 
used previously \cite{Egami19821,Chen19881,Levashov2008B}. 
See Ref.\cite{sphstress1} for the details.

In the following we use the notation (abbreviation) \ssc\, for the ``spherical stress components".

Vice versa from (\ref{eq:to01},\ref{eq:to02},\ref{eq:to03},\ref{eq:to04}) we obtain:
\begin{eqnarray}
&&\sigma^{x}_i = s_{0,i}-\tfrac{1}{\sqrt{3}}s_{5,i} + s_{4,i}\;,\;\;\label{eq:from01}\\
&&\sigma^{y}_i = s_{0,i}-\tfrac{1}{\sqrt{3}}s_{5,i} - s_{4,i}\;,\;\;\;\;\label{eq:from02}\\
&&\sigma^{z}_i = s_{0,i} + \tfrac{2}{\sqrt{3}}s_{5,i}\;,\;\;\label{eq:from03}\\
&&\tau^{xy}_i=s_{1,i}\;,\;\;\;\tau^{xz}_i=s_{2,i}\;,\;\;\;\tau^{yz}_i=s_{3,i} \;.\;\;\;\;\;\;\;\label{eq:from04}
\end{eqnarray}

It was argued in Ref.\cite{Egami19821} that the atomic stress tensor components 
$s_{0,i},\;s_{1,i},\;s_{2,i},\;s_{3,i},\;s_{4,i},\;s_{5,i}$ should be independent 
from each other in the linear approximation. 
This assumption plays an important role in rationalizing why the energies of these 
stress components are equal to each other and
in explaining why the energy of every component in the liquid state is equal 
to $\tfrac{1}{4}k_{\scalebox{0.5}{B}} T$.
The results from numerical simulations presented in Ref.\cite{Chen19881,Levashov2008B} support this assumption. 
In this paper we address the issue of independence of the \ssc\,
from a new perspective, i.e., from the perspective of the probability distributions of the eigenvalues 
of the atomic stresses.

In the eigenframe of its eigenvectors an atomic stress tensor 
is characterised by its 3 eigenvalues
$\lambda_{1}$, $\lambda_{2}$, and $\lambda_{3}$. 
Let us now consider this stress tensor in a different
coordinate frame. It is straightforward to show that in any coordinate frame
the sum of the squares of the shear stress
components in the spherical representation is equal to:  
\begin{eqnarray}
\left(s_{vM,i}\right)^2 &&\equiv \frac{1}{5}\sum_{n=1}^{n=5} \left(s_{n,i}\right)^2 \label{eq;vonMises01}\\
&&=\frac{1}{30}\left[(\lambda_1 - \lambda_2)^2 + (\lambda_1 - \lambda_3)^2 + (\lambda_2 - \lambda_3)^2\right]\;.\nonumber
\end{eqnarray}
Expression (\ref{eq;vonMises01}) is essentially the definition of the von Mises shear stress. It follows from
(\ref{eq;vonMises01}) that the square of the von Mises shear stress is rotationally invariant.
It follows from the comparison of (\ref{eq;vonMises01}) with (\ref{eq;eigenI013}) that the stress tensor\
invariant $I_2$ is related to the von Mises shear stress.

Let us evaluate the value of the square of a shear stress component averaged over all possible
orientations of the reference coordinate frame. 
It follows from formulas (\ref{eq;avesixysjxy1},\ref{eq;avesixysjxy3}), 
that we derive further, that this average value
is equal to the square of the von Mises shear stress from (\ref{eq;vonMises01}):
\begin{eqnarray}
\langle \left(s_{n,i}\right)^2\rangle_{\Omega}=\left(s_{vM,i}\right)^2\;.\;\;\;\;\;\;\;\;\label{eq;squareaverage01}
\end{eqnarray}

In the previous considerations of the atomic stresses in glasses and liquids the atomic level pressure
and von Mises shear stress have been studied \cite{Egami19821,Chen19881,Levashov2008B}.
In particular, studies of the equipartition of the atomic stress energies can be considered 
as the studies of the atomic von Mises shear stresses. 
In contrast, the role of the third invariant $I_3$ from (\ref{eq;eigenI032}) has not been addressed previously.
It is clear from (\ref{eq;eigenI032}) that $I_3$ essentially represents the volume of the ellipsoid with axes 
$\lambda_{1}$, $\lambda_{2}$, $\lambda_{3}$. 
The value of $I_3$ can be used, for example, in order to characterize the relative scale
of the shear deformation. 
For example, one can use the geometric average, 
i.e., $\lambda_{geom} \equiv \left(\lambda_{1}\lambda_{2}\lambda_{3}\right)^{1/3}$ in order
to normalize the von Mises shear stress. 
Similarly the difference between the pressure (arithmetic average of the eigenvalues) and $\lambda_{geom}$
also gives certain measure of the deformation of the local atomic environment from the purely spherical state.

\section{Two Random Reference Models \label{sec:random-reference}}

As we already noted before, 
in the approach based on considerations of the atomic stresses 
the geometry of the nearest neighbour shell, 
if its orientation with respect to the reference coordinate frame is ignored, is characterised by
only three numbers -- three eigenvalues. Alternatively, it is possible to consider three invariants
of the atomic stress tensor (\ref{eq;eigenI01},\ref{eq;eigenI02},\ref{eq;eigenI03}).

In considerations of the eigenvalues the first natural question to ask is: ``What are their distributions?"
The second question to ask is: ``Are there correlations between the eigenvalues of the same atom?"

The issues related to the probability distributions ($PDs$) of the eigenvalues represent large field 
studied in mathematics and physics \cite{randommatrices,Wigner1967}.
One well known application is related to the studies of the \pds of the energy levels which result from 
diagonalization of various Hamiltonians. In the context of supercooled liquids and the glass transition the Hessian 
matrix is routinely diagonalized in order to find the vibrational spectra of the studied systems \cite{amorhpous2001,Matsuoka2012}.

One the other hand, we are familiar with only few papers in which the atomic stress 
tensors were diagonalized and the \pds of their eigenvalues have been investigated \cite{Kust2003a,Kust2003b}. 
It was demonstrated in those studies that there are correlations between 
the eigenvalues for several 2D and 3D mono-atomic and binary systems. 
However, the nature of those correlations is not understood, as the authors themselves point out \cite{Kust2003b}. 

Here we further investigate the correlations between the eigenvalues. 
In some of our considerations we use two modifications of the method suggested in Ref.\cite{Kust2003a,Kust2003b}.

The idea of the first method is the following. 
From MD simulations it is possible to obtain the probability distribution ($PD$) for 
all eigenvalues without making the distinction which eigenvalue is the largest, the middle, or the smallest one.
If the \pd for all eigenvalues is known, then it is possible to 
generate, using Monte Carlo technique, 
independent and random numbers whose \pd 
is the same as the \pd of the eigenvalues.
Let us suppose that we generated three such numbers. 
Then, if needed, we can order them according to their magnitudes.
In order to address correlations between the eigenvalues of the same atom, 
it is possible to compare \emph{the quantities of interest} obtained directly from MD simulations with
the same quantities obtained from the independent and random generation of the eigenvalues.
This method, as far as we understand, is essentially equivalent to the method used in Ref.\cite{Kust2003a,Kust2003b}.
In the following we refer to this method as to the ``\ril approach" (\emph{Random and Independent for $\lambda$}).

Another method that we employed combines the previous method with the idea that atomic stresses 
in the spherical representation (\ref{eq:to01},\ref{eq:to02},\ref{eq:to03},\ref{eq:to04}) should be independent 
from each other in the linear approximation \cite{Egami19821,Chen19881,Levashov2008B}. 
Thus, let us suppose that we obtained the \pds of the atomic stress elements in 
the spherical representation from MD simulations.
Using these \pds we can generate independent and random 
spherical stresses in such a way that their \pds correspond to those obtained 
in MD simulations.
Using a particular random set of pressure and five spherical shear components 
we can form the random stress matrix in the Cartesian 
representation (\ref{eq:from01},\ref{eq:from02},\ref{eq:from03},\ref{eq:from04}) which can be diagonalized.
In this way we can obtain the eigenvalues from the independent and random selection of the \ssc. 
The \pds of the eigenvalues obtained in this way 
can be compared with the \pds of the eigenvalues obtained directly from MD simulations. 
It is also possible to compare the quantities of interest obtained from the independent and random selection of 
the \ssc\, with the same quantities obtained directly from MD simulations.
In the following we refer to this method as to the ``\ris approach" (\emph{Random and Independent for the Spherical Stresses}). 

We conclude this section by describing \emph{the rejection method}, i.e., the well known Monte Carlo algorithm that we used to generate 
random numbers with given \pds \cite{numrecip}.
Let us suppose that some quantity $x$ has such probability distribution, $P(x)$, 
that we always have: $x_{min} \leq x \leq x_{max}$ and $0 \leq P(x) \leq P_{max}$. 
Using a random number generator, which generates homogeneously distributed 
random numbers, we generate
trial numbers $x_{trial}$ and $y_{trial}$ which lie in 
the intervals: $x_{min} \leq x_{trial} \leq x_{max}$ and $0 \leq y_{trial} \leq P_{max}$. 
On the final step $x_{trial}$ is accepted into the randomly generated set of 
interest if $y_{trial} \leq P(x_{trial})$. 
Otherwise $x_{trial}$ is not accepted into the set.

\section{Correlation functions \label{sec:corrfunct}} 

If the parameters of the stress tensor ellipsoids of atoms $i$ and $j$ are known
then it is possible to study all kinds of correlations that take into account the eigenvalues and orientations
of the eigenvectors of atoms $i$ and $j$. However, it is reasonable to study those atomic scale correlations
which are related to the macroscopic physical quantities. For example, in order to study correlations
related to viscosity it is necessary to express the correlation function 
$\left<\sigma_i^{xy}s_j^{xy}\right>$ in terms of the 
eigenvalues and angles that characterise orientations of the atomic stress ellipsoids.

If the atomic stress tensor, $\Sigma_i$, of atom $i$ is known in one (the 1st) coordinate frame then it also can be found
in a different (the 2nd) coordinate frame.  Thus:
\begin{eqnarray}
\tilde{\Sigma}_i= R \Sigma_i R^T\;,\;
\label{eq;tildeS1}
\end{eqnarray}
where the columns in the rotation matrix $R$ are the directional cosines
of the 1st coordinate frame, $\bm{\hat{x}}$, $\bm{\hat{y}}$, $\bm{\hat{z}}$, 
with respect to the 2nd coordinate frame  
$\bm{\hat{\tilde{x}}}$, $\bm{\hat{\tilde{y}}}$, $\bm{\hat{\tilde{z}}}$.
Note that expression (\ref{eq;tildeS1}) is essentially the same as 
the second expression in (\ref{eq;Veigen2}).

Let us suppose that we are interested in correlations between the parameters and orientations
of the atomic stress ellipsoids of atoms $i$ and $j$ separated by radius vector \rij.
If the medium is isotropic then physically meaningful correlations can depend
on distance $r_{ij}$, but should not depend on the direction of \rij.
For this reason it is reasonable to consider for every given pair of atoms $i$ and $j$
the directional coordinate ``\rijF -frame" associated with the direction from atom $i$ to atom $j$.

\subsection{Directional coordinate frame associated with the direction from atom $i$ to atom $j$} 
 
Let us assume that \zax -axis of the directional coordinate ``\rijF-frame" associated with the direction
from atom $i$ to atom $j$ is along \rij.
Thus \xax and \yax -axes of the directional coordinate frame
lie in the plane orthogonal to \rij. 
Their precise directions will not be important to us as we
are going to average over their orientations in the plane.

\subsection{Correlation function $\left<\tau_i^{xy}\tau_j^{xy}\right> \equiv \left<\sigma_i^{xy}\sigma_j^{xy}\right>$ in the directional coordinate frame}

Let us express the product $\tau_i^{xy}\tau_j^{xy} \equiv \sigma_i^{xy}\sigma_j^{xy}$ in terms of the eigenvalues and the directional cosines 
of the eigenvectors in the directional \rijF -frame. 
The expressions derived below provide physical and representation-invariant insight into the
correlations that determine viscosity.

\begin{figure}
\begin{center}
\includegraphics[angle=0,width=2.3in]{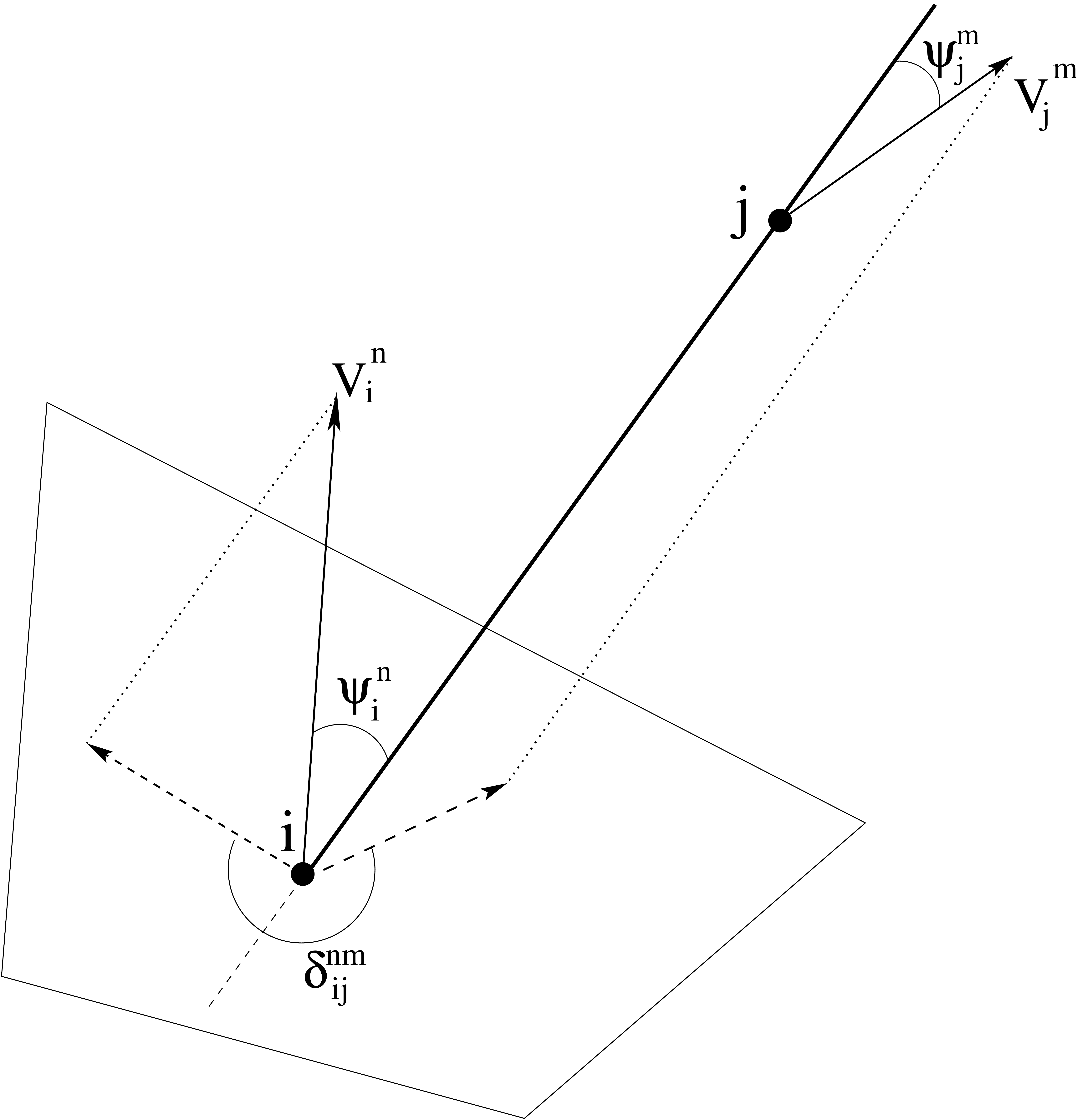}
\caption{Mutual orientation of the eigenvectors $V_i^n$ and $V_j^m$ 
is characterized by the angles $\psi_{i}^n$ and $\psi_{j}^m$ which they form
with the direction $\bm{r}_{ij}$ and by the angle, $\delta_{ij}^{nm}$, 
between their projections on the plane orthogonal to $\bm{r}_{ij}$. 
}\label{fig:lambda-distr-cond-2}
\end{center}
\end{figure}

Further it is assumed that the directional cosines of the eigenvectors of atoms $i$ and $j$ 
in the \rijF -frame are known. Thus:
\begin{eqnarray}
&&\bar{c}_{i}^{n1} = \sin(\psi_i^n)\cos(\varphi_i^n)\;,\;\;\;\;\;
\bar{c}_{j}^{m1} = \sin(\psi_j^m)\cos(\varphi_i^n+\delta_{ij}^{nm})\;,\nonumber\\
&&\bar{c}_{i}^{n2} = \sin(\psi_i^n)\sin(\varphi_i^n)\;,\;\;\;\;\;\;
\bar{c}_{j}^{m2} = \sin(\psi_j^m)\sin(\varphi_i^n+\delta_{ij}^{nm})\;,\nonumber\\
&&\bar{c}_{i}^{n3} = \cos(\psi_i^n)\;,\;\;\;\;\;\;\;\;\;\;\;\;\;\;\;\;\;\;
\bar{c}_{j}^{m3} = \cos(\psi_j^m)\;.
\label{eq;fijcos1}
\end{eqnarray} 
In the expressions above $\psi_i^n$ and $\psi_j^m$ are the angles that $n$-th and $m$-th eigenvectors of atoms
$i$ and $j$ form with the \zax -axis of the \rijF-frame. See Fig.\ref{fig:lambda-distr-cond-2}. 
According to the adopted convention, the angles $\psi_i^n$ and $\psi_j^m$ lie in the interval $\left(0,\tfrac{\pi}{2}\right)$.
The angle $\varphi_i^n$ characterizes the orientation of the projection the $n$-th eigenvector of atom $i$
on the plane orthogonal to \rij with respect to the \xax -axis of the \rijF -frame.
It is also assumed in (\ref{eq;fijcos1}) that the projection of the $m$-th eigenvector of atom $j$ on the
plane orthogonal to \rij forms angle $\delta_{ij}^{nm}$ with the projection of the $n$-th eigenvector of atom $i$. 
The angle $\delta_{ij}^{nm}$ can be found from the scalar product of the eigenvectors' projections on the plane orthogonal to
\rij.
The bars over letters, like in $\bar{c}_{i}^{n1}$, 
signify that the bar-marked parameters are related
to the \rijF-frame.
Thus the matrices that rotate from the \emph{eigenframes} of 
atoms $i$ and $j$ into the \rijF -frame are:
\begin{eqnarray}
\bar{R}_{i} \equiv \begin{pmatrix} 
\bar{c}_{i}^{11} & \bar{c}_{i}^{21} & \bar{c}_{i}^{31} \\ \\
\bar{c}_{i}^{12} & \bar{c}_{i}^{22} & \bar{c}_{i}^{32} \\ \\
\bar{c}_{i}^{13} & \bar{c}_{i}^{23} & \bar{c}_{i}^{33} \\ 
\end{pmatrix}\;\;,\;\;\;\;
\bar{R}_{j} \equiv \begin{pmatrix} 
\bar{c}_{j}^{11} & \bar{c}_{j}^{21} & \bar{c}_{j}^{31} \\ \\
\bar{c}_{j}^{12} & \bar{c}_{j}^{22} & \bar{c}_{j}^{32} \\ \\
\bar{c}_{j}^{13} & \bar{c}_{j}^{23} & \bar{c}_{j}^{33}  \\ 
\end{pmatrix}\;\;.\;\;\;\;
\label{eq;RijRji}
\end{eqnarray}

It follows from (\ref{eq;tildeS1},\ref{eq;RijRji}) that:
\begin{eqnarray}
&&\bar{\sigma}_{i}^{x}=\lambda_i^1\left(\bar{c}_{i}^{11}\right)^2+\lambda_i^2\left(\bar{c}_{i}^{21}\right)^2+\lambda_i^3\left(\bar{c}_{i}^{31}\right)^2\;\;,\;\;\;\;
\label{eq;sijxx}\\
&&\bar{\sigma}_{i}^{y}=\lambda_i^1\left(\bar{c}_{i}^{12}\right)^2+\lambda_i^2\left(\bar{c}_{i}^{22}\right)^2+\lambda_i^3\left(\bar{c}_{i}^{32}\right)^2\;\;,\;\;\;\;
\label{eq;sijyy}\\
&&\bar{\sigma}_{i}^{z}=\lambda_i^1\left(\bar{c}_{i}^{13}\right)^2+\lambda_i^2\left(\bar{c}_{i}^{23}\right)^2+\lambda_i^3\left(\bar{c}_{i}^{33}\right)^2\;\;,\;\;\;\;
\label{eq;sijzz}\\
&&\bar{\tau}_{i}^{xy}=
\lambda_i^1\left(\bar{c}_{i}^{11}\bar{c}_{i}^{12}\right)+
\lambda_i^2\left(\bar{c}_{i}^{21}\bar{c}_{i}^{22}\right)+
\lambda_i^3\left(\bar{c}_{i}^{31}\bar{c}_{i}^{32}\right)\;\;,\;\;\;\;
\label{eq;sijxy}\\
&&\bar{\tau}_{i}^{xz}=
\lambda_i^1\left(\bar{c}_{i}^{11}\bar{c}_{i}^{13}\right)+
\lambda_i^2\left(\bar{c}_{i}^{21}\bar{c}_{i}^{23}\right)+
\lambda_i^3\left(\bar{c}_{i}^{31}\bar{c}_{i}^{33}\right)\;\;,\;\;\;\;
\label{eq;sijxz}\\
&&\bar{\tau}_{i}^{yz}=
\lambda_i^1\left(\bar{c}_{i}^{12}\bar{c}_{i}^{13}\right)+
\lambda_i^2\left(\bar{c}_{i}^{22}\bar{c}_{i}^{23}\right)+
\lambda_i^3\left(\bar{c}_{i}^{32}\bar{c}_{i}^{33}\right)\;\;.\;\;\;\;
\label{eq;sijyz}
\end{eqnarray}
In the expressions above the notation $\lambda_i^3$ is used for the smallest eigenvalue of atom $i$.
Thus the upper index characterizes the order of the eigenvalue (it does not mean that it is $\lambda_i$ in power $3$).
Similar expressions can be written for the stress components of atom $j$. 
For example, for the $\bar{\tau}_{j}^{xy}$ stress component of atom $j$ we have:
\begin{eqnarray}
\bar{\tau}_{j}^{xy}=
\lambda_j^1\left(\bar{c}_{j}^{11}\bar{c}_{j}^{12}\right)+
\lambda_j^2\left(\bar{c}_{j}^{21}\bar{c}_{j}^{22}\right)+
\lambda_j^3\left(\bar{c}_{j}^{31}\bar{c}_{j}^{32}\right)\;\;.\;\;\;\;\;
\label{eq;sjixy}
\end{eqnarray}
Using expressions (\ref{eq;sijxy},\ref{eq;sjixy}) the product
($\bar{\tau}_{i}^{xy}\bar{\tau}_{j}^{xy}$) can be formed. 
In this product there are 9 terms.
All these terms have the form: 
$\lambda_i^n\lambda_j^m\left(\bar{c}_{i}^{n1}\bar{c}_{i}^{n2}\bar{c}_{j}^{m1}\bar{c}_{j}^{m2}\right)$.
It follows from (\ref{eq;fijcos1}) that:
\begin{eqnarray}
&&\left(\bar{c}_{i}^{n1}\bar{c}_{i}^{n2}\bar{c}_{j}^{m1}\bar{c}_{j}^{m2}\right)\label{eq;cccc1}\\
&&=
\sin^2(\psi_i^n)\sin^2(\psi_j^m)\left(\tfrac{1}{4}\right)\sin(2\varphi_i^n)\sin(2\varphi_i^n+2\delta_{ij}^{nm})\;.\;\;\;\;\;\;\;
\label{eq;cccc2}
\end{eqnarray}
The angle $\varphi_i^n$ depends on the choice of the direction of the \xax and \yax axes in the plane orthogonal to \rij.
However, any particular choice of their direction in this plane is irrelevant to the symmetry of the problem for which only the direction of 
\rij is important. Therefore it can be assumed, in performing the averaging of (\ref{eq;cccc1}) over the ensemble, that 
we also average over all possible values of $\varphi_i^n$ in (\ref{eq;cccc2}). 
Thus, for the terms associated with the correlation function $\langle \tau_i^{xy}\tau_j^{xy}\rangle$  in the \rijF-frame we get:
\begin{eqnarray}
\langle\bar{\tau}_i^{xy}\bar{\tau}_j^{xy}\rangle &&\rightarrow \langle\lambda_i^n\lambda_j^m\bar{c}_{i}^{n1}\bar{c}_{i}^{n2}\bar{c}_{j}^{m1}\bar{c}_{j}^{m2}\rangle\label{eq;cccc3}\\
&&=
\tfrac{1}{8}\langle\lambda_i^n\lambda_j^m\sin^2(\psi_i^n)\sin^2(\psi_j^m)\cos(2\delta_{ij}^{nm})\rangle\;.\label{eq;cccc31}\;\;\;\;
\end{eqnarray}
With respect to (\ref{eq;cccc31}) note the following. Let us assume that there are no correlations in the orientations of the projections
of the eigenvectors on the plane orthogonal to the direction of \rij. This means that the angles $\delta_{ij}^{nm}$ are homogeneously 
distributed in the interval $(-\pi/2, \pi/2)$ and correspondingly the correlation function in (\ref{eq;cccc3}) is zero. 

Note that expressions (\ref{eq;cccc3},\ref{eq;cccc31}) suggest that a particularly simple 
organization of the eigenvectors provides a maximum value to $\bar{\tau}_i^{xy}\bar{\tau}_j^{xy}$. 
This is the organization when the eigenvectors of the smallest $\lambda$-s of atoms $i$ and $j$ are directed along
\rij, while two others eigenvectors of both atoms lie in the plane orthogonal to \rij. 
Moreover, the eigenvectors of atom $i$, that lie in the plane orthogonal to \rij,
should be aligned with those eigenvectors of atom $j$ that also lies in the plane
orthogonal to \rij. Essentially this means that identical orientations of the eigenframes
of atoms $i$ and $j$ provide a maximum to $\bar{\tau}_i^{xy}\bar{\tau}_j^{xy}$.

Similarly it can be shown that in considerations of the following correlation function there appear the terms of the form:
\begin{eqnarray}
\langle\bar{\tau}_i^{xz}\bar{\tau}_j^{xz}\rangle &&\rightarrow \langle\lambda_i^n\lambda_j^m\bar{c}_{i}^{n1}\bar{c}_{i}^{n3}\bar{c}_{j}^{m1}\bar{c}_{j}^{m3}\rangle \nonumber \\
&&=
\tfrac{1}{8}\langle\lambda_i^n\lambda_j^m\sin(2\psi_i^n)\sin(2\psi_j^m)\cos(\delta_{ij}^{nm})\rangle\;.\;\;\;\;
\label{eq;cccc4}
\end{eqnarray}
Note that in (\ref{eq;cccc4}) there is $\cos(\delta_{ij}^{nm})$, while in (\ref{eq;cccc3}) there is $\cos(2\delta_{ij}^{nm})$.
Also:
\begin{eqnarray}
\langle\bar{\sigma}_i^{x}\bar{\sigma}_j^{x}\rangle &&\rightarrow \langle\lambda_i^n\lambda_j^m\bar{c}_{i}^{n1}\bar{c}_{i}^{n1}\bar{c}_{j}^{m1}\bar{c}_{j}^{m1}\rangle\label{eq;cccc5}\\
&&=\tfrac{1}{8}\langle\lambda_i^n\lambda_j^m\sin^2(\psi_i^n)\sin^2(\psi_j^m)\left[2 + \cos(2\delta_{ij}^{nm})\right]\rangle\;.\nonumber\;\;\;\;\\
\langle\bar{\sigma}_i^{x}\bar{\sigma}_j^{y}\rangle &&\rightarrow \langle\lambda_i^n\lambda_j^m\bar{c}_{i}^{n1}\bar{c}_{i}^{n1}\bar{c}_{j}^{m2}\bar{c}_{j}^{m2}\rangle\label{eq;cccc6}\\
&&=\tfrac{1}{8}\langle\lambda_i^n\lambda_j^m\sin^2(\psi_i^n)\sin^2(\psi_j^m)\left[2 - \cos(2\delta_{ij}^{nm})\right]\rangle\;.\nonumber\;\;\;\;\\
\langle\bar{\sigma}_i^{x}\bar{\sigma}_j^{z}\rangle &&\rightarrow \langle\lambda_i^n\lambda_j^m\bar{c}_{i}^{n1}\bar{c}_{i}^{n1}\bar{c}_{j}^{m3}\bar{c}_{j}^{m3}\rangle\label{eq;cccc7}\\
&&=\tfrac{1}{2}\langle\lambda_i^n\lambda_j^m\sin^2(\psi_i^n)\cos^2(\psi_j^m)\rangle\;.\nonumber\;\;\;\;\\
\langle\bar{\sigma}_i^{z}\bar{\sigma}_j^{z}\rangle &&\rightarrow \langle\lambda_i^n\lambda_j^m\bar{c}_{i}^{n3}\bar{c}_{i}^{n3}\bar{c}_{j}^{m3}\bar{c}_{j}^{m3}\rangle\label{eq;cccc8}\\
&&=\langle\lambda_i^n\lambda_j^m\cos^2(\psi_i^n)\cos^2(\psi_j^m)\rangle\;.\nonumber\;\;\;\;
\end{eqnarray}

\subsection{Correlation function $\left<\tau_i^{xy}\tau_j^{xy}\right> \equiv \left<s_{i}^{xy}s_{j}^{xy}\right>$ in an arbitrary frame}

If the values of the stress tensor components are known in the \rijF-frame then
the stress tensor components can be found in any other frame.  
In order to find the stress tensor components in a new frame it is necessary to know the directional cosines
of the axes of the \rijF-frame with respect to the axes of the new frame, i.e., it is necessary to know
the rotation matrix. 
In this subsection we assume the notations ``$A_{ij}$" for this rotation matrix and ``$B_{ij}$" 
for its transpose. Thus: $b_{ij}^{\beta \alpha} \equiv a_{ij}^{\alpha \beta}$.
With the adopted notations, the expressions connecting the stress tensor elements in the new frame with the stress tensor elements in the
directional frame are:
\begin{eqnarray}
\sigma_i^{\alpha \beta} = a_{ij}^{\alpha \gamma} \bar{\sigma}_i^{\gamma \delta} b_{ij}^{\delta \beta}\;,\;\;\;\;
\sigma_j^{\alpha \beta} = a_{ij}^{\alpha \xi} \bar{\sigma}_j^{\xi \zeta} b_{ij}^{\zeta \beta}\;,\;\;\;
\label{eq;Rxi1}
\end{eqnarray}
where the summation over the repeating upper indices is assumed.
Correspondingly, for example:
\begin{eqnarray}
\langle \sigma_i^{xy}\sigma_j^{xy} \rangle  = 
\left(a_{ij}^{x \gamma} a_{ij}^{x \xi}\right)
\langle \left(\bar{\sigma}_i^{\gamma \delta}\bar{\sigma}_j^{\xi \zeta}\right) \rangle
\left(b_{ij}^{\delta y} b_{ij}^{\zeta y}\right)\;.\;\;\;
\label{eq;Rxi2}
\end{eqnarray}
\emph{It is necessary to realize that in isotropic medium the average,
\begin{eqnarray}
\langle \left(\bar{\sigma}_i^{\gamma \delta}\bar{\sigma}_j^{\xi \zeta}\right) \rangle\;\;,\;\;
\label{eq;Rxi3}
\end{eqnarray}
should not depend on the direction of \rij. 
It is only necessary to ensure that the values of the stress tensor components in (\ref{eq;Rxi3}) are 
associated with the directional coordinate frame whose \zax -axis is along \rij.} 

Therefore, as follows from (\ref{eq;Rxi2}), if the values of the correlation functions between different 
stress tensor components are known in the directional frame,
then the values of the correlation functions in any other frame also can be found. 
Note that the values of the correlation functions in the new rotated frame depend 
on the direction of \rij with respect to the new rotated frame.

\subsection{Correlation function invariants}

It follows from the previous considerations that the value of the product
$\tau_i^{xy}\tau_j^{xy} \equiv \sigma_i^{xy}\sigma_j^{xy}$ depends on the orientation of the observation coordinate
frame with respect to the direction of \rij  [see formulas (\ref{eq;Rxi1},\ref{eq;Rxi2})]. 
Therefore it is reasonable to ask what is the 
value of $\tau_i^{xy}\tau_j^{xy}$ averaged over all
possible orientations of the observation coordinate frame.
This average value, expressed in terms of the stress components in a particular frame,
should be rotationally invariant. In an isotropic medium the averaging over all possible 
directions of the observation frame is equivalent to the averaging over all possible
orientations of a ``rigid dumbbell" associated with the eigenframes of 
atoms $i$ and $j$ connected by \rij.

The details of the derivation are given in Appendix \ref{app:angle-averaging}.
The final answer for the value of $\tau_i^{xy}\tau_j^{xy}$  averaged over all directions
of the observation frame is:
\begin{eqnarray}
\left<\tau_i^{xy}\tau_j^{xy}\right>_{\Omega} = G_{ij}^1 + G_{ij}^2 + G_{ij}^3\;\;\;,\;\;\;\;\;
\label{eq;avesixysjxy1}
\end{eqnarray}
where
\begin{eqnarray}
&&G_{ij}^1=-\left(\tfrac{3}{10}\right)p_i p_j,\;\;\;\;\;\;\label{eq;avesixysjxy21}\\
&&G_{ij}^2 =\tfrac{1}{10}\left[ \sigma_i^{x}\sigma_j^{x}+\sigma_i^{y}\sigma_j^{y}+\sigma_i^{z}\sigma_j^{z} \right],\;\;\label{eq;avesixysjxy22}\\
&&G_{ij}^3 =\tfrac{1}{5}\left[\tau_i^{xy}\tau_j^{xy} + \tau_i^{xz}\tau_j^{xz} +\tau_i^{yz}\tau_j^{yz}\right],\;\;\;\;\;
\label{eq;avesixysjxy2}
\end{eqnarray}
and $p_i$ is given by (\ref{eq:to01}). 
Note that, by construction, the sum $G_{ij}^1+G_{ij}^2+G_{ij}^3$ is rotationally invariant. On the other hand,
$p_i$ and $p_j$ are by themselves rotationally invariant. Thus we have to conclude that the sum
$G_{ij}^2 + G_{ij}^3$ is rotationally invariant. 
It is not difficult to realize that the value of the sum $G_{ij}^2 + G_{ij}^3$ should depend on
the eigenvalues and also on the mutual orientations of the eigenvectors (eigenframes) of atoms $i$ and $j$.

Let us evaluate the value of the sum $G_{ij}^2 + G_{ij}^3$ in the eigenframe of atom $i$.
It is assumed that the directional cosines of all eigenvectors of atom $j$ with
respect to all eigenvectors of atom $i$ are known. 
For the evaluation it is necessary to rotate the diagonal stress tensor of atom $j$ 
in its own eigenframe into the eigenframe of atom $i$. 
This rotation is described by formulas which are totally analogous
to the formulas (\ref{eq;sijxx},\ref{eq;sijyy},\ref{eq;sijzz},\ref{eq;sijxy},\ref{eq;sijxz},\ref{eq;sijyz}).
In the eigenframe of atom $i$ we have $G_{ij}^3=0$ because in this frame $\tau_i^{xy}=0$, $\tau_i^{xz}=0$ and $\tau_i^{yz}=0$.
Thus, using (\ref{eq;sijxx},\ref{eq;sijyy},\ref{eq;sijzz},\ref{eq;sijxy},\ref{eq;sijxz},\ref{eq;sijyz}), we get:
\begin{eqnarray}
G_{ij}^2 = \left(\tfrac{1}{10}\right)\sum_{n=1}^{n=3}\sum_{m=1}^{m=3}\lambda_i^{n}\lambda_j^{m}\left(c_{ij}^{nm}\right)^2\;,\;
\label{eq:eigenGij201}
\end{eqnarray}
where $c_{ij}^{nm}$ is the cosine between the $n$-th eigenvector of atom $i$ and $m$-th eigenvector of atom $j$.

From the previous considerations in this section, it follows that:
\begin{eqnarray}
\left<\tau_i^{xy}\tau_j^{xy}\right>_{\Omega} =-\left(\tfrac{3}{10}\right)p_i p_j + 
\left(\tfrac{1}{10}\right)\sum_{n,m=1}^{n,m=3}\lambda_i^{n}\lambda_j^{m}\left(c_{ij}^{nm}\right)^2\;.\;\;\;\;\;\;\;
\label{eq;avesixysjxy3}
\end{eqnarray}
It is obvious that expression (\ref{eq;avesixysjxy3}) is rotationally invariant, as it depends only on rotation-invariant
parameters. Note also that expression (\ref{eq;avesixysjxy3}) does not depend on the direction of \rij.
In order to get some more insight into the meaning of expression (\ref{eq;avesixysjxy3}) let us imagine that all
eigenvalues are equal to 1 (one). In this case $p_i = p_j = 1$. 
It is also easy to realize that the sum over the squares of all cosines in the second term
should be equal to 3, as this sum, in this case, is just the sum of the lengths of the three unit eigenvectors. 
Thus, in the case when all eigenvalues are equal, expression (\ref{eq;avesixysjxy3}) is equal to zero.
It is also not difficult to see that if all eigenvalues of just one atom are equal to each other then
expression (\ref{eq;avesixysjxy3}) is also equal to zero. 
Let us also consider the orientational ordering of the eigenframes of atoms $i$ and $j$.
It is clear that the sum of the squares of the directional cosines of any eigenvector of atom $j$ with respect to the
eigenframe of atom $i$ is equal to 1. Thus, if there is no orientational ordering between the eigenframes, 
it is reasonable to assume that the average value of the square of every directional cosine in (\ref{eq;avesixysjxy3})
is $1/3$. If we assume that the square of every directional cosine in (\ref{eq;avesixysjxy3}) is equal to $1/3$ then
we find that $\left<\tau_i^{xy}\tau_j^{xy}\right>_{\Omega}=0$. \emph{Thus we come to the conclusion that 
$\left<\tau_i^{xy}\tau_j^{xy}\right>_{\Omega}$ is not equal to zero only if the stress ellipsoids of atoms $i$ and $j$ 
both have shear distortions and also if there is orientational ordering between the
eigenframes of atoms $i$ and $j$.}

It is of interest to compare expression (\ref{eq;avesixysjxy3}) with expressions (\ref{eq;cccc3},\ref{eq;cccc31}).
See also the second paragraph after (\ref{eq;cccc3},\ref{eq;cccc31}). It is easy to see that expression (\ref{eq;avesixysjxy3})
also suggests that similar orientations of the eigenframes of atoms $i$ and $j$ provide 
a maxim to $\tau_i^{xy}\tau_j^{xy}$.

It also can be shown that:
\begin{eqnarray} 
(1/4)\left<\left(\sigma_i^{xx}-\sigma_i^{yy}\right)
\left(s_j^{xx}-s_j^{yy}\right)\right>_{\Omega}
=\left<\tau_i^{xy}\tau_j^{xy}\right>_{\Omega}.
\end{eqnarray}

On the other hand, the averaging over some other combinations leads to zero. For example:
\begin{eqnarray}
&&\left<\tau_i^{xy}\left(s_j^{xx}-s_j^{yy}\right)\right>_{\Omega}=0\;,\;\;\; 
\left<\tau_i^{xz}\left(s_j^{xx}-s_j^{yy}\right)\right>_{\Omega}=0\;,\;\;\;\nonumber\\
&&\left<p_i\left(s_j^{xx}-\sigma_j^{yy}\right)\right>_{\Omega}=0\;,\;\;\; \left<p_i\tau_j^{xy}\right>_{\Omega}=0.\nonumber
\end{eqnarray} 

\begin{figure}
\begin{center}
\includegraphics[angle=0,width=3.0in]{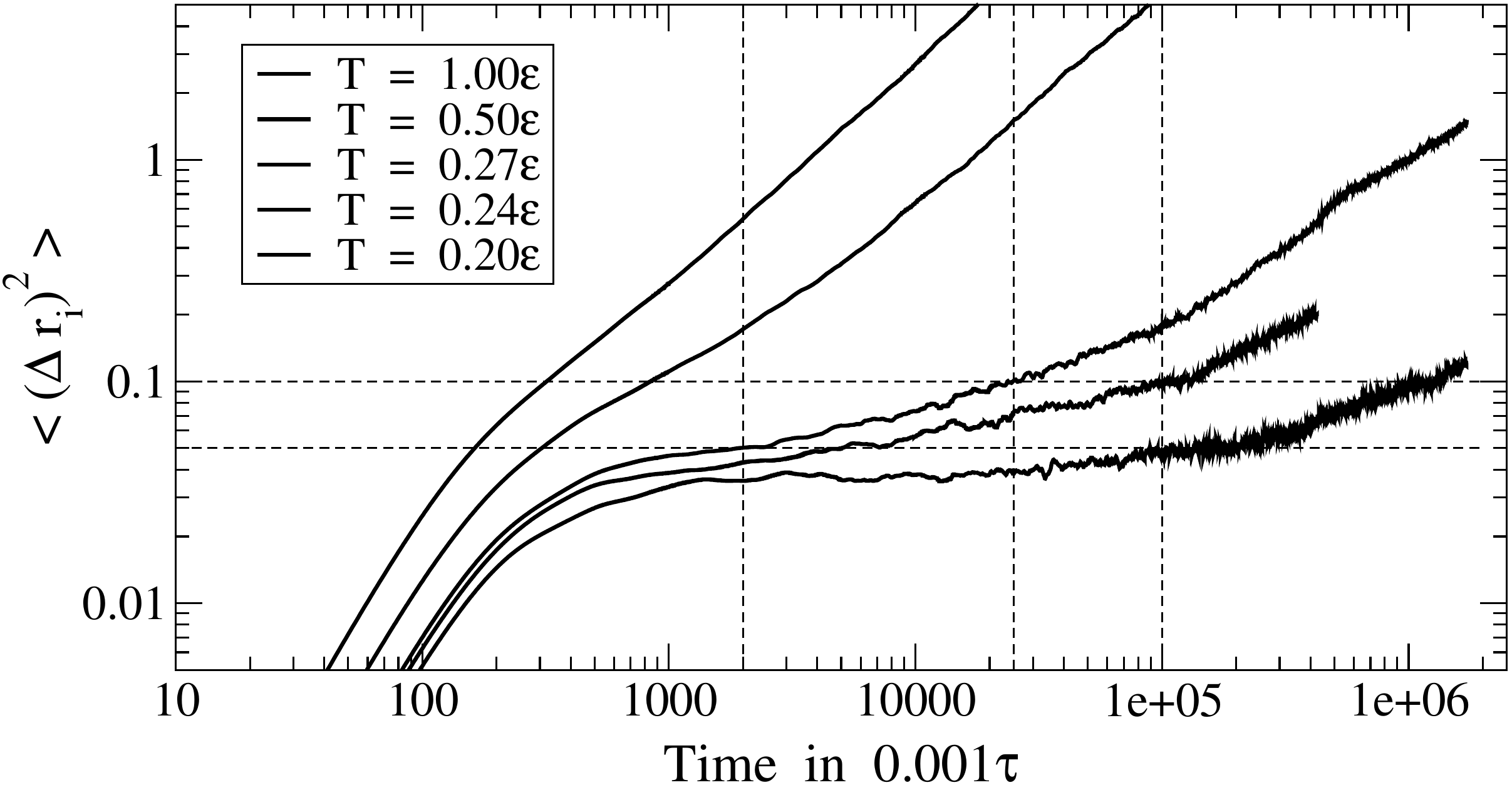}
\caption{The dependencies of the mean square particles' displacements 
on time for five studied temperatures. 
The data were collected on the system of 5324 particles at the same number 
density that was used in simulations of the large system.
}\label{fig:msd-vs-time}
\end{center}
\end{figure}

\section{Results of Simulations \label{sec:ressim}}

\subsection{Studied system}
  
We studied a system consisting of $50\%$ of particles A and $50\%$ of particles $B$. 
The particles interact through the pairwise repulsive potential:
\begin{eqnarray} 
U_{ab}(r_{ij})=\epsilon\left(\frac{\sigma_{ab}}{r_{ij}}\right)^{12} \;.\;\;\;\;
\label{eq:mdpotential}
\end{eqnarray} 
In (\ref{eq:mdpotential}) $\sigma_{ab}$ is the length that determines the characteristic 
interaction range. The indices $a$ and $b$ stand for the types of particles: $A$ or $B$.
In the following we measure the temperature, $T$, in units of $\epsilon$.
Further:
\begin{eqnarray} 
&&\sigma_{AA} = 1.0\;,\;\sigma_{BB} = 1.2\;,\;\sigma_{AB}=\frac{\sigma_{AA}+\sigma_{BB}}{2}=1.1\;,\label{eq:mdpotential2}\;\;\;\;\;\;\;\\
&&m_{B} = 2m_{A}\;,\;\tau =\sqrt{\frac{m_A \sigma_{AA}^2}{\epsilon}}\;,\;\;N_A=N_B=31250\;,\label{eq:mdpotential3}\;\;\;\\
&&L_x=L_y=L_z=42.7494\sigma_{AA}\;,\;\;\rho_o = 0.80/\sigma_{AA}^{3}\;,\label{eq:mdpotential4}
\end{eqnarray} 
where $m_{A}$ and $m_{B}$ are the masses of particles $A$ and $B$.
The time unit is $\tau$. 
$N_A$ and $N_B$ are the numbers of particles. 
The lengths of the sides of the cubic simulation box are $L_x,L_y,L_z$.
The number density is $\rho_o$. Periodic boundary conditions were used.

This system of particles was extensively studied 
previously \cite{Hansen19881,Miyagawa19911,Mizuno20101,Mizuno20111,Matsuoka2012,Mizuno20131}.
However, this model was not studied previously from the perspective of atomic level stresses.

We used LAMMPS molecular dynamics (MD) package in our simulations \cite{Plimpton1995,lammps}.
Initial particles' configuration was created as FCC lattice with alternating planes
of $A$ and $B$ particles. 
The system was melted and equilibrated at temperature $T = 2$ in the NVT ensemble.
The equilibration was controlled by the absence of change in the average value of potential energy and 
by the absence of change in the partial pair density functions. 
Equilibration is achieved when particles are well mixed. 
Further we reduced the temperature in the NVT ensemble to $T=1.5$ and again equilibrated the system.
Then we reduced the temperature to $T=1$. 
After the equilibration we switched to the NVE ensemle. 
After the equlibration we collected structural configurations.
Similar algorithm was used to collect configurations at 
temperatures $T=0.5$ and $T=0.27$.
We also produced inherent structures by applying conjugate gradient relaxation 
to the structures (restart files) collected at the temperatures $T=0.27$ and $T=1$.

\subsection{Mean square particles' displacement and the partial pair density correlation functions}

Figure \ref{fig:msd-vs-time} shows how  the mean square particles' displacement, \emph{msd}, 
depends on time at different temperatures. 
The \emph{msd}s for Fig.\ref{fig:msd-vs-time}
were calculated without making the distinction between the particles of type ``A" and ``B".
The purpose of Fig.\ref{fig:msd-vs-time} is to remind about the characteristic 
temperature scales \cite{Hansen19881,Miyagawa19911,Mizuno20101,Mizuno20111,Matsuoka2012,Mizuno20131}.

Panels (a,b,c) of Fig.\ref{fig:pdfone} show the partial pair density functions, \ppdfs,\; at temperatures
$T=1$, $T=0.5$, $T=0.27$, and the \ppdfs\; calculated on the inherent structures. 
As temperature is reduces from $T=1$ to $T=0.27$ the \ppdfs\; do not show pronounced changes.
The famous splitting of the second peak \cite{EMa20111,Pan20111} becomes noticeable, 
but it is not pronounced in comparison
with the well expressed splitting observed on the inherent structures.
Another thing to notice is that the \ppdfs\; for the pairs of particles of different types exhibit
qualitatively similar behaviour at all temperatures. 
Figure \ref{fig:pdfone} will be useful in the following as it shows the characteristic length scales, such
as the positions of the first peak, first minimum, second peak, and the position where the splitting of the second
peak occurs. 
It also will be useful because it shows the amplitudes of the changes on the \emph{y}-axis on decrease of temperature.

\subsection{Distributions of the Atomic Level Stresses \label{sec:dals}}

Using formula (\ref{eq;stressdef1}) the Cartesian components of the atomic stress tensor of every atom can be calculated
if the atomic configuration is known. Then, using formulas (\ref{eq:to01},\ref{eq:to02},\ref{eq:to03},\ref{eq:to04}) 
for every atom, the \ssc\, of the atomic stress tensors can be obtained.
We performed these calculations for several temperatures and obtained the 
probability distributions (\emph{PDs}) of the \ssc\,
by averaging over all relevant atoms and over 100 different configurations for every temperature. 

\begin{figure}
\begin{center}
\includegraphics[angle=0,width=3.3in]{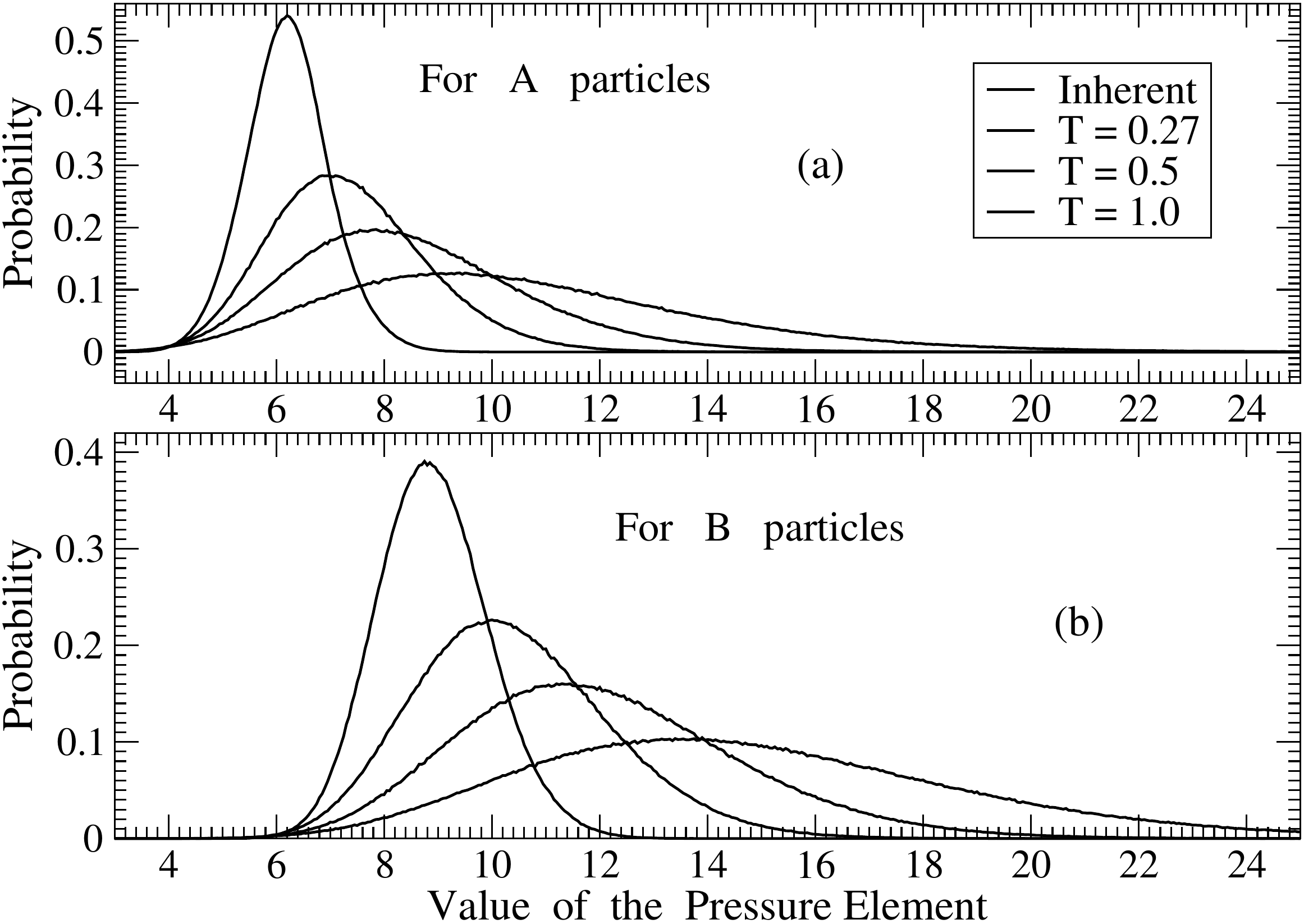}
\caption{The \pds of the pressure elements 
for the particles of type ``A" in panel (a) and for the particles of type ``B" in panel (b) for different temperatures.
The lower the temperature the more pronounced is the peak in the $PD$.
At lower temperatures the average pressure is smaller. 
Thus the peaks corresponding to the lower temperatures
are located to the left with respect to the higher temperature peaks.
}\label{fig:pressure-distributions-AB-1x}
\end{center}
\end{figure}

The \pds of the atomic pressure, $p_{i} \equiv s_{0,i}$, for ``A" and ``B" particles are
shown in panels (a) and (b) of Fig.\ref{fig:pressure-distributions-AB-1x}. 
The finite widths of the \pds of the pressure calculated on inherent structures, 
i.e., at $T=0$, are caused by the structural disorder only.
At non-zero temperatures there are structural and vibrational contributions to the \pds of  
the pressure. Since in the studied system particles interact through the purely repulsive potentials
the pressure on every atom has to be positive. As temperature increases the average pressure also increases.
The widths of the \pds also increase with increase of temperature. 

Panel (b) shows that the average pressure on ``B"-particles is larger than the average pressure on ``A" particles.
Note that, according to the definition (\ref{eq;stressdef1}), the contribution of every neighbour to the
\emph{diagonal} stress tensor components is always positive for the purely repulsive potentials.
Therefore the pressure, which is proportional to the sum of the diagonal components, 
on average becomes proportional to the average number of the neighbours. 
Thus, since larger atoms on average have more neighbours, 
they also tend to have larger pressure. These considerations, however, do not take
into account the fact that larger atoms also have larger atomic volume.
If the difference in the atomic volumes is taken into account then the atoms of both types
should have similar values of the pressure.
Thus, the discrepancy in the values of the atomic pressure 
in Fig.\ref{fig:pressure-distributions-AB-1x} is caused by the identical 
values of the atomic volume $\langle V_i \rangle$ 
used for both types of particles in (\ref{eq;stressdef1}). 

It is possible to introduce artificially the atomic volume which would account
for the difference in sizes between ``A" and ``B" particles \cite{Egami19802,Egami19821,Chen19881,Levashov2008B}. 
This should lead to the similar values of the average
atomic pressure for both types of particles. Note that at $T=0.27$ the maximum in the \pd of pressure
for ``A"-particles occurs at $s_{0,i}^{max}(A) \approx 7$, while for ``B"-particles at $s_{0,i}^{max}(B) \approx 10$.
Thus  $s_{0,i}^{max}(B)/s_{0,i}^{max}(A) \approx 10/7 \approx 1.43$. 
In order to have the same pressure on both types of particles it is necessary to assume 
that the volume of ``B"-particle
is $\approx 1.43$ times larger than the volume of ``A"-particle. 
This means that the radius of ``B"-particles should be 
$\approx 1.13$ times larger than the radius of ``A"-particles. 
This result can be compared with (\ref{eq:mdpotential2}).
In our present considerations we use the same value 
of $\langle V_i \rangle = 1/\rho_o$ for both types of particles.
We prefer do not use the atom-dependent atomic volume because it is not present
in the Green-Kubo formula for viscosity \cite{Levashov20111,Levashov2013}.  

The \pds of the Cartesian components $\sigma_{i}^{xx}$, $\sigma_{i}^{yy}$, and $\sigma_{i}^{zz}$ 
for ``B"-particles at $T=1$ are shown in panel (a) of Fig.\ref{fig:components-lambdas-distributions-1x0} 
with the blue curves (these curves coincide into one).
The \pds of the Cartesian components $\sigma_{i}^{xx}$, $\sigma_{i}^{yy}$, and $\sigma_{i}^{zz}$ 
for ``A"-particles at $T=0.27$ are shown in panel (a) of 
Fig.\ref{fig:components-lambdas-distributions-0x27} with the blue curves.
\nobreak
\begin{figure}
\begin{center}
\includegraphics[angle=0,width=3.3in]{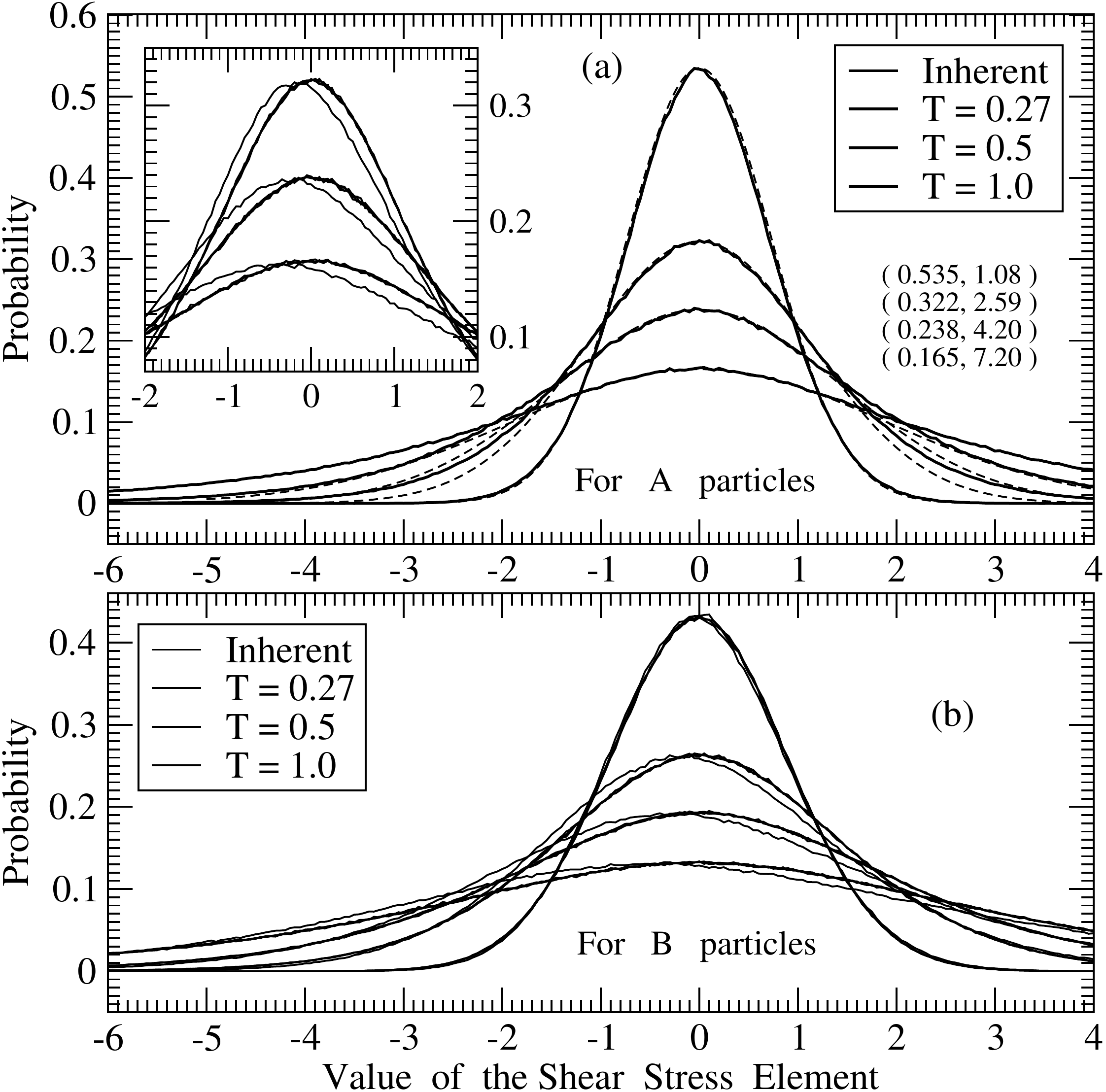}
\caption{Panel (a): The solid curves in the main plot show 
the $PDs$ of the \ssc\; (except $s_{5,i}$) for ``A" particles
at different temperatures. See formulas (\ref{eq:to02},\ref{eq:to03},\ref{eq:to04}). 
The lower the temperature the narrower and taller are the corresponding peaks.
The dashed curves are the Gaussian functions with which \emph{the peaks} of the \pds
are fitted. 
The numbers in brackets show the parameters  $a$ and $b$ of the fitting 
functions: $f(s) = a\exp\left[-(s^2/b^2)\right]$.
Since the tails of the \pds are not well fitted, 
while the peaks are fitted, it is clear that the \pds for 
the \ssc\; are not Gaussians.
The inset shows the peaks of the \pds for $\sigma^{xy}_i$, $\sigma^{xz}_i$, $\sigma^{yz}_i$, 
$\tfrac{1}{2}(\sigma^{xx}_i - \sigma^{yy}_i)$, and 
$\tfrac{1}{\sqrt{3}}\left[\sigma^{zz}_i-\tfrac{1}{2}(\sigma^{xx}_i + \sigma^{yy}_i)\right]$
at the temperatures $T=1.0$, $T=0.5$, and $T=0.27$. 
Note that the \pds 
of $\tfrac{1}{\sqrt{3}}\left[\sigma^{zz}_i-\tfrac{1}{2}(\sigma^{xx}_i + \sigma^{yy}_i)\right]$ are noticeably
different from the other \pds. 
In our view, this difference indicates that different \ssc\, are not (completely) independent.
Panel (b): The \pds of the \ssc\; at different temperatures for ``B" particles.
Note that the \pds of the $\tfrac{1}{\sqrt{3}}\left[\sigma^{zz}_i-\tfrac{1}{2}(\sigma^{xx}_i + \sigma^{yy}_i)\right]$
component are noticeably different from the \pds of the other components. 
Also note that this difference is nearly absent in the curves obtained from the inherent structures.
}\label{fig:shear-stresses-distributions-AB-1}
\end{center}
\end{figure}

Figure \ref{fig:shear-stresses-distributions-AB-1} shows the \pds of 
the \emph{shear} \ssc\; for ``A" and ``B" particles at different temperatures. 
The means of these \pds naturally occur at zero value.
The widths of the \pds calculated on the inherent structures originate from the structural disorder only. 
As temperature increases the \pds become wider. 
At non-zero temperatures there are structural and vibrational contributions to the shear 
\ssc, as for the pressure.
Note that the \pds for ``B" particles are wider than for ``A"-particles. 
They are wider because particles of ``B"-type on average have more neighbours than ``A"-particles and thus the
shear \ssc\; for ``B"-particles fluctuate in a wider range than the stress components for ``A"-particles.
This difference can be taken into account by assuming that ``A" and ``B" particles have 
different atomic volumes, $\langle V_i \rangle$. 
However, in the present paper we use the same value of the atomic volume in (\ref{eq;stressdef1}).
The dashed curves in panel (a) show the Gaussians  
whose parameters were adjusted to fit \emph{the peaks} of the \pds obtained in MD simulations.
It is clear that these fits do not match the tails of the \pds obtained in MD simulations.
Thus, the \pds of the shear stresses are not described well by the Gaussian functions. 

The inset in panel (a) of Fig.\ref{fig:shear-stresses-distributions-AB-1} shows that the \pds of the shear
component $(1/\sqrt{3})\left[\sigma_{i}^{zz} - \tfrac{1}{2}\left(\sigma_i^{xx}+\sigma_i^{yy}\right)\right]$ are noticeably different
from the distributions of the other shear stress components. 
This effect is present for ``A" and ``B" particles, as follows from the 
comparison of the inset in panel (a) with the panel (b) itself. 
We interpret this difference as an indication that different \ssc\; 
are not completely independent from each other.
This issue is discussed more in the following.

\subsection{Eigenvalues of the atomic stress matrices and correlations between the eigenvalues}   

The real and symmetric $ 3\times 3$ matrix of the atomic stress tensor (\ref{eq;stressdef1}) 
can be diagonalized and its eigenvalues and eigenvectors can be found \cite{KoppJ20081}.
For purely repulsive potentials all eigenvalues should be positive. This can be demonstrated as follows.
Let us assume that $\bf{b}$ is an arbitrary column vector, while $\bf{b}^T$ is the transpose of $\bf{b}$, 
i.e., the raw-vector.
Let us consider the inner product:
\begin{eqnarray}
\bm{b}^T \Sigma_i \bm{b}&&= \sum_{\alpha \beta} b^{\alpha}\sigma_{i}^{\alpha \beta}b^{\beta}=
\sum_{\alpha \beta} b^{\alpha}\left(-\sum_j \frac{\partial U}{\partial r_{ij}}\frac{r_{ij}^{\alpha}r_{ij}^{\beta}}{r_{ij}}\right)b^{\beta}\;\;\;\;\;\;\;\;\nonumber\\ 
&&=
-\sum_j \left(\frac{\partial U}{\partial r_{ij}}\right)\frac{\left(\bm{b}\bm{r}_{ij}\right)^2}{r_{ij}}\;.\;\;\label{eq:posdefinite}
\end{eqnarray} 
For a purely repulsive potential the derivative of the potential is negative and thus (\ref{eq:posdefinite}) is positive.
Thus the atomic stress matrices for the purely repulsive potentials are \emph{positive-definite matrices}.
Let us now assume that vector $\bm{b}$ is not an arbitrary vector, but an eigenvector of the matrix $\Sigma_{i}$. 
In this case we get: $\bm{b}^T \Sigma_i \bm{b}=\lambda (\bm{b}\bm{b})=\lambda b^2$. 
The comparison of this result with (\ref{eq:posdefinite}) leads to the conclusion that all eigenvalues, $\lambda$, should be positive.

Thus the atomic stress matrix for every particle has 3 positive eigenvalues $\lambda_1$, $\lambda_2$, and $\lambda_3$ which describe the geometry of the local atomic environment. 
Further we assume that the eigenvalues are ordered: $\lambda_1 \ge \lambda_2 \ge \lambda_3$. 

\begin{figure}
\begin{center}
\includegraphics[angle=0,width=3.3in]{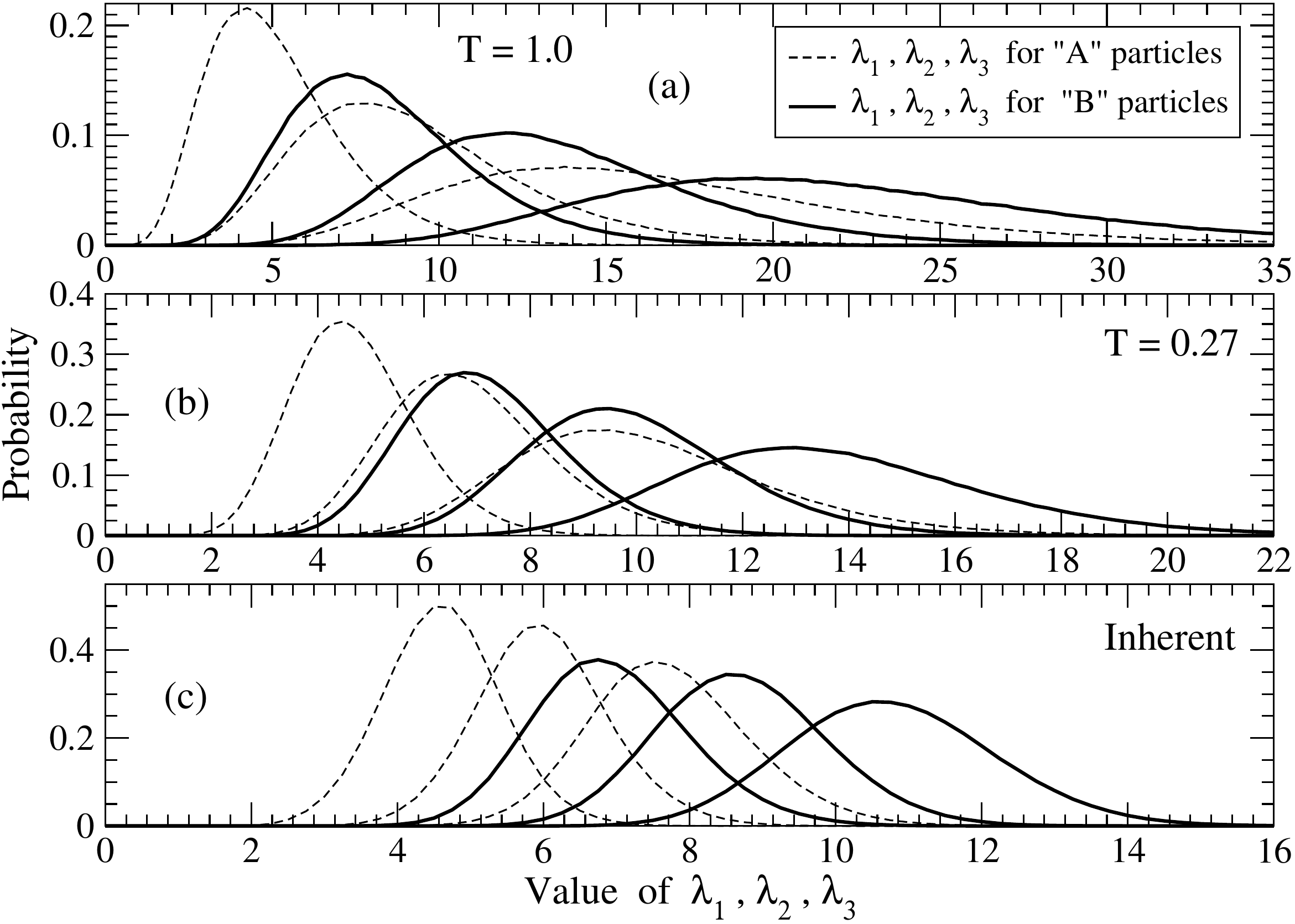}
\caption{The \pds of the eigenvalues for particles of type ``A" and ``B" at 
temperatures $T=1.0$ (panel (a)), $T=0.27$ (panel (b)), 
and in the inherent structures obtained from the structures at $T=0.27$ (panel (c)).
Note that the scales on the axes in all panels are different.
Since, for every particle, $\lambda_1 \geq \lambda_2 \geq \lambda_3$ 
the peaks of the \pds corresponding 
to the larger eigenvalues are located to the right with respect to the peaks of 
the curves corresponding to the smaller eigenvalues.
}\label{fig:lambdas-D-AB3T-1}
\end{center}
\end{figure}
Figure \ref{fig:lambdas-D-AB3T-1} shows the \pds of the ordered eigenvalues 
for ``A" and ``B" particles at different temperatures.
As temperature decreases the eigenvalues, in general, become smaller and their \pds become narrower. 
This is the expectable behaviour. 

\begin{figure}
\begin{center}
\includegraphics[angle=0,width=3.3in]{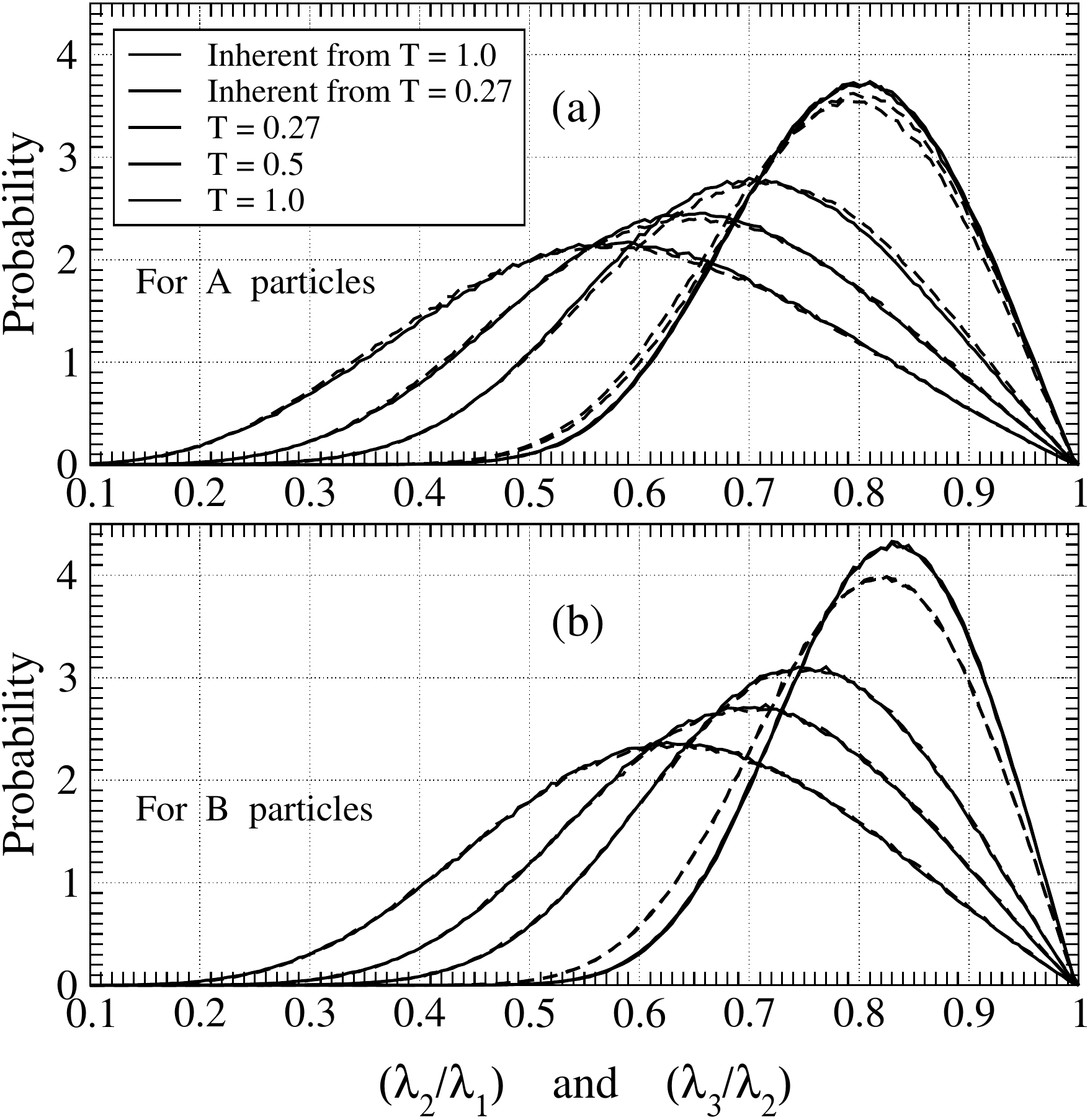}
\caption{The solid curves in both panels show the \pds of the ratio $(\lambda_2/\lambda_1)$, while
the dashed curves show the \pds of the ratio $(\lambda_3/\lambda_2)$.
Panel (a) shows the results for ``A" particles at different temperatures, while panel (b) shows the results for ``B" particles.
Note from panel (a) that the \pds of \lto and \ltt
for the particles of type ``A" essentially coincide at all temperatures in the liquid state. 
On the other hand, in the inherent states, there appears a more noticeable difference 
in the \pds of \lto and \ltt.
In panel (b) the differences in the \pds of \lto and \ltt 
are even smaller than in panel (a) for the temperatures corresponding to the liquid states. 
On the other hand, in the inherent states, the difference between the two \pds becomes quite significant.
The coincidence of the \pds of \lto and \ltt is not a trivial point. 
As we demonstrate further this coincidence is not a general property. 
Note also that all curves arrive to the point $(\lambda_2/\lambda_1)=(\lambda_3/\lambda_2)=1$ in a linear fashion.
See Appendix (\ref{sec:app-l2overl1}).
}\label{fig:lambda-distr-3}
\end{center}
\end{figure}

Correlations between the eigenvalues of the same atomic stress tensor 
were studied in Ref.\cite{Kust2003a,Kust2003b}
for several model systems, but not for the system that we study.
It was demonstrated that there are correlations between the eigenvalues.
However, the real understanding of the nature of these correlations has not been achieved.
Thus we decided to elaborate further on these correlations.
In particular, we considered the \pds of 
the ratios \lto and \ltt. 
These \pds are shown in Fig.\ref{fig:lambda-distr-3}.
\begin{figure}
\begin{center}
\includegraphics[angle=0,width=3.3in]{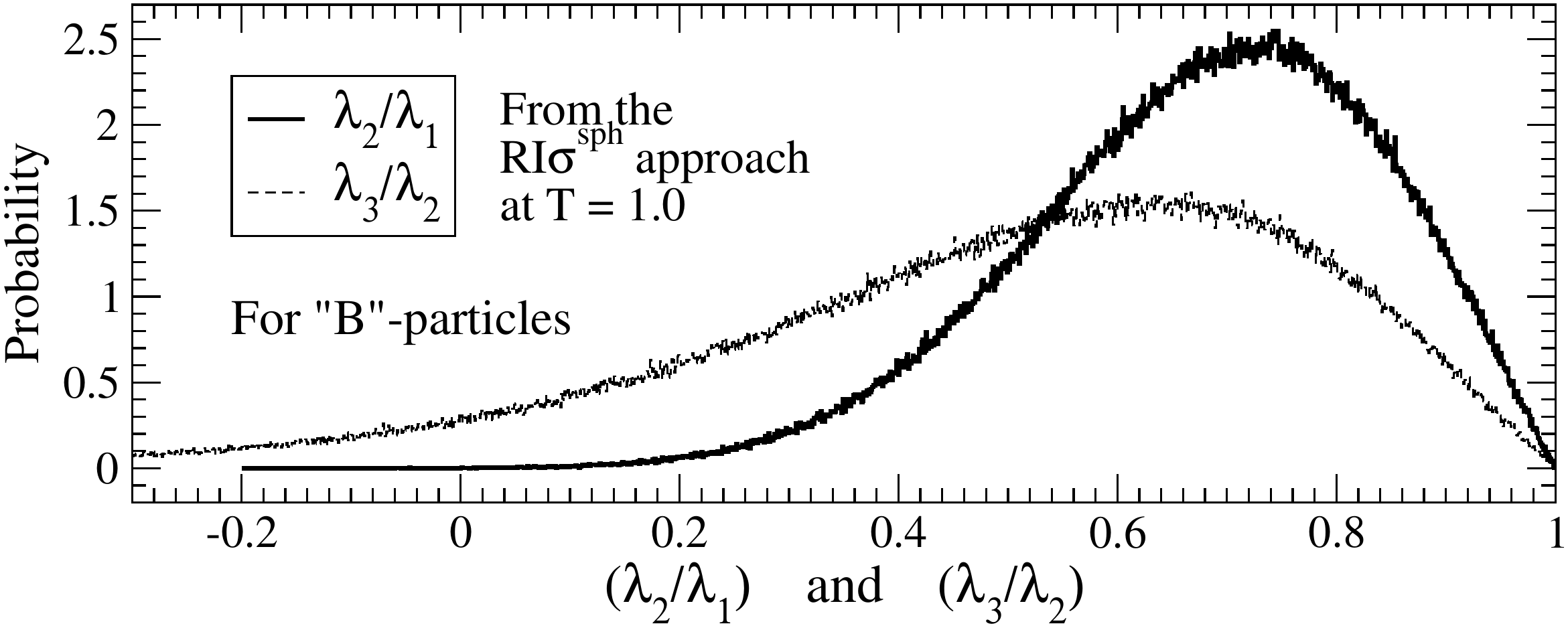}
\caption{The \pds of \lto and \ltt obtained 
using the \ris approach for ``B" particles at $T=1.0$. 
Thus, in the random and independent case for the spherical stresses, there is no coincidence in the \pds
of \lto and $(\lambda_3/\lambda_2)$, while both distributions obtained directly from MD simulations  
are essentially the same, as shown in Fig.\ref{fig:lambda-distr-3}(b).
}\label{fig:components-lambdas-distributions-2}
\end{center}
\end{figure}

Note in Fig.\ref{fig:lambda-distr-3} that the \pds 
of $\left(\lambda_2/\lambda_1\right)$ and $\left(\lambda_3/\lambda_2\right)$ are
very similar to each other for ``A" and ``B" particles, if the system is in a liquid state. 
The \pds of $\left(\lambda_2/\lambda_1\right)$ 
and $\left(\lambda_3/\lambda_2\right)$ for ``B"-particles are essentially identical to each other.
This finding is not expectable, and, in our view, it is rather surprising. 
The similarity in these \pds can not be a general property, 
as a difference between the \pds can be observed in the results obtained on the inherent structures.
For ``A"-particles the difference in the \pds obtained on the inherent structures is noticeable, 
while the results for ``B"-particles show a very clear difference.

The fact that the coincidence of the \pds of \lto and \ltt is something
unusual can also be demonstrated using the \ris approach described in section \ref{sec:random-reference}.
Thus Fig.\ref{fig:components-lambdas-distributions-2} shows the \pds of \lto and \ltt
calculated using the \ris approach on the \pds of the \ssc\, corresponding to ``B" particles
at $T=1$. We see in Fig.\ref{fig:components-lambdas-distributions-2} that the \pds of 
\lto and \ltt from the \ris approach are completely different, 
while on MD data, as Fig.\ref{fig:lambda-distr-3}(b) shows, they are simply identical.

\begin{figure}
\begin{center}
\includegraphics[angle=0,width=2.6in]{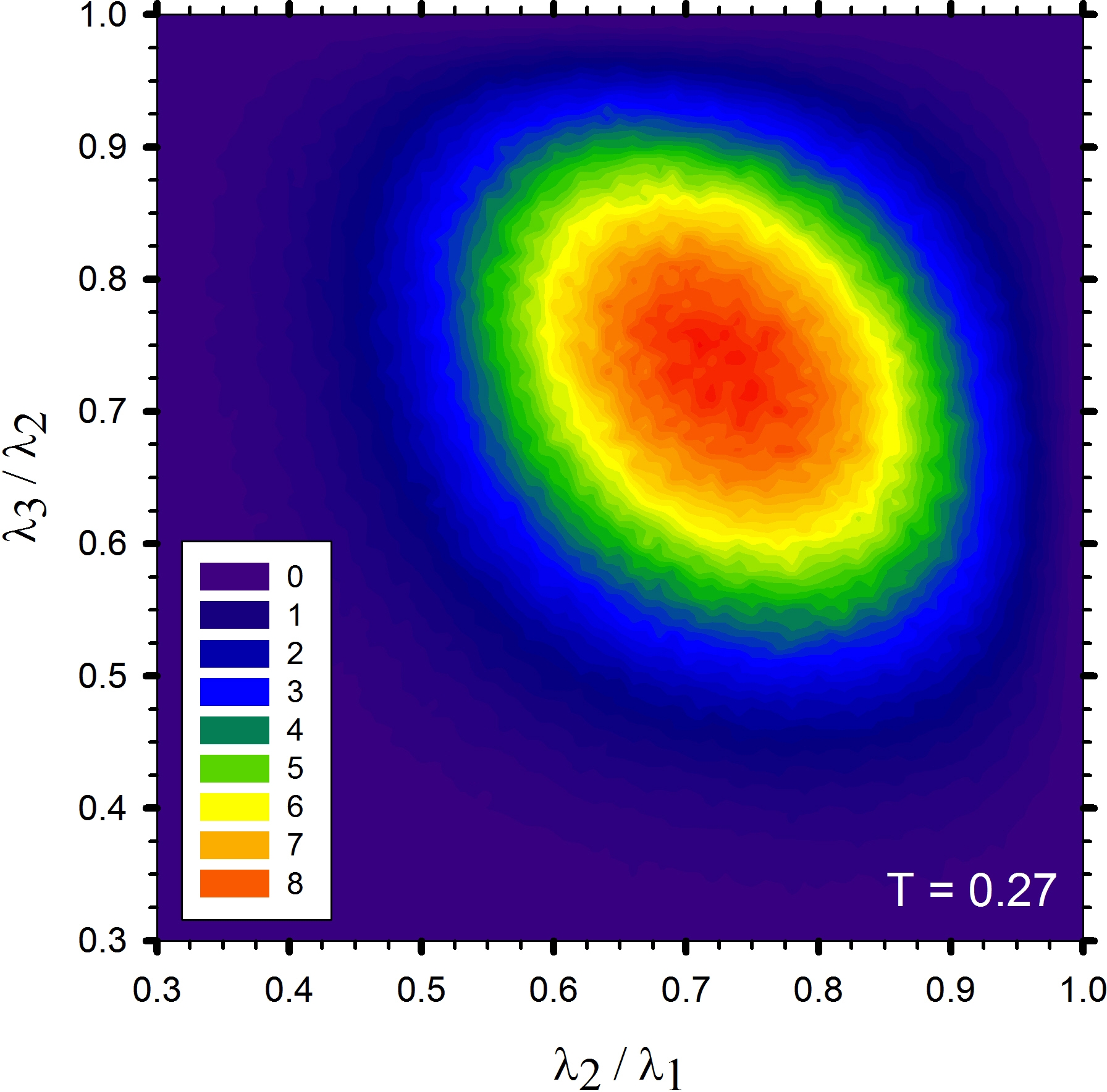}
\caption{2D contour plot of the \pd for the ratios 
of eigenvalues $[(\lambda_2/\lambda_1),(\lambda_3/\lambda_2)]$ 
for ``A" particles at $T=0.27$.
The amplitude of the probability (on $z$-axis) was multiplied by 10000.
Note that the distribution appears to be symmetric with respect to the diagonal
``$(\lambda_2/\lambda_1)=(\lambda_3/\lambda_2)$". 
This point is verified in Fig.\ref{fig:lambdas-2D-cuts}. 
This means that the probability function 
$W\left[(\lambda_2/\lambda_1),(\lambda_3/\lambda_2)\right]$ is symmetric
with respect to the interchange of its 
arguments: $W\left(x_1,x_2\right)=W\left(x_2,x_1\right)$.
We observed similarly symmetric distributions for the other temperatures in 
the liquid states for the particles of both types.
}\label{fig:lambdas-2D-0x27}
\end{center}
\end{figure}

\begin{figure}
\begin{center}
\includegraphics[angle=0,width=3.3in]{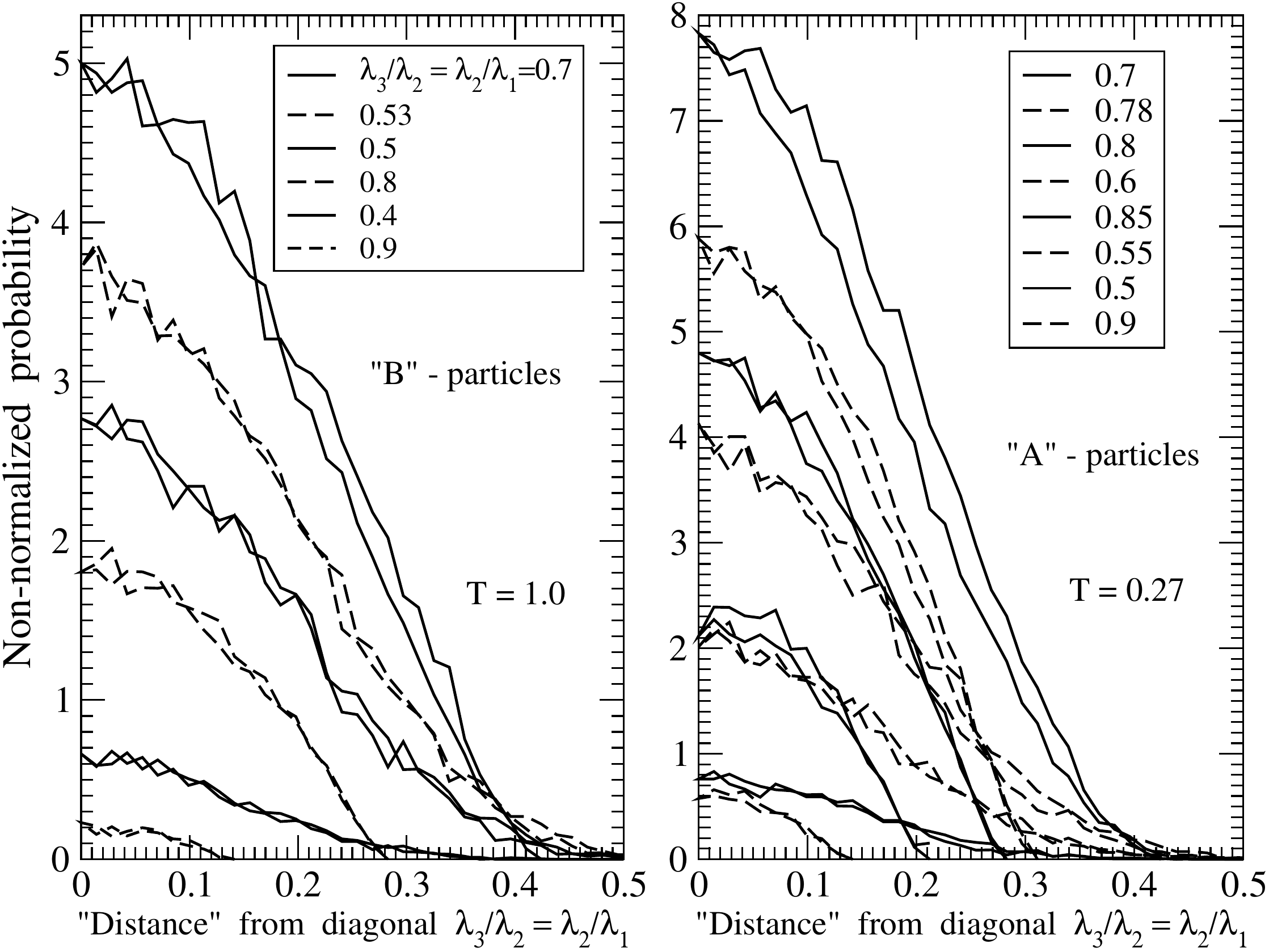}
\caption{The cuts of the \emph{PDs}, analogous to those 
in Fig.\ref{fig:lambdas-2D-0x27}, along the lines orthogonal
to the diagonal ``$(\lambda_2/\lambda_1)=(\lambda_3/\lambda_2)$".
Such lines intersect the diagonal 
at points $(\xi,\xi)$, where $\xi = (\lambda_2/\lambda_1)=(\lambda_3/\lambda_2)$.
The values of $\xi$ for the intersection points are shown as legends in the panels.
The legends from top to the bottom correspond to the curves from top to the bottom.
The distances from the intersection points along the orthogonal cuts are shown on the $x$-axis.
Two curves for every intersection point correspond to the two directions from the diagonal (North-West and South-East).
Thus the figure shows that the probability distributions $W\left(x_1,x_2\right)$ are indeed quite symmetric 
with respect to the interchange of its arguments.
}\label{fig:lambdas-2D-cuts}
\end{center}
\end{figure}

In order to get further insight into the \pds of $\left(\lambda_2/\lambda_1\right)$ and $\left(\lambda_3/\lambda_2\right)$
we considered the \pd of the pair of values $\left(\lambda_2/\lambda_1\right)$ and $\left(\lambda_3/\lambda_2\right)$.
The contour plot of this \pd is shown in Fig.\ref{fig:lambdas-2D-0x27}.
This figure suggests that the probability distribution, $W(\lambda_2/\lambda_1,\lambda_3/\lambda_2)$, 
of $\left(\lambda_2/\lambda_1\right)$ and $\left(\lambda_3/\lambda_2\right)$ is symmetric 
with respect to the diagonal 
``$\left(\lambda_2/\lambda_1\right) = \left(\lambda_3/\lambda_2\right)$". 
This means that $W(x_1,x_2)=W(x_2,x_1)$.
This point is verified in Fig.\ref{fig:lambdas-2D-cuts} that shows the cuts of $W(\lambda_2/\lambda_1,\lambda_3/\lambda_2)$
along the lines orthogonal to 
the diagonal ``$\left(\lambda_2/\lambda_1\right) = \left(\lambda_3/\lambda_2\right)$" 
at two different temperatures.
Because of this symmetry it might be tempting to assume 
that $W\left(x_1,x_2\right) = W(x_1)W(x_2)$. 
However, we checked this point and found that this last assumption is incorrect. 

We verified that the contour plots for the other non-zero temperatures also appear to be 
symmetric with respect to the diagonal from ``South-West" to ``North-East".

\begin{figure*}
\begin{center}
\includegraphics[angle=0,width=6.0in]{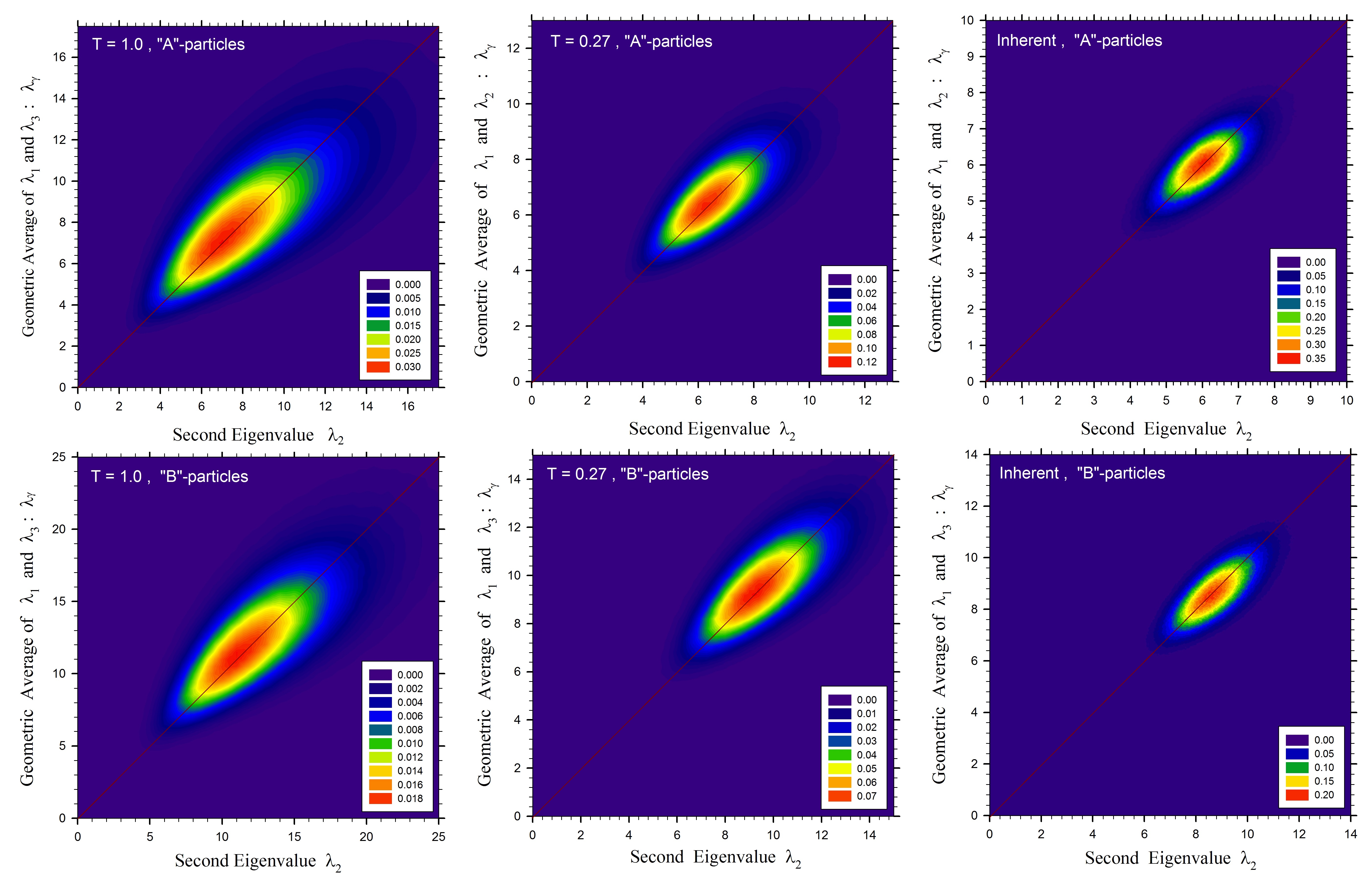}
\caption{The 2D \pds for the occurrence of pairs $(\lambda_2,\sqrt{\lambda_1\lambda_3})$ for ``A" and
``B" particles for temperatures $T=1.0$, $T=0.27$, and for the inherent structures obtained from $T=0.27$.
The concentration of probability along the diagonal ``$\lambda_2 = \lambda_{geom}$",
as well as the symmetry of the \pds with respect to this diagonal, suggest that 
indeed $\lambda_2$ tends to be $\sqrt{\lambda_1 \lambda_3}$.
}\label{fig:L2Lgeom-correlations-1}
\end{center}
\end{figure*}

The observed symmetry with respect to the diagonal ``$(\lambda_2/\lambda_1)=(\lambda_3/\lambda_2)$" lead us to think
that $\lambda_2$ tends to be the geometric average of $\lambda_1$ and $\lambda_3$. 
Indeed, if $\lambda_2 = \sqrt{\lambda_1 \lambda_3}$ then $(\lambda_2/\lambda_1)=(\lambda_3/\lambda_2)$.

In order to verify the assumption that $\lambda_2$ tends to be the geometric average of $\lambda_1$ and $\lambda_3$
we calculated the 2D contour plots of the \pds of the occurrence of pairs $(\lambda_2,\lambda_{geom})$,
where $\lambda_{geom} \equiv \lambda_{\gamma} \equiv \sqrt{\lambda_1 \lambda_3}$. 
These plots for ``A" and ``B" particles for temperatures $T=1.0$, $T=0.27$, 
and for the inherent structures (produced from the configurations at $T=0.27$) 
are shown in Fig.\ref{fig:L2Lgeom-correlations-1}.
It follows from Fig.\ref{fig:L2Lgeom-correlations-1} that the maximums of the \pds
indeed occur at $\lambda_2 \approx \lambda_{geom}$. 
Note also the symmetry of the \pds with
respect to the diagonal ``$\lambda_2 = \lambda_{geom}$".

\begin{figure}
\begin{center}
\includegraphics[angle=0,width=3.3in]{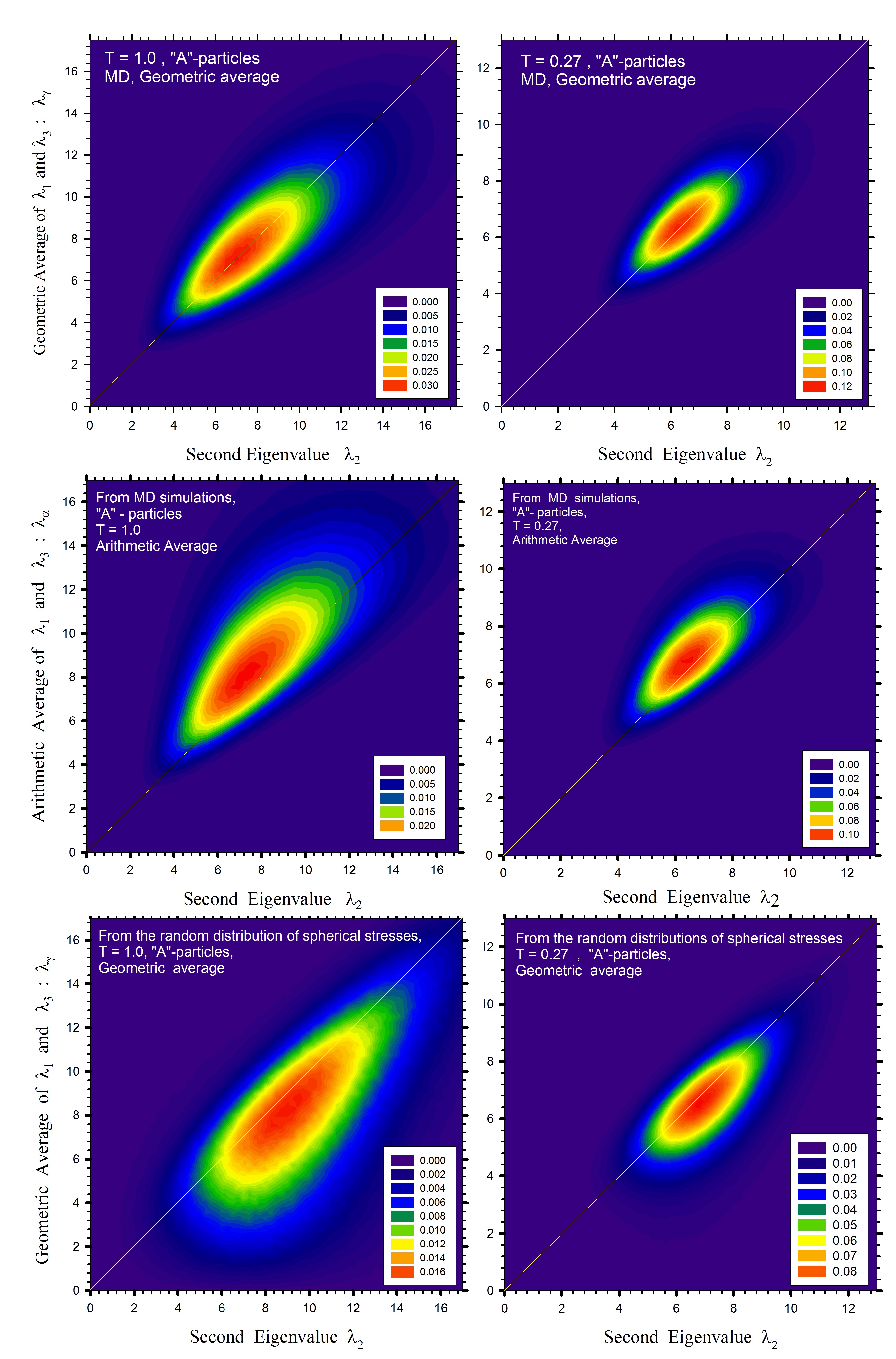}
\caption{
The top two panels show the 2D \pds of $\lambda_{2,i}$ and the
geometric average of $\lambda_{1,i}$ and $\lambda_{3,i}$, 
i.e., $\lambda_{\gamma,i} \equiv \sqrt{\lambda_{1,i}\lambda_{3,i}}$, at temperatures $T=1.0$ and $T=0.27$ obtained
directly from the MD configurations. Thus the top two panels show how good is $\lambda_{\gamma,i}$ 
as an approximation for $\lambda_{2,i}$. In the middle two panels the arithmetic average,
$\lambda_{\alpha,i}=(1/2)(\lambda_{1,i}+\lambda_{2,i})$, was used instead of $\lambda_{\gamma,i}$. The
comparison of the top and middle panels shows that $\lambda_{\gamma,i}$ is a better approximation for $\lambda_{2,i}$ than
$\lambda_{\alpha,i}$.
The bottom two panels show the \pds for $\lambda_{2,i}$ and $\lambda_{\gamma,i}$ obtained using the \ris approach.
The comparison of the top two panels with the bottom two panels
shows that the results obtained directly from MD simulations and from the \ris approach 
are quite different.
This demonstrate the presence of correlations between the \ssc.
}\label{fig:L-geom-L2-different-1-1.jpg}
\end{center}
\end{figure}

It is possible also to ask the following question. 
Is the geometric average of $\lambda_1$ and $\lambda_3$, as an approximation 
for $\lambda_2$, is much better than the arithmetic average of $\lambda_1$ and $\lambda_3$?
The answer to this question follows from the comparison of 
the upper four panels in Fig.\ref{fig:L-geom-L2-different-1-1.jpg}.
Thus $\sqrt{\lambda_1 \lambda_3}$ is indeed a better approximation for $\lambda_2$ than
$\tfrac{1}{2}(\lambda_1 + \lambda_3)$.

In the considerations of correlations between the eigenvalues it is of interest to understand
the scale of the existing correlations. 
We used \ris approach described in section \ref{sec:random-reference} in order 
to estimate the effect of correlations 
between the \ssc\, on the \pds 
presented in Fig.\ref{fig:L-geom-L2-different-1-1.jpg}. 

We proceeded as follows. 
Using the \pds of the \ssc\, obtained
in MD simulations, we generated the atomic pressure and 
five atomic \ssc\; with the random number generator, 
as described in section \ref{sec:random-reference}. 
Thus, all six \ssc\, were generated independently.
Then, using these six numbers and 
formulas (\ref{eq:from01},\ref{eq:from02},\ref{eq:from03},\ref{eq:from04}),
we produced six Cartesian stress components, i.e., the Cartesian stress matrix.
Then, by diagonalising this stress matrix, we produced three eigenvalues.
Then we calculated the geometric average of the largest and the smallest eigenvalues.
The lower two panels in Fig.\ref{fig:L-geom-L2-different-1-1.jpg} show the probability
distributions for the pairs $(\lambda_2,\sqrt{\lambda_1 \lambda_3})$ obtained by means 
of this random generation procedure.
The comparison of the upper two panels in Fig.\ref{fig:L-geom-L2-different-1-1.jpg}
with the lower two panels suggests the presence of correlations between the \ssc.

In order to measure the magnitude of the correlations we calculated quantities,
\begin{eqnarray} 
&&\gamma = \left< \tfrac{\sqrt{(\lambda_{geom,i}-\lambda_{2,i})^2}}{\lambda_{2,i}} \right>_i ,\;\;\;
\alpha = \left< \tfrac{\sqrt{(\lambda_{arith,i}-\lambda_{2,i})^2}}{\lambda_{2,i}} \right>_i .\;\;\;\;\;\;\;\label{eq:gamma-alpha-01}
\end{eqnarray} 
which show how well the geometric and arithmetic averages of $\lambda_1$ and
$\lambda_3$ approximate $\lambda_2$. 
We calculated these quantities using the eigenvalues obtained directly from
MD simulations. 
We also calculated $\gamma$ from (\ref{eq:gamma-alpha-01}) using the \ris approach 
described in section \ref{sec:random-reference}.
The results of the calculations are presented in Table (\ref{table:geom-vs-arith}).
\begin{center}
\begin{table}
\caption{The values of $\gamma$ and $\alpha$ from (\ref{eq:gamma-alpha-01}) 
for ``A" and ``B" particles at different temperatures.}\label{table:geom-vs-arith}
  \begin{tabular}{| c | c | c | c | c | }
    \hline
Calculated quantity &  T = 1.0 & T = 0.5 & T = 0.27 & T = 0 \\ \hline
$\gamma$ for ``A"-MD & 0.20 & 0.16 & 0.13 & 0.09 \\ \hline
$\alpha$ for ``A"-MD & 0.303 & 0.211 & 0.157 & 0.09 \\ \hline
$\gamma$ for ``A"-\ris & 0.301 & --- & 0.156 & --- \\ \hline
$\gamma$ for ``B"-MD & 0.17 & 0.13 & 0.11 & 0.07 \\ \hline
$\alpha$ for ``B"-MD & 0.228 & 0.162 & 0.125 & 0.08 \\ \hline
$\gamma$ for ``B"-\ris & 0.229 & --- & --- & --- \\ \hline
\end{tabular}
\end{table}  
\end{center}
It follows from Table (\ref{table:geom-vs-arith}) that the geometric average, 
i.e., $\sqrt{\lambda_1 \lambda_3}$, approximation to $\lambda_2$
is noticeably better than the arithmetic average on the data obtained
from MD simulations. It also follows from the table that the geometric average from
the \ris approach is approximately as good as the arithmetic average from
MD simulations.

As we demonstrated that there are correlations between the eigenvalues 
of the atomic stress tensors there 
arises the question about the importance of these correlations 
from the macroscopic perspective.
In our view, the observed correlations between the eigenvalues can 
be related to the atomistic origin of the Poisson ratio (effect). 
Thus, for macroscopic samples elongation in one direction usually leads to contraction 
in the directions orthogonal to the direction of elongation.
The magnitude of this effect is determined by the Poisson ratio.
Thus, for a given sample, there is a correlation between its length and its width. 
In our view, this effect can originate
from the correlations between the eigenvalues of the atomic stresses.

\subsection{Correlations between the eigenvalues and the random and independent approximations
\label{sec:random-and-independent}}

In the previous subsection we demonstrated that there are correlations between the eigenvalues of the atomic
stress tensors. 
We also speculated that these correlations might be related to the Poisson ratio effect. 
Thus it is important to gain a better understanding of the
correlations between the eigenvalues. Therefore in this subsection we further elaborate on this issue.

In particular, it is of interest to study correlations between the eigenvalues from the following perspective.
In the previous considerations of the atomic stresses it was assumed that 6 components of the atomic stress tensor in the
spherical representation are independent \cite{Egami19821,Chen19881,Levashov2008B}
in the linear approximation. 
This assumption allows to introduce the concept of the atomic stress energies 
and rationalize the values of these energies \cite{Egami19821,Chen19881,Levashov2008B}. 
In particular, it was argued that every atom in the liquid with its nearest neighbour shell  
(in the linear approximation) is equivalent to a 3-dimensional 
harmonic oscillator \cite{Egami19821,Chen19881,Levashov2008B}.
Further, it was assumed that the \emph{potential} energy of this oscillator is equally divided between 
6 independent atomic stress components. 
This assumption is supported by the result from MD simulations.
Thus, in MD simulations the potential energy of every spherical stress component can be calculated 
independently and it was demonstrated that the potential energy of every component depends on temperature as 
``$\tfrac{1}{6}\cdot3\cdot\tfrac{1}{2}k_B T=\tfrac{1}{4}k_B T$".

It follows from the previous paragraph that the assumption about independence of 
\ssc\; plays an important role in the considerations based on the concept of atomic level stresses. 
Thus, it is reasonable to address the issue of independence of the \ssc. 
In the following we consider several examples that provide certain 
insights in the relevant correlations and into their magnitudes.

\subsubsection{From Independent and Random Spherical Stress Components to the Cartesian Stress Components and Eigenvalues} 

As we already discussed, the \pds of the \ssc\, can be obtained
from the atomic configurations that were generated in MD simulations. 
Figures \ref{fig:pressure-distributions-AB-1x},\ref{fig:shear-stresses-distributions-AB-1} 
provide the examples of such distributions. 
Using these \pds random and independent \ssc\, can be generated.
It is important that generated in this way \ssc\; are independent from each other.

Then, using formulas (\ref{eq:from01},\ref{eq:from02},\ref{eq:from03},\ref{eq:from04}), 
these random and independent \ssc\, can be transformed into the Cartesian stress components.
After that, the eigenvalues of the obtained stress tensor in the Cartesian representation can also be calculated.

\begin{figure}
\begin{center}
\includegraphics[angle=0,width=3.3in]{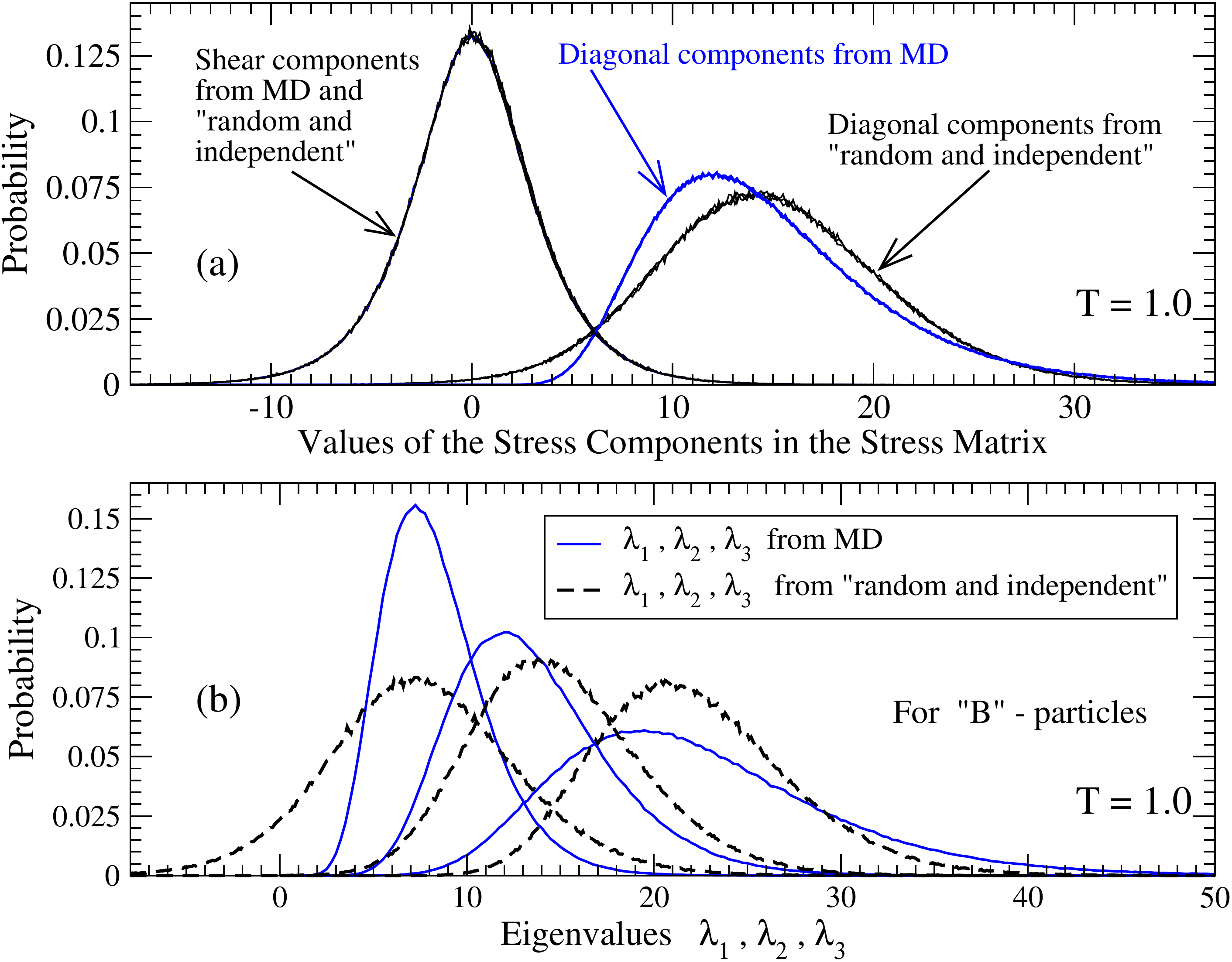}
\caption{Panel (a) shows the \pds of the Cartesian stress elements 
for ``B" particles obtained
directly from MD simulations and from the \ris approach.
There is a perfect agreement between the \pds for the off-diagonal stress elements.
This agreement is not surprising -- it is expectable from the relations (\ref{eq:to02},\ref{eq:from04}) 
between the spherical and Cartesian stress elements. 
The disagreement between the \pds of the diagonal components suggests that the atomic stress components
of the same atom are not independent. Panel (b) shows the \pds of the eigenvalues obtained directly from
MD simulations and from the \ris approach.  
It is clear that the \pds obtained through these two methods are completely different.
This again suggests the presence of correlations between the \ssc\, 
of the same atom and also the presence of correlations between the eigenvalues. 
Note that the eigenvalues from MD simulations
are always positive, while the eigenvalues from the \ris approach can be negative.
}\label{fig:components-lambdas-distributions-1x0}
\end{center}
\end{figure}

\begin{figure}
\begin{center}
\includegraphics[angle=0,width=3.3in]{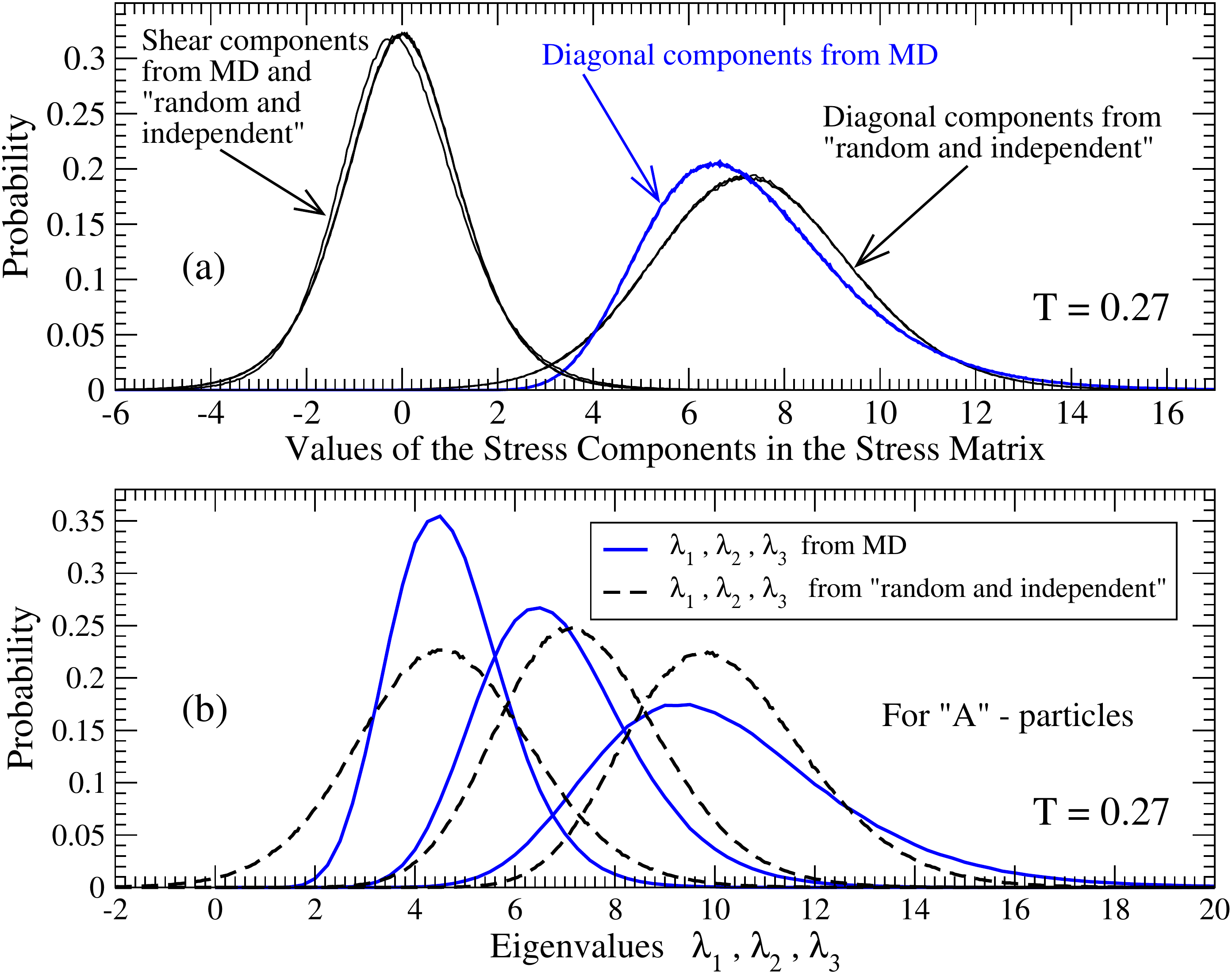}
\caption{Similar to Fig.\ref{fig:components-lambdas-distributions-1x0}, but for ``A" particles at $T=0.27$.
}\label{fig:components-lambdas-distributions-0x27}
\end{center}
\end{figure}

Thus, the \pds of the Cartesian stress components generated using the described \ris approach 
can be compared
with the \pds of the Cartesian stress components obtained directly from MD simulations. 
These comparisons are presented in 
Fig.\ref{fig:components-lambdas-distributions-1x0}(a),\ref{fig:components-lambdas-distributions-0x27}(a).
These figures clearly suggest the presence of correlation between the \ssc.
They also demonstrate the scale of the influence of these correlations on the distributions of the Cartesian stress
components.

The \pds of the eigenvalues obtained via the \ris approach
also can be compared to the \pds of the eigenvalues obtained directly from the MD data.
Figures \ref{fig:components-lambdas-distributions-1x0}(b),\ref{fig:components-lambdas-distributions-0x27}(b)
show that the \pds of the eigenvalues obtained directly from MD simulations are quite different 
from the \pds obtained via the \ris approach. 
Thus, while in certain situations
the \ris might lead to the reasonable results, 
it is clear that, after all, it is just an approximation.


As we already discussed, the geometry of the atomic environment
can be describe by six \ssc. However,
these six components also describe the orientation of the atomic environment
with respect to the reference coordinate frame. 
If this orientation is irrelevant then the geometry of the atomic environment
is described by only 3 numbers, i.e., by eigenvalues.
\emph{The relation between the \pds for the eigenvalues and 
the \pds
for the \ssc\, is actually counter-intuitive.} 
Naively, one can expect that independent
\pds for the eigenvalues should lead to the independent \pds for the \ssc\; 
and vice versa. However, this is not the case. 
Thus, in Appendix (\ref{sec:eigen-to-spherical}) we consider an example
that demonstrates how \emph{independent} \pds for the eigenvalues lead to 
\emph{dependent} \pds for the spherical stresses. 
Vice versa in Appendix (\ref{sec:spherical-to-eigen}) we show how \emph{independent} probability
distributions for the \ssc\, lead to \emph{dependent} \pds
for the eigenvalues.

\subsubsection{Random Distributions of Eigenvalues vs. the Distributions of Eigenvalues from MD Simulations}

\begin{figure}
\begin{center}
\includegraphics[angle=0,width=3.3in]{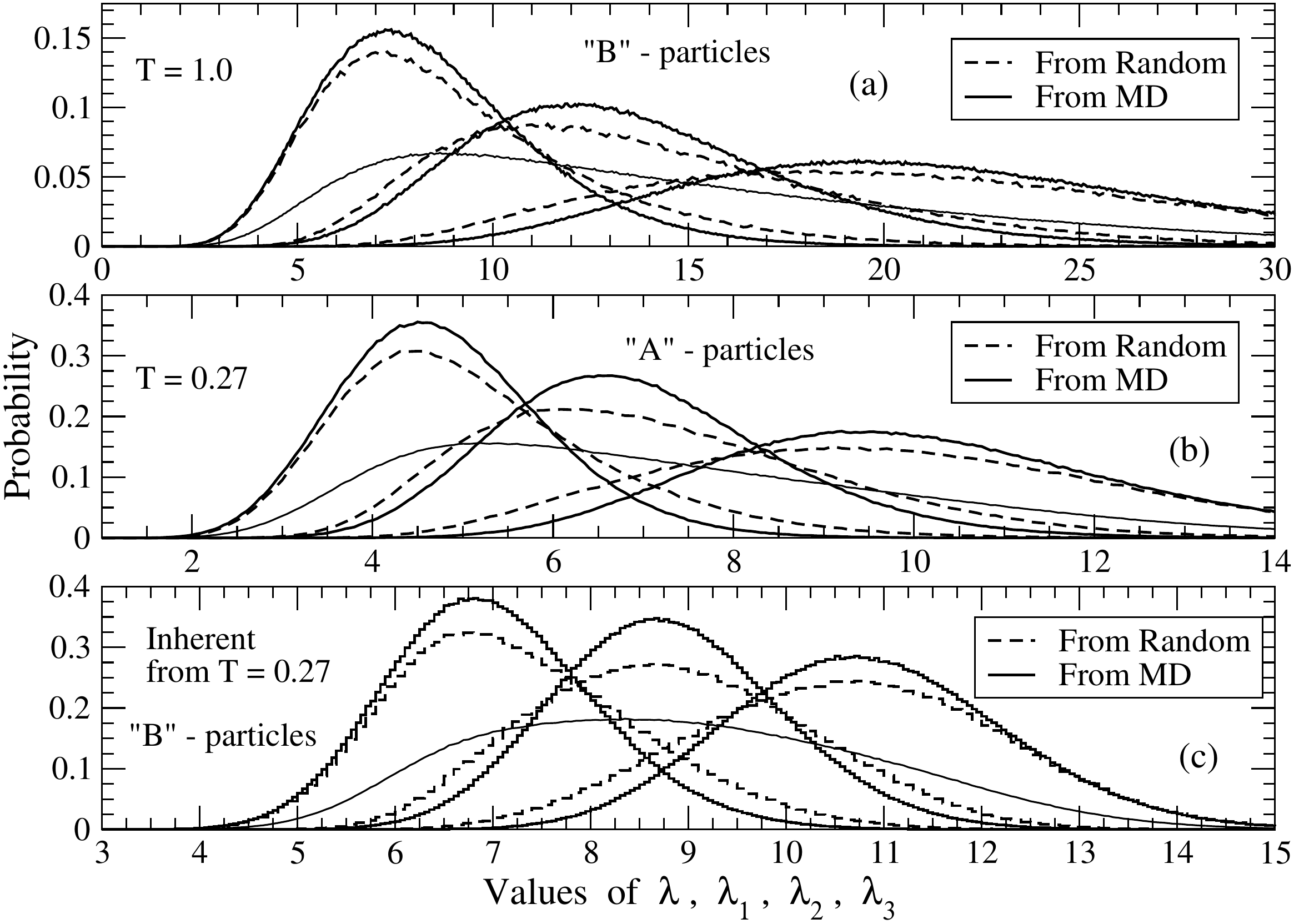}
\caption{The \pds of $\lambda_1,\;\lambda_2,\;\lambda_3$ from MD simulations and from 
the selection of three eigenvalues from the total random distribution of eigenvalues (\ril approach). 
The thin solid curves in all panels show the total \pds for all eigenvalues irrespective of their magnitudes.
The thick solid curves in all panels show the \pds for $\lambda_1,\;\lambda_2,\;\lambda_3$ obtained from MD simulations.
The dashed curves show the \pds for the randomly 
and independently generated $\lambda_1^r,\;\lambda_2^r,\;\lambda_3^r$. 
It is clear that the distributions obtained from MD simulations 
are noticeably different from the distributions obtained from the ``random" approach.
This again suggests the presence of correlations between the eigenvalues. 
}\label{fig:LambdaDist-MD-vs-Random-1xyz}
\end{center}
\end{figure}

We used the \ril approach described in section \ref{sec:random-reference} in order to generate
the \pds of the three random and independent magnitude-ordered eigenvalues which can be compared to
the \pds of the three magnitude-ordered eigenvalues obtained directly from MD simulations.
The results of this procedure are shown in Fig.\ref{fig:LambdaDist-MD-vs-Random-1xyz}.

\begin{figure}
\begin{center}
\includegraphics[angle=0,width=3.3in]{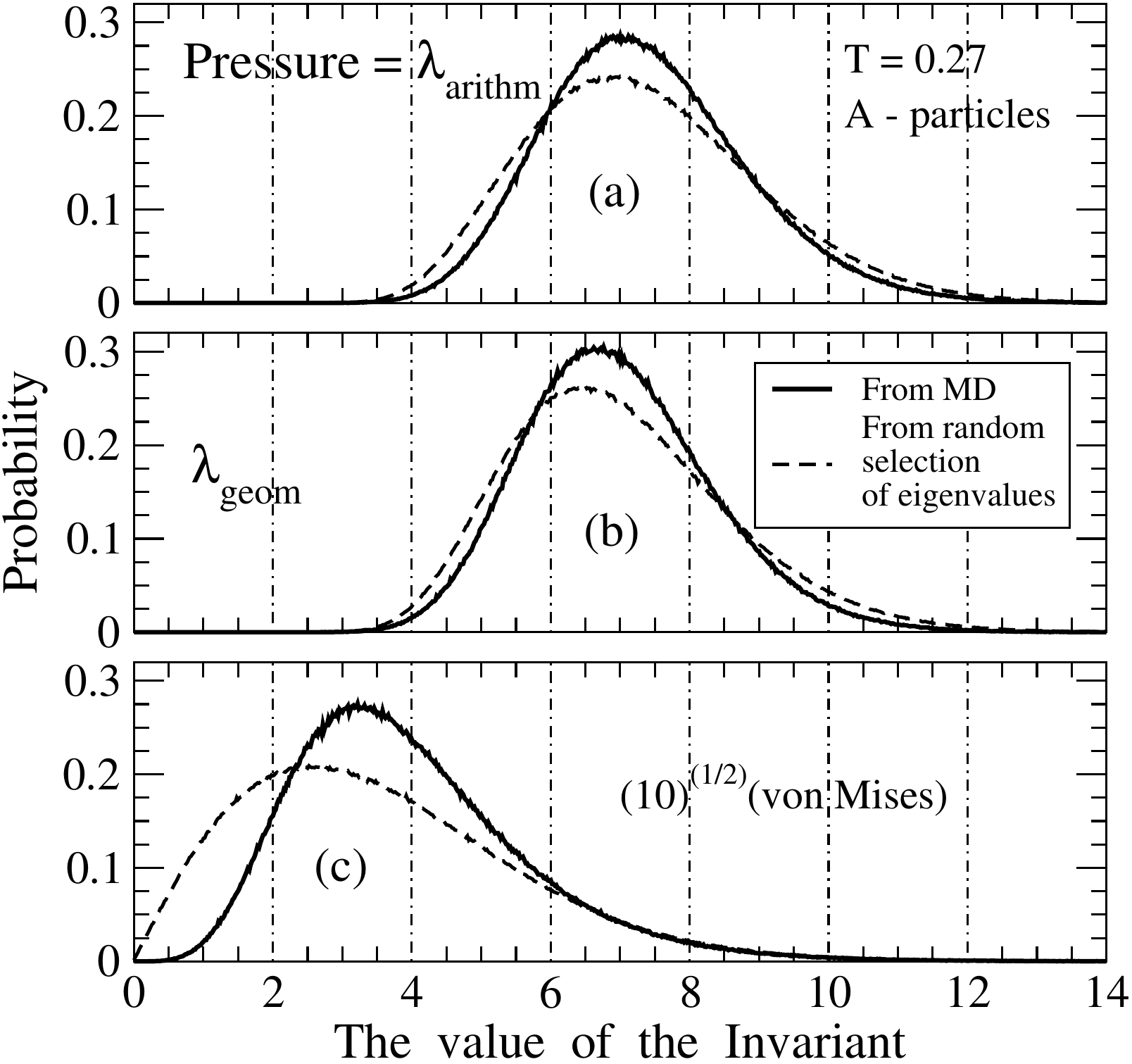}
\caption{The \pds of the stress tensor invariants for ``A"-particles at $T=0.27$.
The solid lines in all panels show the results obtained from MD simulations, i.e., 
by diagonalization of the atomic stress matrices.
The dashed curves in all panels show the \pds of the invariants
obtained using the \ril approach.  
The (a)-panel shows the \pds for the pressure, i.e., 
for the arithmetic average of the eigenvalues. 
The (b) panel shows the \pds for the geometric average of the eigenvalues
$\lambda_{geom}=\left(\lambda_{1} \lambda_{2} \lambda_{3}\right)^{1/3}$. 
The (c) panel shows the \pds for the scaled value of the von Mises shear stress:
$\left[\left((\lambda_1 - \lambda_2)^2 + (\lambda_1 - \lambda_3)^2 + (\lambda_2-\lambda_3)^2 \right)/3\right]^{(1/2)}$.
All panels suggest the presence of correlations between the eigenvalues 
of the same atomic stress tensor and demonstrate how these
correlations affect the \pds of the atomic stress tensor invariants.  
}\label{fig:distrib-invariants-3}
\end{center}
\end{figure}

It is also of interest to consider the distributions of the stress tensor invariants obtained
directly from MD simulations and also by means of the \ril approach.
The results are presented in Fig.\ref{fig:distrib-invariants-3}.

It follows from Fig.\ref{fig:distrib-invariants-3}(a,b) that the \pds 
of the local atomic pressure and the cubic root from the ``volume" of the stress tensor 
ellipsoids generated in two ways are clearly different. This difference however is not very large. 
On the other hand, the \pds of the scaled square roots from the von Mises shear stresses 
generated in two ways show more significant difference. 
Note, in particular, different behaviours of the two distributions in the
region of zero von Mises stress. 
Thus von Mises stresses calculated from the MD data avoid being zero more strongly than the von Mises stresses obtained
from the \ril approach. This means, as we discuss in more details in the following, that
the eigenvalues obtained from MD simulations avoid being equal, while there is no (there should not be) 
such behaviour in the randomly generated eigenvalues.

\begin{figure}
\begin{center}
\includegraphics[angle=0,width=3.4in]{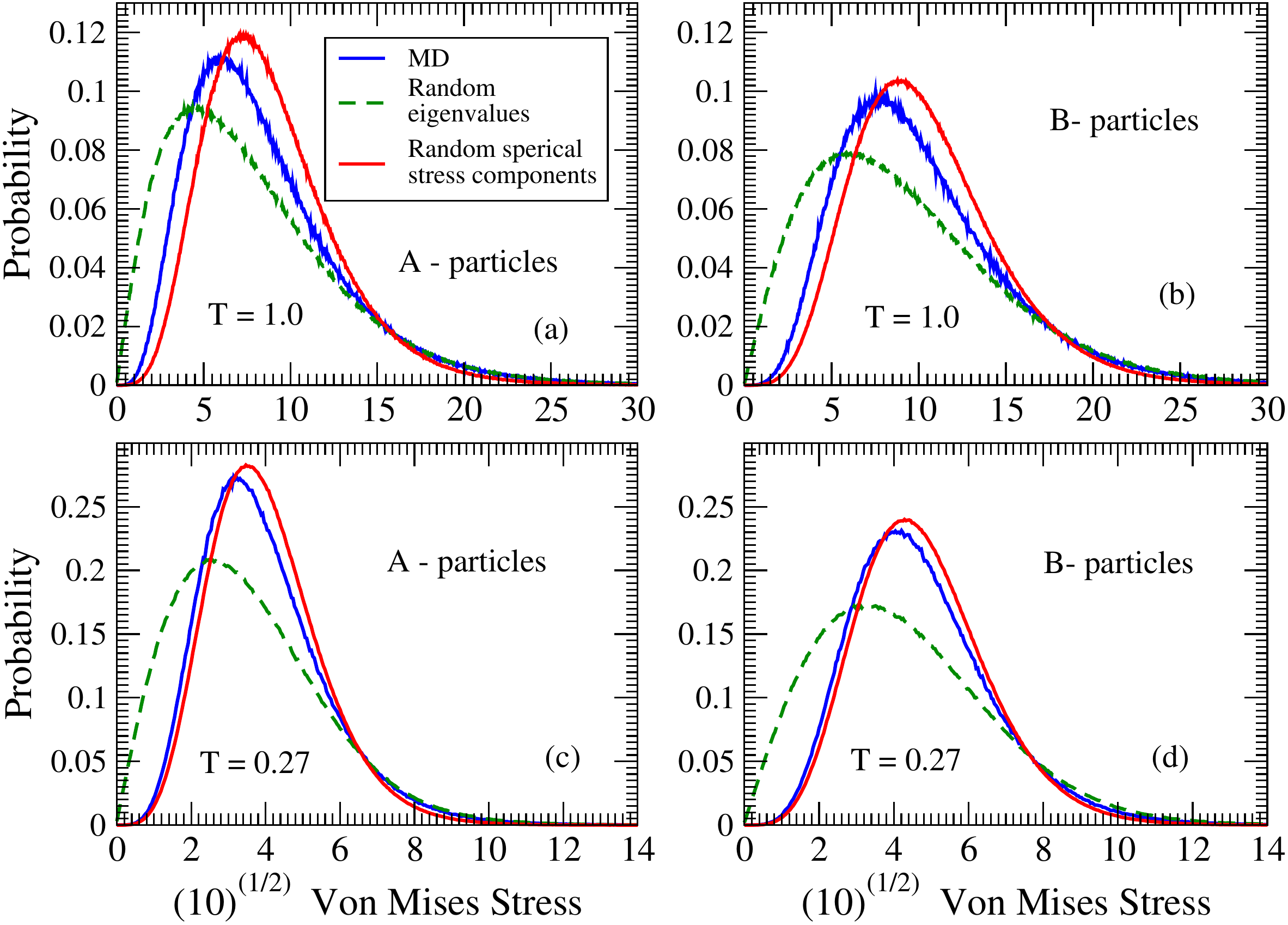}
\caption{The \pds of the von Mises stresses for ``A" and ``B" particles at $T=1.0$ and $T=0.27$.
The blue curves in each panel show the results from MD simulations. 
The results from the \ril approach are shown with the dashed green curves. 
The results from the \ris approach are shown with the red curves.
Note that the results from the \ril approach
are always quite different from the results of MD simulations.
This suggests the presence of correlations between the eigenvalues. 
On the other hand, the results from the \ris approach are relatively close
to the results of MD simulations.
}\label{fig:pressure-distributions-AB-1}
\end{center}
\end{figure}

Figure \ref{fig:pressure-distributions-AB-1} shows the \pds 
of the von Mises stresses for ``A" and ``B" particles at temperatures $T=1.0$ and $T=0.27$ 
calculated in three different ways. 
First, there are the \pds obtained directly from MD simulations. 
It follows from these curves that 
from a qualitative perspective the results for both types of particles are similar.
Then there are curves produced by \ril and \ris approaches 
described in section \ref{sec:random-reference}.
The curves produced by the \ris approach are of particular interest because 
they behave near zero value of the von Mises stress in a way similar to the
curves obtained from MD simulations. 
Thus random generation of the \ssc\, preserves the repulsion between the 
eigenvalues. 

In order to address the correlations between 
$\lambda_1$, $\lambda_2$, and $\lambda_3$ of the same atomic stress tensor we also considered, 
as it is usually done, the averaged products
$\langle \lambda_1 \lambda_2 \rangle$, $\langle \lambda_1 \lambda_3 \rangle$, $\langle \lambda_2 \lambda_3 \rangle$.
It is necessary to realize that the distributions of $\lambda_1$, $\lambda_2$, $\lambda_3$ are not independent by definition.
This is because of the convention that for every  atomic stress tensor $\lambda_1 \geq \lambda_2 \geq \lambda_3$.
Thus, in order to obtain $\lambda_1$, $\lambda_2$, and $\lambda_3$
three eigenvalues of every atomic stress tensor should be ordered according to their magnitudes.
This ordering procedure makes $\lambda_1$, $\lambda_2$, and $\lambda_3$ dependent on each other.
We also evaluated 
$\langle \lambda_1 \lambda_2 \rangle$, 
$\langle \lambda_1 \lambda_3 \rangle$, 
and $\langle \lambda_2 \lambda_3 \rangle$
within the \ril approach.
The results of the described calculation are presented in Table \ref{table:avproducts}.
It follows from these data that the differences between the averages obtained 
directly from MD and the averages obtained using the \ril approach are very small.
This demonstrates, in our view, that the most traditional approach to study correlations 
between two quantities does not really work for the eigenvalues of the atomic stress tensors.
\begin{center}
\begin{table}
\caption{The values of the correlators, 
$\langle \lambda_1 \lambda_2 \rangle$, $\langle \lambda_1 \lambda_3 \rangle$, $\langle \lambda_2 \lambda_3 \rangle$,
between the eigenvalues calculated directly from the MD structures and by means of the \ril approach. 
Note that the values of the correlators obtained in these two ways are very similar.
This suggests that the considered correlators are not sensitive to the \emph{existing} correlations between the eigenvalues.\label{table:avproducts}}
  \begin{tabular}{| c | c | c | c | }
    \hline
  Method &  $\langle \lambda_1 \lambda_2 \rangle$ & $\langle \lambda_1 \lambda_3 \rangle$ & $\langle \lambda_2 \lambda_3 \rangle$ \\ \hline
MD for ``A" at $T = 0.27$    & 74.34 & 50.79 & 34.90 \\ \hline
\ril for ``A" at $T=0.27$  & 72.84 & 51.41 & 37.63 \\ \hline \hline
MD for ``B" at $T = 1.0$    & 325.6 & 198.2 & 123.1 \\ \hline
\ril for ``B" at $T=1.0$  & 321.8 & 202.6 & 135.7 \\ \hline\hline
MD for ``B" at $T = 0$    & 97.8 & 77.9 & 63.4 \\ \hline
\ril for ``B" at $T=0$  & 95.3 & 77.8 & 65.1 \\ \hline
  \end{tabular}
\end{table}  
\end{center}

\subsubsection{The Distributions of the Spacings Between the Eigenvalues}

In considerations of the eigenvalues' spectra of the random matrices it is common to consider the distributions
of the spacings between the eigenvalues \cite{randommatrices,Wigner1967}. We also performed this analysis.

\begin{figure}
\begin{center}
\includegraphics[angle=0,width=3.3in]{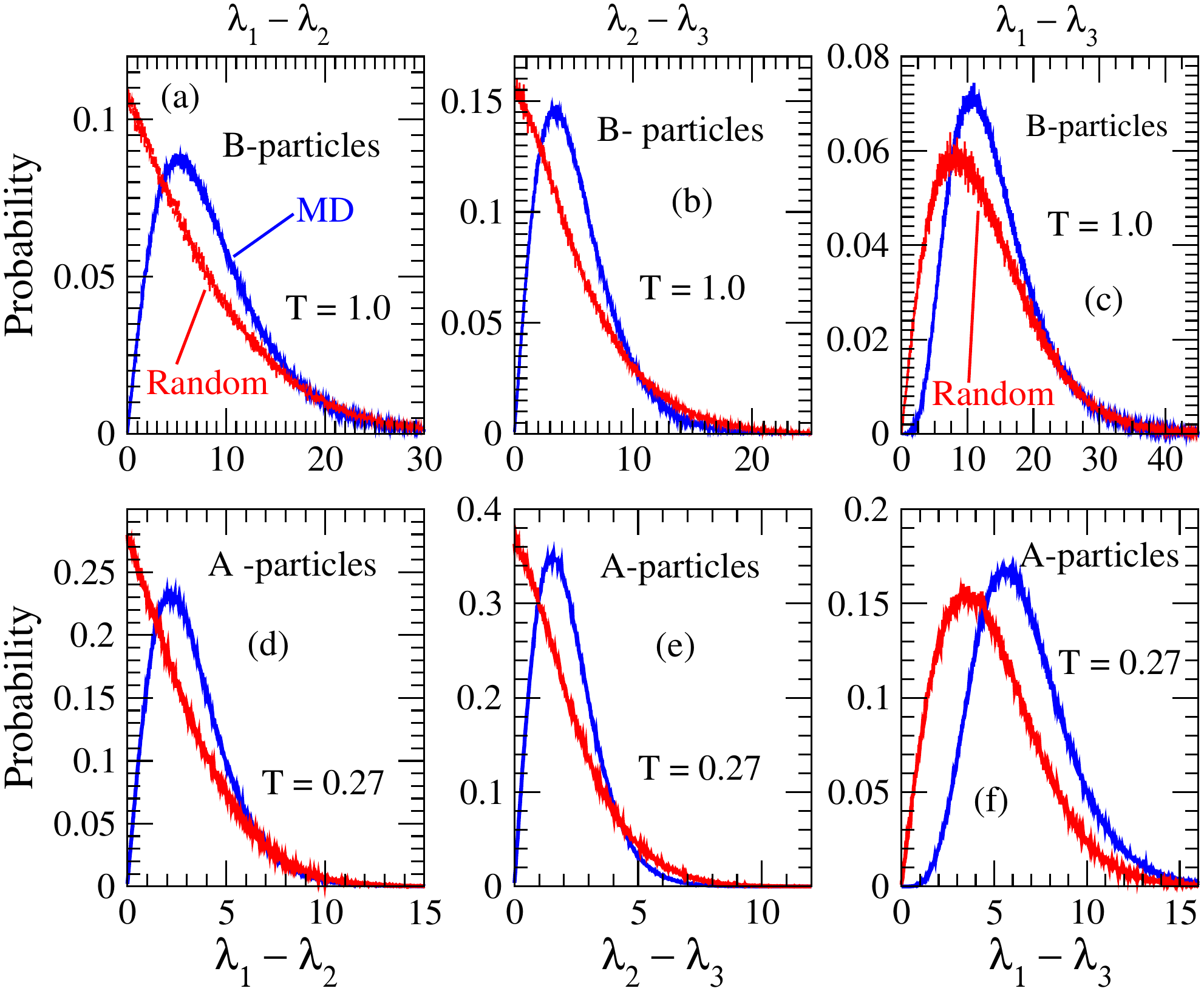}
\caption{The blue curves in all panels show the \pds of the spacings $(\lambda_1 - \lambda_2)$,
$(\lambda_2 - \lambda_3)$, $(\lambda_1 - \lambda_3)$ obtained directly from MD simulations.
The red curves in all panels show the \pds of the spacings obtained using the \ril approach.
Panels (a,b,c) show the results for ``B"-particles at temperature $T=1.0$. 
Panels (d,e,f) show the results for ``A"-particles at temperature $T=0.27$.
The curves obtained from MD simulations suggest that local atomic configurations 
with equal eigenvalues essentially never happen.
This follows from the observation that the \pds from MD go to zero, as spacings 
$(\lambda_1-\lambda_2)$ and $(\lambda_2 - \lambda_3)$ go to zero. 
The curves obtained using the \ril approach, in contrast, have maximums at zero values of the spacings.
It is easy to realise that the results from the \ril approach are not surprising.
\label{fig:MD-vs-Random-from-Lambda-1}}
\end{center}
\end{figure}

The panels (a,b,c) of Fig.\ref{fig:MD-vs-Random-from-Lambda-1} show the \pds of the spacings
between the eigenvalues for ``B"-particles at $T=1.0$. The blue curves were obtained directly from the MD data.
The red curves were obtained using the \ril approach.  
Zero probability at zero spacing, observed in panels (a,b) in the MD data, suggests that 
atomic stress tensors avoid having two eigenvalues of the same magnitude. 
It is likely that this effect originates from the vanishing volume of the phase space
associated with the corresponding values of the \ssc\, \cite{randommatrices}.
Thus, due to purely probabilistic reason, there essentially no atoms whose environment is almost spherical 
from the perspective of atomic stresses. 
Such spherical environments would lead to three eigenvalues which are all equal to each other. 
However, the data presented in Fig.\ref{fig:MD-vs-Random-from-Lambda-1} suggest that this essentially never happens.

The red curves in panels (a,b,c) show the \pds obtained using the \ril approach.
In panels (a) and (b) which correspond to the neighbour eigenvalues there is 
no ``repulsion" between them.
This result is expectable for the randomly selected eigenvalues. 
On the other hand, panel (c) shows ``repulsion" between the randomly selected $\lambda_1$ and $\lambda_3$.
It is easy to realize that the ``repulsion" between the non-neighbouring eigenvalues has a trivial probabilistic origin. 
The panels (d,e,f) of Fig.\ref{fig:MD-vs-Random-from-Lambda-1} show the results for ``A"-particles at $T=0.27$.
From a qualitative perspective the \pds for ``A"-particles are rather similar to the
\pds for ``B"-particles. 
Note also by comparing panels (a) with (b) and (d) with (e) that the \pds of 
$(\lambda_1 - \lambda_2)$ are wider than the \pds of $(\lambda_2 - \lambda_3)$.
 
\begin{figure}
\begin{center}
\includegraphics[angle=0,width=3.3in]{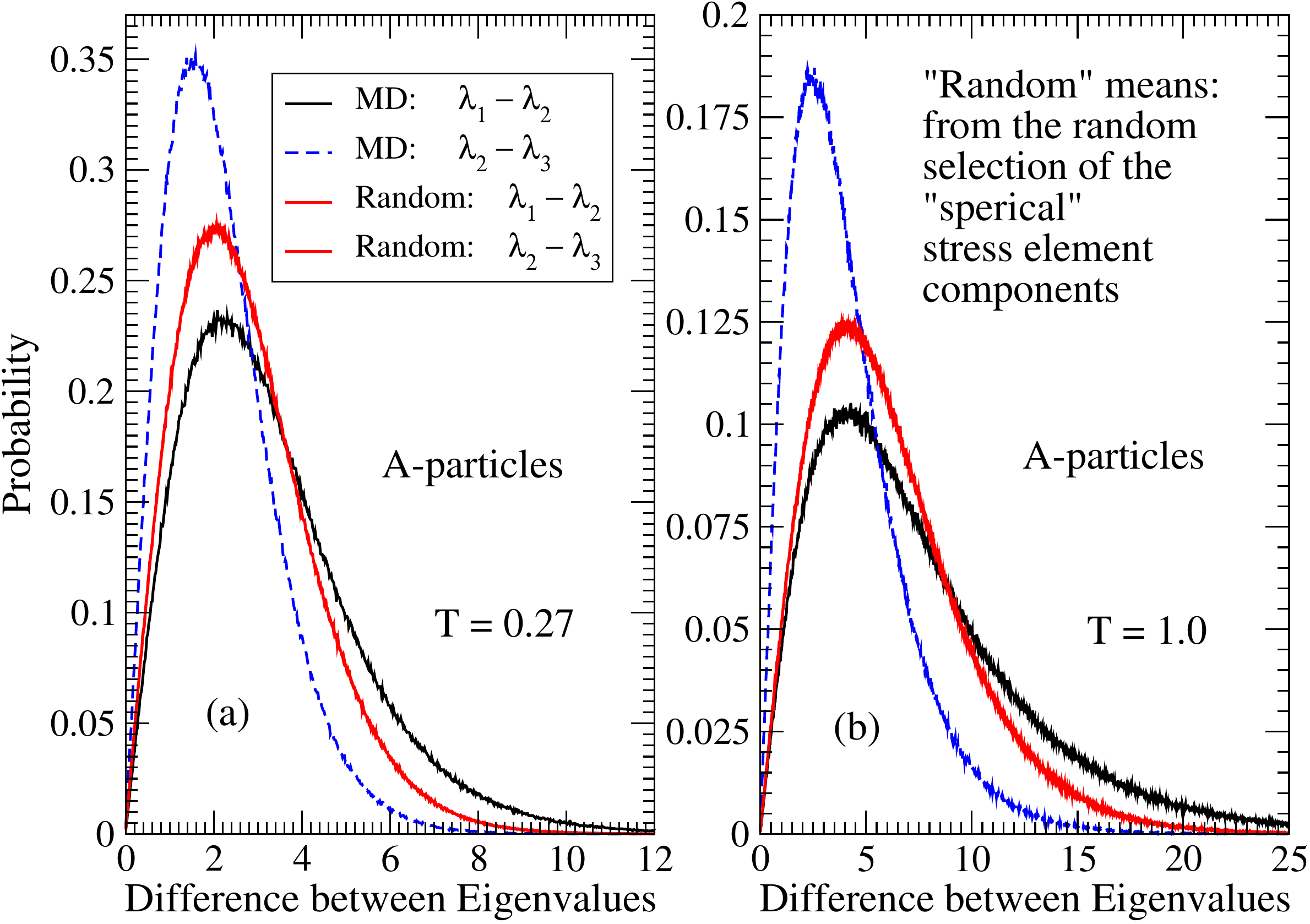}
\caption{The \pds of the spacings between the eigenvalues obtained in MD simulations 
and using the \ris approach. 
The legends for the curves in both panels are the same.
The black curves in both panels show the results for the \pds 
of $(\lambda_1 - \lambda_2)$ from MD simulations, while
blue curves show the \pds of $(\lambda_2 - \lambda_3)$ also from MD simulations. 
The \pds for $(\lambda_1 - \lambda_2)$ and $(\lambda_2 - \lambda_3)$ obtained 
using the \ris approach coincide, i.e., there are two red curves in each red curve. 
The \pds of $(\lambda_1 - \lambda_2)$ and of $(\lambda_2 - \lambda_3)$ from the \ris approach 
coincide at $T=0.27$ and at $T=1.0$. 
}\label{fig:lamdas-spacing-MD-vs-Cartesian-random-1}
\end{center}
\end{figure}

We also calculated the \pds of the spacings between the eigenvalues using 
the \ris approach. 
The corresponding \pds are presented 
in Fig.\ref{fig:lamdas-spacing-MD-vs-Cartesian-random-1}.
Again note that in MD data in both panels the \pds
of $(\lambda_1 - \lambda_2)$ are wider than the \pds
of $(\lambda_2 - \lambda_3)$. 
On the other hand, the \pds of
$(\lambda_1 - \lambda_2)$ and of $(\lambda_2 - \lambda_3)$ obtained
using the \ris approach are identical to each other. 
Thus there are two red curves in panel (a), i.e., 
one red curve in panel (a) is the \pd of $(\lambda_1 - \lambda_2)$ while
another red curve is the \pd of $(\lambda_2 - \lambda_3)$.
Both curves coincide. The situation is similar for the panel (b).

Finally we calculated the average values of the spacings between the eigenvalues
using the data from MD simulations, using the \ril approach, and
using the \ris approach. 
The results are presented in table \ref{table:avspacings}.
Note that in MD data $\langle \lambda_1 - \lambda_2 \rangle > \langle \lambda_2 - \lambda_3 \rangle$.
Also note that in the \ris approach $\langle \lambda_1 - \lambda_2 \rangle \approx \langle \lambda_2 - \lambda_3 \rangle$.
It is of interest also that $\langle \lambda_1 - \lambda_3 \rangle$ is the same for the
MD data and for the \ris approach.
On the other hand, $\langle \lambda_1 - \lambda_3 \rangle$ in the \ril approach
is smaller than in the other two approaches.
\begin{center}
\begin{table}
\caption{The average spacings between the eigenvalues for ``A"-particles at $T=0.27$.\label{table:avspacings}}
  \begin{tabular}{| c | c | c | c | }
    \hline
  Method &  $\langle \lambda_1 -\lambda_2 \rangle$ & $\langle \lambda_2 -\lambda_3 \rangle$ & $\langle \lambda_1 -\lambda_3 \rangle$ \\ \hline
MD  & 3.3 & 2.2 & 5.5 \\ \hline
\ris  & 2.8 & 2.8 & 5.5 \\ \hline
\ril  &  2.8 & 2.0 & 4.8 \\
    \hline
  \end{tabular}
\end{table}  
\end{center}

\section{Correlation Functions Between Different Atoms} 

In this section we describe the results of our analysis of atomic stress correlations
between different atoms.
In this analysis, it is natural to compare the results for the stress correlation functions
with the results for the partial pair density functions. 
In particular, it is useful to compare the positions of the characteristic features 
and also the relative changes in both types of functions on decrease of temperature. 

\begin{figure*}
\begin{center}
\includegraphics[angle=0,width=6.8in]{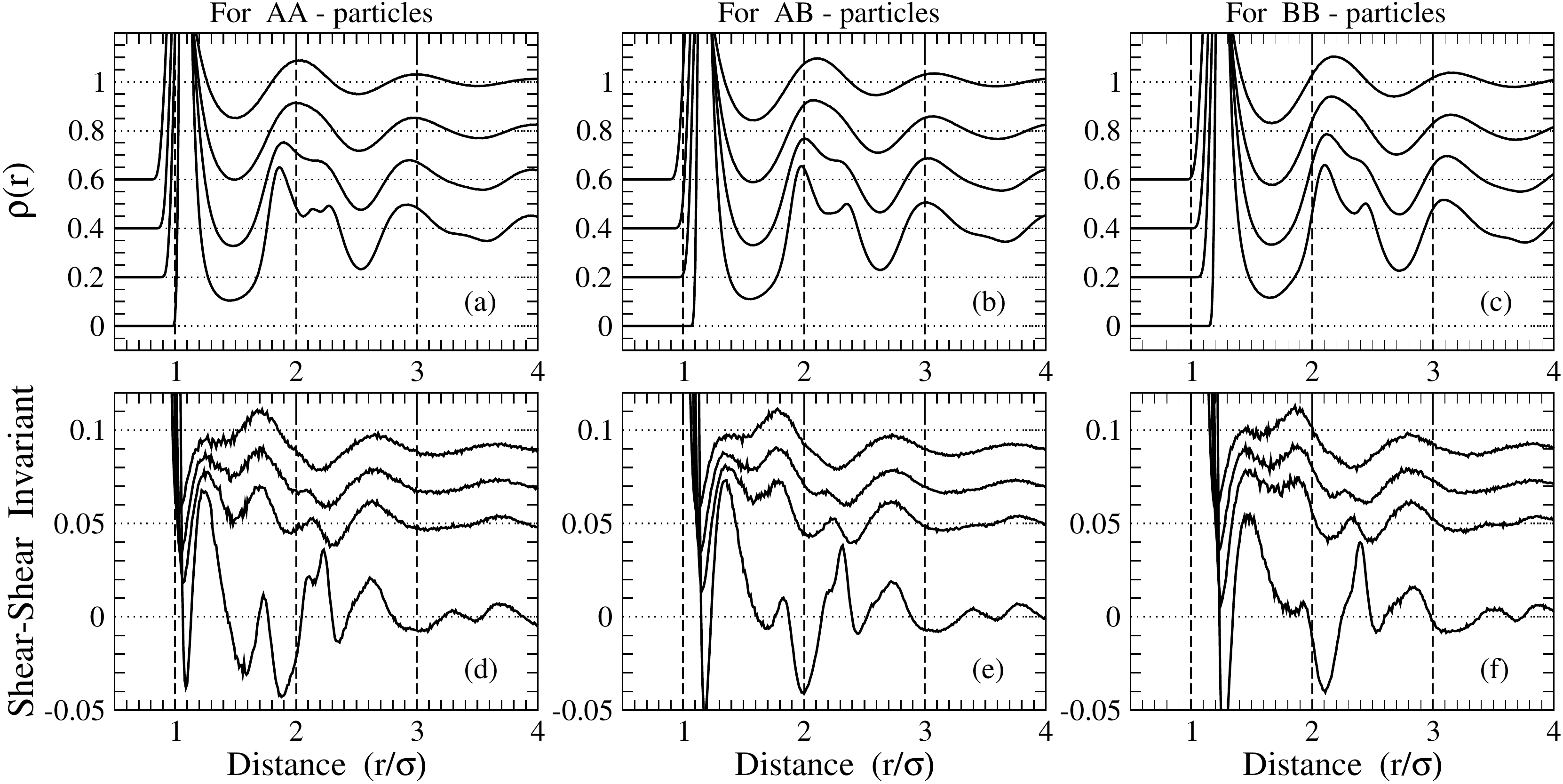}
\caption{Panels (a,b,c) show the partial pair density functions,  $\rho(r)$, for temperatures (from top to the bottom)
$T=1.0$, $T=0.5$, $T=0.27$, and $T=0$. The curves for $T=1.0$, $T=0.5$, and $T=0.27$ were shifted upward 
by 0.6, 0.4, and 0.2 correspondingly. The ``inherent" $T=0$ curve was not shifted.
Panels (d,e,f) show the invariant shear-shear correlation functions per pair of atoms, defined by formulas 
(\ref{eq;avesixysjxy1},\ref{eq;avesixysjxy21},\ref{eq;avesixysjxy22},\ref{eq;avesixysjxy2},\ref{eq;avesixysjxy3}), 
normalized to the
average square of the spherical shear stress component, i.e., $\langle \left(\sigma_i^{xy}\right)^2\rangle$.
The curves for the invariant shear-shear correlation functions at temperatures $T=1.0$, $T=0.5$, and $T=0.27$ 
were shifted upward by 0.09, 0.07, and 0.05 correspondingly. The ``inherent" curve was not shifted.
The curves in panels (d,e,f) show how the average \emph{correlation state} for the pairs of atoms depends on distance 
and temperature.
}\label{fig:pdfone}
\end{center}
\end{figure*}

\subsection{Partial pair density functions and the stress-stress correlation function invariants related to viscosity}

Panels (a,b,c) of Fig.\ref{fig:pdfone} show the partial pair density functions for ``AA", ``AB", and ``BB"-particles at different temperatures.
In each panel the curves from the top to the bottom correspond to the temperatures $T=1.0$, $T=0.5$, $T=0.27$, and $T=0$. 
The $T=0$ curves were calculated using the inherent structures produced by the conjugate gradient relaxation of the structures at
$T=0.27$. The results for the pairs of particles of different types are similar from a qualitative point of view.
Since the particles of type ``B" are larger than the particles of type ``A", the curves 
corresponding to the ``AB" pairs [see panel (b)] are shifted to the
right with respect to the curves corresponding to the ``AA" pairs [see panel (a)]. 
Similarly the curves corresponding to the ``BB"-particles [see panel (c)]
are shifted to the right with respect to the curves corresponding to the ``AB"-particles. 
As the temperature of the liquid is reduced there appears in all partial pair density curves a 
noticeable precursor of the famous \cite{EMa20111,amorhpous2001,Pan20111} splitting of the second peak.
This splitting becomes well pronounced in the $T=0$ state. 
Overall, the changes in the partial pair density functions (beyond the first peak) observed on 
decrease of temperature are not very pronounced.

Panels (d,e,f) of Fig.\ref{fig:pdfone} show the shear invariant stress correlation function
given by formulas (\ref{eq;avesixysjxy1},\ref{eq;avesixysjxy21},\ref{eq;avesixysjxy22},\ref{eq;avesixysjxy2},\ref{eq;avesixysjxy3}) normalized to the
average square of the spherical shear stress component, i.e., to $\langle \left(\sigma_i^{xy}\right)^2\rangle$.
All curves were calculated using the expressions (\ref{eq;avesixysjxy1},\ref{eq;avesixysjxy21},\ref{eq;avesixysjxy22},\ref{eq;avesixysjxy2}).
However, we checked that formula (\ref{eq;avesixysjxy3}) leads to the same results.

Note that while the results for the stress correlation functions are presented next 
to the results for the partial pair density functions the meaning of 
the stress correlation function curves is quite different. 
Thus the value of a stress correlation curve at distance $r$ describes 
the average \emph{correlation state} of two particles separated by distance $r$.

Note that on decrease of temperature the changes in the stress correlation curves are
significantly more pronounced than the changes in the partial pair density curves.
In particular, as temperature changes from $T=0.27$ to $T=0$, there is a very abrupt change.
Thus the invariant stress correlation function turns out to be quite sensitive
to the structural changes that happen to the liquid as it goes into an inherent state.
Also note that the range of this abrupt change is limited to the third nearest neighbours.
In our view, this means that the stress correlation function is sensitive to the widely discussed
formation of the intermediate range order (It is known that the pair density function 
is not sensitive to the formation of the intermediate range order) \cite{MaE20061,EMa20111,Tanaka20131}. 
In our view, the most abrupt changes in the stress correlation functions (in the stress correlation state) 
happen for (the pairs of atoms separated by) the distance which approximately corresponds to the
first minimum in the pair density function. 
This indicates, in our view, formation of a definite correlated state for weakly connected atoms.

\nobreak
\begin{figure*}
\begin{center}
\includegraphics[angle=0,width=6.8in]{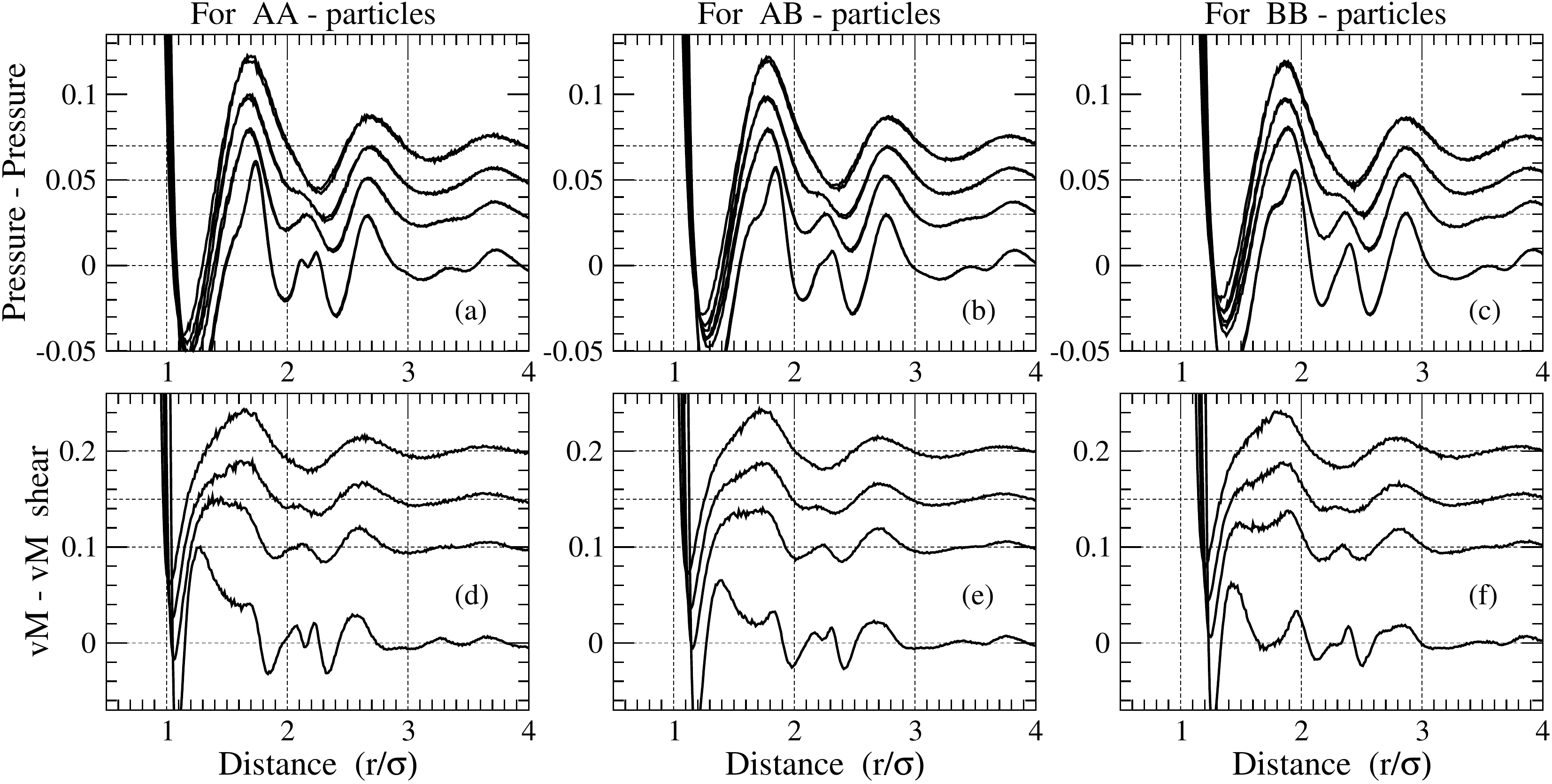}
\caption{Panels (a,b,c) show the normalized correlation functions between the pressure-pressure invariants.
See the first formula in (\ref{eq:I1I3-corr-funct}). The same panels also show the correlation functions
between the third stress tensor invariants, i.e., between the geometric averages of the stress tensor eigenvalues.
See the second formula in (\ref{eq:I1I3-corr-funct}).
For the pairs of particles of the same type and at the same temperature these two correlation functions 
essentially coincide. For this reason every curve in panels (a,b,c) consists of two curves corresponding to two
different correlation functions.
In all panels in this figure the curves from top to the bottom correspond to the temperatures
$T=1.0$, $T=0.5$, $T=0.27$, and $T=0$.
Panels (d,e,f) show the normalized (von Mises stress)--(von Mises stress) correlation functions.
\label{fig:inv-all-in-one}}
\end{center}
\end{figure*}

\begin{figure*}
\begin{center}
\includegraphics[angle=0,width=6.8in]{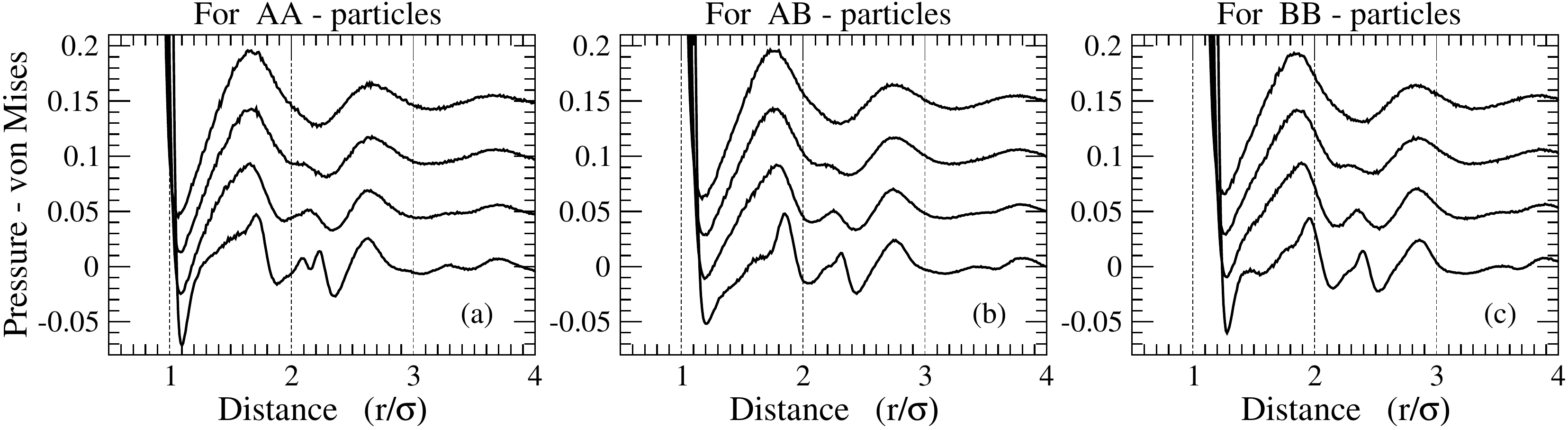}
\caption{The normalized correlation functions between the pressure and the von Mises stress invariant, i.e.,
$\left(\langle p_i s_{vM,j} \rangle / \langle p_i \rangle \langle s_{vM,j} \rangle\right) - 1$.
In all panels the curves from the top to the bottom correspond to the temperatures
$T=1.0$, $T=0.5$, $T=0.27$, and $T=0$.
\label{fig:inv-pressure-von-Mises}}
\end{center}
\end{figure*}

Panels (a,b,c) of Fig.\ref{fig:inv-all-in-one} show the normalized correlation functions between the 
stress tensor invariants of different atoms.
Thus the correlations curves in panels (a,b,c) were defined as:
\begin{eqnarray}
\frac{\langle I_1(i)I_1(j) \rangle}{\langle I_1(i)\rangle \langle I_1(j) \rangle} - 1\;\;\;\;\text{and}\;\;\;\;
\frac{\langle \lambda_{geom}(i)\lambda_{geom}(j) \rangle}{\langle\lambda_{geom}(i)\rangle \lambda_{geom}(j) \rangle} - 1\;,\;\;\;\;\;\;\;\;
\label{eq:I1I3-corr-funct}
\end{eqnarray}
where $\lambda_{geom}(i) = \left[\lambda_1(i) \lambda_2(i) \lambda_3(i)\right]^{1/3}$.
See formulas (\ref{eq;eigenI012},\ref{eq;eigenI032}).
Thus panels (a,b,c) show the normalized (pressure)-(pressure) and (stress-volume)-(stress-volume) 
correlation functions. 
We found that for the pairs of particles of a particular type (``AA" or ``AB" or ``BB")
and at the same temperature the (pressure)-(pressure) and the (stress-volume)-(stress-volume) normalized correlation
curves are essentially identical. Thus every curve in panels (a,b,c) consists of two curves---one of these two curves
shows the correlation function between the arithmetic averages of the eigenvalues, while another curve shows the
correlation function between the geometric averages of the eigenvalues. 
Note that the relative scale of the changes in these stress correlation functions, 
as temperature is reduced, is comparable
to the scale of the changes in the pair density functions in Fig.\ref{fig:pdfone}.

Panels (d,e,f) of Fig.\ref{fig:inv-all-in-one} show the normalized correlation functions between the von Mises
shear stresses of different atoms. 
As temperature is reduced there appear more features in the correlations functions.
The comparison with the pair density functions in Fig.\ref{fig:pdfone} suggests that the relative changes in the
(von Mises stress)---(von Mises stress) correlation functions are somewhat more noticeable than 
the changes in the pair density functions. 
However, these differences do not appear to be very significant.

Figure \ref{fig:inv-pressure-von-Mises} shows the normalized correlation functions between the pressure on one atom
and the von Mises shear stress on another atom. 
Overall, the presented curves exhibit the behaviour which is qualitatively similar
to the behaviour of the curves in Fig.\ref{fig:inv-all-in-one}.  

In our view, it is important to notice the significant difference in the scales of the relative changes 
between the shear stress tensor invariant correlation functions in panels (d,e,f) of Fig.\ref{fig:pdfone} and
the scales of the relative changes in the (von Mises stress)---(von Mises stress) correlation functions in panels
(d,e,f) of Fig.\ref{fig:inv-all-in-one}. 
It is obvious that the changes in the invariant shear stress correlation 
functions in panels (d,e,f) of Fig.\ref{fig:pdfone} are more significant.
In our view, the reason for this difference is the following.
The invariant shear stress correlation function, defined by formulas  
(\ref{eq;avesixysjxy1},\ref{eq;avesixysjxy21},\ref{eq;avesixysjxy22},\ref{eq;avesixysjxy2},\ref{eq;avesixysjxy3})
takes into account the mutual orientation of the eigenframes of atoms $i$ and $j$.
On the other hand, the (von Mises stress)---(von Mises stress) correlation function is the correlation
function between the scalar quantities.
Thus, the comparison of the two types of the correlation functions suggests that in order to describe
the stress state of the liquid it is necessary to take into account the mutual orientations of the
eigenframes of atoms $i$ and $j$. 
This conclusion is in agreement with the multiple previous suggestions (conclusions) about 
the importance of the angular correlations for the proper description of the 
supercooled liquids and glasses \cite{Tanaka20131}.

\section{Conclusions \label{sec:concl}}

In this paper we were developing an approach for the atomic scale description of the stress states of liquids and glasses.
The approach is based on considerations of the eigenvalues and eigenvectors of the atomic stress tensors.
Thus, it is possible to associate an atomic stress tensor with every atom in a liquid or 
in a glass \cite{Egami19801,Egami19802,Egami19821}.
This tensor can be diagonalized and its eigenvalues and eigenvectors can be found.
Therefore with every atom can be associated 3 eigenvalues which describe its local atomic environment 
(without taking into account the orientation of this environment with respect to the reference
coordinate frame).
We studied correlations between the eigenvalues of the same atomic tensor.
We also studied correlations between the eigenvalues and eigenvectors of different atoms.

In our studies we investigated a binary model of particles interacting through the repulsive part of 
the Lennard-Jones pair potential. In this system all eigenvalues are positive. 
Thus the convention $\lambda_{1,i} \geq \lambda_{2,i} \geq \lambda_{3,i}$ was adopted for every 
atom $i$.

\emph{With respect to the correlations between the eigenvalues of the same atom our main findings are the following.}\\
{\bf (a)} We found that there are correlations between the eigenvalues of the same atomic stress tensor. 
In our view (it is a speculation), the presence of correlations between the eigenvalues of the same atomic stress tensor
is essentially the ``Poisson ratio effect" on the atomic scale.\\ 
{\bf (b)} We found that the probability distributions ($PDs$) of the ratios $(\lambda_{2,i}/\lambda_{1,i})$ 
and $(\lambda_{3,i}/\lambda_{2,i})$ for the particles of the same type 
are essentially identical in the liquid state. 
This is so for the ratios associated with both types of particles.
We also found that in the inherent state there is a noticeable difference between these
two $PDs$.\\ 
{\bf (c)} We found that the 2D probability distributions, 
$W[(\lambda_{2,i}/\lambda_{1,i}), (\lambda_{3,i}/\lambda_{2,i})]$, in the liquid state 
are symmetric with respect to the diagonal
``$(\lambda_{2,i}/\lambda_{1,i}) = (\lambda_{3,i}/\lambda_{2,i})$" 
with rather high precision for both types of particles.\\
{\bf (d)} We found that the middle eigenvalue, $\lambda_{2,i}$, tends to be the geometric average of the largest and 
the smallest eigenvalues, i.e., $\lambda_{2,i} \approx \sqrt{\lambda_{1,i}\lambda_{3,i}}$.\\
{\bf (e)} We investigated the quality of the two independent random approximations that can be used to model
the \pds of the eigenvalues and related quantities. We found that the approximation based
on the independent and random selection of the spherical stress components is better than more direct approximation
based on independent and random selection of the eigenvalues. 
However, in our view, both methods provide poor approximations to the data obtained directly from MD simulations.

\emph{With respect to the correlations between the eigenvalues and eigenvectors of different 
atoms our findings are the following.}\\
{\bf (a)} We studied changes with temperature in the correlation functions between 
the invariants, $I_1$, $I_2$, and $I_3$, of the atomic stress tensors. 
These are the correlations functions between the scalar quantities. 
We found that the relative magnitudes of the changes in the correlation functions between these scalar
quantities on decrease of temperature 
are similar to the relative magnitudes of the changes in the partial pair density functions.\\
{\bf (b)} We also studied changes with temperature in the rotationally invariant part of 
the correlation function 
$\langle \tau_{i}^{xy}\tau_{j}^{xy} \rangle$ which is directly related to viscosity. 
This correlation function takes into account the mutual orientations of the eigenvectors 
of the stress tensors of atoms $i$ and $j$ and thus \emph{it is not} a correlation 
function between the scalar quantities. We found that on decrease of temperature this
non-scalar correlation function exhibits changes which are clearly more pronounced than 
the changes in the partial pair density functions. This finding suggests that in considerations
of the structures of supercooled liquids and glasses it is important to take into account angular correlations.
This is so even for systems whose interaction potentials do not explicitly depend on angles.
This view, is in agreement with other publications.\\
{\bf (c)} We found that the most pronounced changes in the stress correlation functions happen within the range of distances
limited to the 3rd nearest neighbours. 
Thus our data indicate formation of an intermediate range order in liquids on supercooling and the existence
of such order in glasses. This again is in agreement with other publications.

Finally we note that all results presented in this paper are related to the \emph{same-time} correlation functions and thus
they describe instant structural properties. 
It is of interest to investigate the behaviour of the time-dependent correlation functions 
analogous to those discussed in this paper. 

\section{Acknowledgements}

We would like to thank M.G. Stepanov for several very useful discussions.

\appendix

\section{Averaging of $\sigma_i^{xy}\sigma_j^{xy}$ over the orientations of the
observation frame \label{app:angle-averaging}}

In the main text we did not explain how to perform the averaging 
of $\sigma_i^{xy}\sigma_j^{xy}$
over the orientations of the coordinate frame 
$\bm{\hat{\tilde{x}}},\bm{\hat{\tilde{y}}},\bm{\hat{\tilde{z}}$}.
In order to perform this averaging it is necessary
to provide the expression for the rotation matrix $R$.
The columns in this matrix are the directional cosines
of $\bm{\hat{\tilde{x}}},\bm{\hat{\tilde{y}}},\bm{\hat{\tilde{z}}}$ with respect
to $\bm{\hat{x}},\bm{\hat{y}},\bm{\hat{z}}$.

We performed the relevant calculations under Linux/Ubuntu using the 
program wxMaxima \cite{wxMaxima}. This program can perform analytical and numerical 
operations.

The definitions of the $\bm{\hat{\tilde{x}}},\bm{\hat{\tilde{y}}},\bm{\hat{\tilde{z}}}$ given
below allow to perform the required averaging. 
We define $\bm{\hat{\tilde{x}}},\bm{\hat{\tilde{y}}},\bm{\hat{\tilde{z}}}$ through  
the angles $\tilde{\theta},\tilde{\varphi},\xi$ in such a way that integrations over
these angles lead to the required average value. 
We define  $\bm{\hat{\tilde{x}}},\bm{\hat{\tilde{y}}},\bm{\hat{\tilde{z}}}$ 
in several steps:\\
1) We assume that the coordinates of $\bm{\hat{\tilde{x}}}$ with respect to
$\bm{\hat{x}},\bm{\hat{y}},\bm{\hat{z}}$ are: 
$\bm{\hat{\tilde{x}}}=\left[\sin(\tilde{\theta})\cos(\tilde{\varphi}),
\sin(\tilde{\theta})\sin(\tilde{\varphi}),
\cos(\tilde{\theta})\right]$.\\
2) We introduce a unit vector $\bm{\hat{k}_3}$ which is orthogonal 
to $\bm{\hat{\tilde{x}}}$ and we assume that
$\bm{\hat{k}_3}$ has the same value of $\varphi$ as $\bm{\hat{\tilde{x}}}$, 
but its value of $\theta$ is $(\tilde{\theta}-\pi/2)$. Thus:\\ 
$\bm{\hat{k}_3}=\left[-\cos(\tilde{\theta})\cos(\tilde{\varphi}),
-\cos(\tilde{\theta})\sin(\tilde{\varphi}),
\sin(\tilde{\theta})\right]$.\\
3) We define vector $\bm{\hat{k}_2}$ as the cross product of $\bm{\hat{k}_3}$ and $\bm{\hat{\tilde{x}}}$. 
Thus we get:
$\bm{\hat{k}_2}=\left[-\sin(\tilde{\varphi}),
\cos(\tilde{\varphi}),0\right]$.\\
4) We define vector $\bm{\hat{\tilde{y}}}$ as a linear combination 
of $\bm{\hat{k}_2}$ and $\bm{\hat{k}_3}$:
$\bm{\hat{\tilde{y}}}=\bm{\hat{k}_2}\cos(\xi)+\bm{\hat{k}_3}\sin(\xi)$.\\
Thus vector $\bm{\hat{\tilde{y}}}$ lies in the plane of $\bm{\hat{k}_2}$ and $\bm{\hat{k}_3}$ which 
is orthogonal to $\bm{\hat{\tilde{x}}}$.
As $\xi$ goes from $0$ to $2\pi$ vector $\bm{\hat{\tilde{y}}}$ goes over all possible
orientations orthogonal to $\bm{\hat{\tilde{x}}}$. Note that fixed directions of 
$\bm{\hat{\tilde{x}}}$ and $\bm{\hat{\tilde{y}}}$ completely determine 
the direction of $\bm{\hat{\tilde{z}}}$.\\
5) We define $\bm{\hat{\tilde{z}}}$ as the cross 
product of $\bm{\hat{\tilde{x}}}$ and $\bm{\hat{\tilde{y}}}$.

Then, since we know the orientations of $\bm{\hat{\tilde{x}}},\bm{\hat{\tilde{y}}},\bm{\hat{\tilde{z}}}$ 
with respect to $\bm{\hat{x}},\bm{\hat{y}},\bm{\hat{z}}$, we can write the rotation matrix $R$.
By applying the rotation matrix to the stress tensors $\Sigma_i$ and $\Sigma_j$ we obtain matrices
$\tilde{\Sigma}_i$ and $\tilde{\Sigma}_j$. 
Taking from these matrices the elements $\tilde{\sigma}_i^{xy}$ and $\tilde{\sigma}_j^{xy}$ we form the product
$\tilde{\sigma}_i^{xy} \tilde{\sigma}_j^{xy}$. 
This product $\tilde{\sigma}_i^{xy} \tilde{\sigma}_j^{xy}$ 
is a function of angles $\tilde{\theta},\tilde{\varphi},\xi$. 
It follows from the definitions of these angles 
that in performing the averaging over them the angles $\tilde{\varphi}$ and $\xi$ 
go from $0$ to $2\pi$ with weight 1 (unity), while the integration 
over $\tilde{\theta}$ goes from $0$ to $\pi$ with weight $\sin(\tilde{\theta})$. 
Thus effectively we calculate:
\begin{eqnarray}
\left<\tilde{\sigma}_i^{xy}\tilde{\sigma}_j^{xy}\right> = \frac{1}{8\pi^2}\int_0^{\pi}\int_0^{2\pi}\int_0^{2\pi} 
\left[\tilde{\sigma}_i^{xy}\tilde{\sigma}_j^{xy}\right]\sin(\tilde{\theta})d\tilde{\theta}d\tilde{\varphi}d\xi\;.\;\;\;\;\;\;\;\;
\label{eq;avesixysjxy1apx}
\end{eqnarray}
Note that, by construction, expression (\ref{eq;avesixysjxy1apx}) should be rotationally invariant.
The averaging procedure described above leads to the result presented in formulas
(\ref{eq;avesixysjxy1},\ref{eq;avesixysjxy21},\ref{eq;avesixysjxy22},\ref{eq;avesixysjxy2}).

\section{From the \pds of the eigenvalues to the \pds of the spherical stresses.
The 2D Gaussian case of equal centres and widths for both eigenvalues. \label{sec:eigen-to-spherical}} 

Let us suppose that the \pds of both eigenvalues of some 2D stress tensor 
are the Gaussian functions centred around $\bar{\lambda}$ with width $\sigma_o$. 
We are interested in finding the \pds for the spherical components of the stress tensor.

We have:
\begin{eqnarray}
&&dW(\lambda_1,\lambda_2,\varphi) \equiv W_{\lambda}(\lambda_1)W_{\lambda}(\lambda_2)d\lambda_1 d\lambda_2 \left(\frac{d\varphi}{2\pi}\right)\;\label{eq;2D-from-L-to-ss-01}\\
&&=\frac{d\lambda_1 d\lambda_2 d\varphi}{(2\pi)^2 \sigma_o^2}
\exp\left[-\frac{\left(\lambda_1-\bar{\lambda}\right)^2+\left(\lambda_2-\bar{\lambda}\right)^2}{2\sigma_o^2}\right]
.\;\;\;\;\;\label{eq;2D-from-L-to-ss-02}
\end{eqnarray}
In order to solve the problem it is necessary to express the eigenvalues through the \ssc.
It is also necessary to express, using the Jacobian determinant, 
the volume element $d\lambda_1 d\lambda_2d\varphi$ in terms of the volume element built from the \ssc. 

The relations between the Cartesian and the \ssc\, for the 2D case are:
\begin{eqnarray}
\hat{\Sigma}_i = \begin{pmatrix} 
\sigma_i^{xx} & \sigma_i^{xy} \\ 
\sigma_i^{yx} & \sigma_i^{yy} \\  
\end{pmatrix}\;
=\;
\begin{pmatrix} 
p_i+s_{1,i} & s_{2,i} \\ 
s_{2,i} & p_i-s_{1,i} \\  
\end{pmatrix}\;, 
\label{eq;s2Dmatrix1}
\end{eqnarray}
where:
\begin{eqnarray}
p_i \equiv \tfrac{\sigma_i^{xx} + \sigma_i^{yy}}{2}\;,\;s_{1,i} \equiv \tfrac{\sigma_i^{xx}-\sigma_i^{yy}}{2}\;,\;s_{2,i} \equiv \sigma_i^{xy}=\sigma_i^{yx}\;.\;\;\;\;\;\;\;
\label{eq;s2Dmatrix2}
\end{eqnarray}
For a particular set of $p=p_i$, $s_{1,i}$ and $s_{2,i}$ the determinant equation
for the eigenvalues $\lambda_{1,i}$ and $\lambda_{2,i}$ is (for briefness of the notations we omit index $i$ in the following):
\begin{eqnarray}
\begin{vmatrix} 
p+s_1-\lambda & s_2 \\ 
s_2 & p-s_1-\lambda \\  
\end{vmatrix}=0\;.\;\; 
\label{eq;s2Dmatrix3}
\end{eqnarray}
Equation (\ref{eq;s2Dmatrix3}) leads to:
\begin{eqnarray}
&&\lambda_1 = p+\sqrt{s_1^2 + s_2^2}\;,\;\;\;\;\;\;\;\lambda_2 = p-\sqrt{s_1^2 + s_2^2}\;.\;\;\;\label{eq;s2Dmatrix4}\\
&&p=\tfrac{1}{2}\left(\lambda_1 + \lambda_2\right),\;\;\;\;\;
\varphi=\tfrac{1}{2}\arctan\left(s_2/s_1\right),\;\label{eq;2D-from-L-to-ss-021}\\
&&s_1 = \tfrac{1}{2}\left(\lambda_1 - \lambda_2\right)\cos(2\varphi),\;\;s_2 = \tfrac{1}{2}\left(\lambda_1 - \lambda_2\right)\sin(2\varphi).\;\;\;\;\;\;\;\;\label{eq;2D-from-L-to-ss-022}
\end{eqnarray}
Using expressions 
(\ref{eq;s2Dmatrix4},\ref{eq;2D-from-L-to-ss-021},\ref{eq;2D-from-L-to-ss-022}) 
it is straightforward to obtain the expression for the absolute value of the determinant of the Jacobian matrix. 
This expression is:
\begin{eqnarray}
|\det(J_{\lambda \rightarrow s})| =1/\sqrt{s_1^2+s_2^2}\;,\label{eq:jacob}
\end{eqnarray}
Using expressions (\ref{eq;s2Dmatrix4},\ref{eq;2D-from-L-to-ss-021},\ref{eq;2D-from-L-to-ss-022}) for the eigenvalues 
and the Jacobian determinant (\ref{eq:jacob}) we rewrite (\ref{eq;2D-from-L-to-ss-01},\ref{eq;2D-from-L-to-ss-02}) as:
\begin{eqnarray}
&&dW(p,s_1,s_2)=W_p(p)W_{ss}(s_1,s_2)dp ds_1 ds_2\;\label{eq;2D-from-L-to-ss-031}\\
&&=\frac{1}{(2\pi)^2 \sigma_o^2}
\exp\left[-\frac{\left(p-\bar{\lambda}\right)^2 + s_1^2 + s_2^2}{\sigma_o^2}\right]\frac{dp ds_1 ds_2}{\sqrt{s_1^2+s_2^2}}\;.\;\;\;\;\;\;\;\;\label{eq;2D-from-L-to-ss-032}
\end{eqnarray}
Thus the Gaussian distributions for the eigenvalues lead to the Gaussian distributions for 
$p$ and $s \equiv \sqrt{s_1^2 + s_2^2}$ (note that $2\pi s ds \sim ds_1 ds_2$).
However, note that the \pds for $s_1$ and $s_2$ are not independent
(because of the Jacobian determinant), i.e., it is
impossible to rewrite $W_{ss}(s_1,s_2)$ as $W_{s}(s_1)W_s(s_2)$.
\emph{The last conclusion is important, as it means that independent distributions of 
eigenvalues do not, in general, lead to the independent distributions for the \ssc.}

Expression (\ref{eq;2D-from-L-to-ss-032}) suggests that the spread of the pressure component 
is the same as the spread of the shear components.
This contradicts to the previous observations that relate the spread of 
the pressure component to the bulk modulus, while the spreads of the shear
components are related to the shear modulus \cite{Egami19821,Chen19881,Levashov2008B}. 
Thus the assumption that both eigenvalues are independent and have the same \pds 
leads to the result that contradicts to the known facts and thus this is not a good assumption.

\section{From the Gaussian \pds for the spherical stresses to the \pds
for the eigenvalues \label{sec:spherical-to-eigen}}

Let us now consider the problem which is the inverse with respect to the previous one.
Thus we assume that the \pd for the pressure is the Gaussian function centred around $p_o$ with width $\sigma_p$,
while the \pds for both spherical shear components are the Gaussian functions centred around zero with widths $\sigma_s$.
We assume that the \pds for the spherical
stress components, $s_1$, $s_2$, and $p$ are independent.
We are interested in finding the \pd for the eigenvalues. 
Thus we have:
\begin{eqnarray}
W(p,s_1,s_2) \equiv W_p(p)W_s(s_1)W_s(s_2)\;,\;\;\;\;\;\;\;
\label{eq;s2Dmatrix6}
\end{eqnarray}
where
\begin{eqnarray}
&&W_p(p)dp \equiv \frac{1}{\sqrt{2\pi\sigma_p^2}}\exp{\left[-\frac{(p-p_o)^2}{2\sigma_p^2}\right]}dp\;,\label{eq;p-distr-01}\;\;\;\\
&&W_s(s_n)ds_n \equiv \frac{1}{\sqrt{2\pi\sigma_s^2}}\exp{\left[-\frac{s_n^2}{2\sigma_s^2}\right]}ds_n\;,\label{eq;s-distr-01}\;\;\;\;
\end{eqnarray}
and $s_n = s_1$ or $s_n = s_2$.

In order to find the \pds for $\lambda_1$ and $\lambda_2$ from 
(\ref{eq;s2Dmatrix4},\ref{eq;2D-from-L-to-ss-021},\ref{eq;2D-from-L-to-ss-022}) 
it is necessary to express the pressure and shear components in 
(\ref{eq;s2Dmatrix6},\ref{eq;p-distr-01},\ref{eq;s-distr-01}) through the eigenvalues 
and it is necessary to express the ``volume" element in terms of the eigenvalues
using the Jacobian determinant. 
It follows from (\ref{eq:jacob},\ref{eq;s2Dmatrix4}) that the absolute value of the
Jacobian determinant is $\tfrac{1}{2}|\lambda_1 - \lambda_2|$.
The result, after the integration over $\varphi$, is:
\begin{widetext}
\begin{eqnarray}
W(\lambda_1,\lambda_2)d\lambda_1 d\lambda_2 = 
\frac{1}{\left(\sqrt{2\pi}\right)\sigma_p \sigma_s^2}
\exp{\left[-\frac{p_o^2}{2\sigma_p^2}
+\frac{p_o(\lambda_1+\lambda_2)}{2\sigma_p^2}
-\frac{(\lambda_1+\lambda_2)^2}{8\sigma_p^2}
-\frac{(\lambda_1-\lambda_2)^2}{8\sigma_s^2}
\right]}|\lambda_1 - \lambda_2|d\lambda_1 d\lambda_2\;.\;\;\;\;\;\;\;
\label{eq;pstol1l1}
\end{eqnarray}
\end{widetext}
Note that expression (\ref{eq;pstol1l1}) \emph{can not} be split into two independent distributions
for $\lambda_1$ and $\lambda_2$. \emph{This last observation is important as it means that independent
\pds for the \ssc\, do not, in general, lead to the independent 
\pds for the eigenvalues.}

\section{Analytical calculation of the \pd of \lto in a model
2D case \label{sec:app-l2overl1}}

Here we provide an insight on why the probability distribution $W(\lambda_2/\lambda_1)$ 
decreases \emph{linearly} to zero as \lto approaches one.
We provide this argument for a two-dimensional case.
It is assumed that different shear stress components are independent and distributed around zero with the same Gaussian probabilities.
It is also assumed that the pressure is distributed with the Gaussian probability around 
some sufficiently large positive value--so that all eigenvalues
of the stress tensor are positive. 

We start from the consideration of the atomic stress tensor and 
express the pressure and shear components in terms of the eigenvalues.
The obtained expression is used to transform the \pd 
for the stress components into the \pd for 
the ratio of the eigenvalues.

Keeping in mind expressions 
(\ref{eq;s2Dmatrix1},\ref{eq;s2Dmatrix2},\ref{eq;s2Dmatrix4},\ref{eq;2D-from-L-to-ss-021},\ref{eq;2D-from-L-to-ss-022})
we introduce the following notations:
\begin{eqnarray}
&&l \equiv \left(\frac{\lambda_2}{\lambda_1}\right)\;,\;\;\;l=1-\delta\;,\;\;\;\label{eq;s2Dmatrix51}\\
&&s \equiv \sqrt{s_1^2+s_2^2} =p\left(\frac{1-l}{1+l}\right)=\left(\frac{p\delta}{2-\delta}\right)\;,\;\;\;\;\;\;\;
\label{eq;s2Dmatrix52}
\end{eqnarray}
where the last two equalities in (\ref{eq;s2Dmatrix52}) follow from (\ref{eq;s2Dmatrix4}).

Expressions (\ref{eq;s2Dmatrix6},\ref{eq;p-distr-01},\ref{eq;s-distr-01}) can be rewritten 
in terms of $p$ and $s$ and then in terms of $p$ and $\delta$:
\begin{eqnarray}
&&W(p,\delta)d\pd\delta=W(p)dp\label{eq;s2Dmatrix9}\\
&&\cdot\left(\frac{p^2}{\sigma_s^2}\right)\exp{\left[-\frac{\delta^2}{2(2-\delta)^2}\left(\frac{p^2}{\sigma_s^2}\right)\right]}\frac{2\delta}{(2-\delta)^3} d\delta\;.\;\;\;\;\;\;\;\nonumber
\end{eqnarray}
It can be seen from (\ref{eq;s2Dmatrix9}) that for small $\delta$ function $W(p,\delta)$ linearly increases with $\delta$. 
This provides an insight into the linear decay of the probability of \lto as this ratio approaches one.
Also note that the rate of increase of the probability with increase of $\delta$ is determined by the ratio $(p^2/\sigma_s^2)$.

\begin{figure}
\begin{center}
\includegraphics[angle=0,width=3.3in]{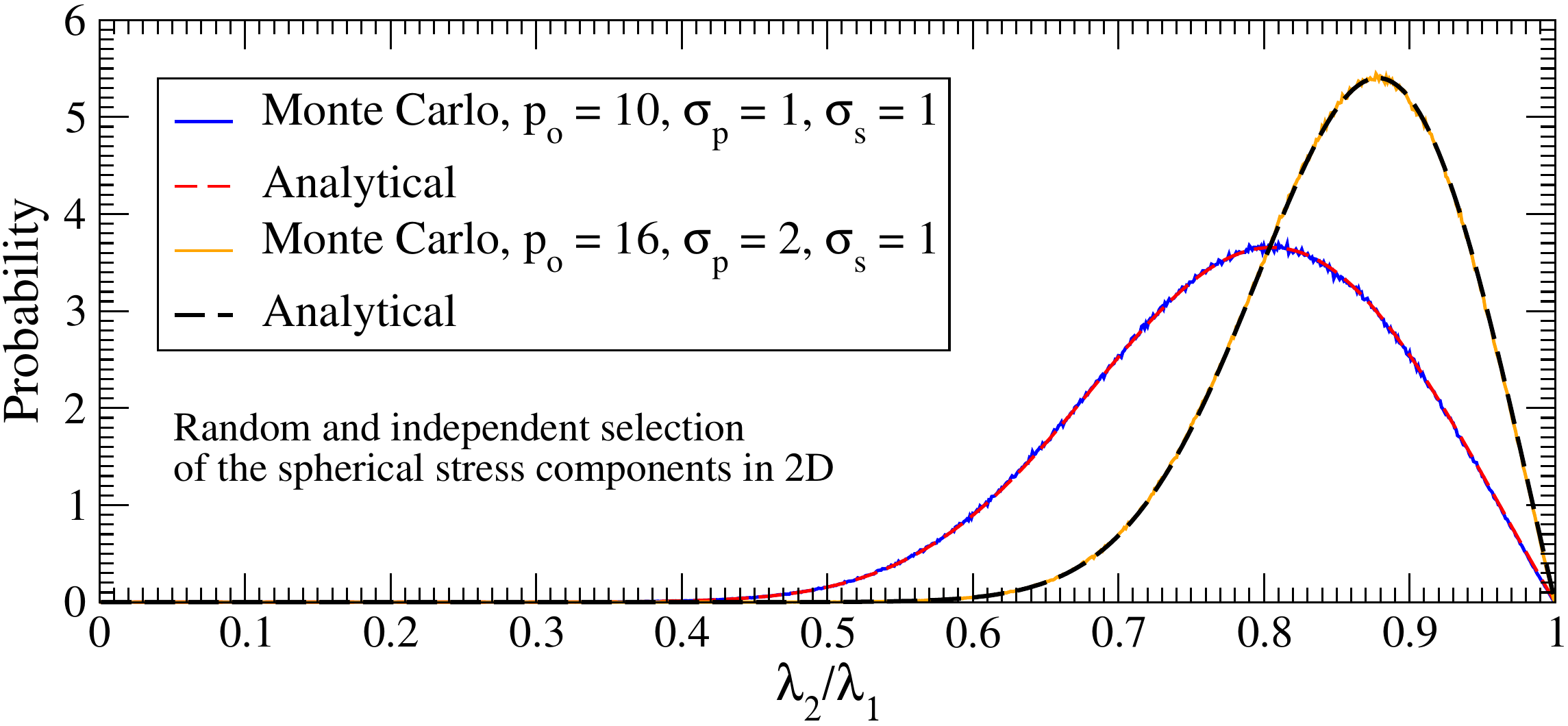}
\caption{Probability of the ratio $l = (\lambda_2/\lambda_1)$ for the two 2D cases
of the pressure and shear stress Gaussian distributions. 
The curves in the figure were calculated using
Monte Carlo simulations and also using formula (\ref{eq;s2Dmatrix13}).
}\label{fig:shear-shear-invariant-AA-1}
\end{center}
\end{figure}

For the Gaussian \pd of pressure it is possible to obtain 
the analytical expression for the probability of $l \equiv (\lambda_2/\lambda_1)$ in the whole
range of $l \in (0:1)$. 
Thus, in the case of the Gaussian \pd for pressure, 
expression (\ref{eq;s2Dmatrix9}) rewritten in terms of $l$ is:
\begin{eqnarray}
&&W(p,l) dp dl
=\frac{1}{\sqrt{2\pi\sigma_p^2}}\exp{\left[-\frac{(p-p_o)^2}{2\sigma_p^2}\right]}dp\;\;\;\;\label{eq;s2Dmatrix10}\\
&&\cdot\left(\frac{p^2}{\sigma_s^2}\right)\exp{\left[-\frac{(1-l)^2}{2(1+l)^2}\left(\frac{p^2}{\sigma_s^2}\right)\right]}\frac{2(1-l)}{(1+l)^3} dl\;.\;\;\;\;\;\;\;\nonumber
\end{eqnarray}
It is possible to integrate (\ref{eq;s2Dmatrix10}) over $p$ and obtain the analytical expression for $W(l)$.
It is convenient to introduce the notations,
\begin{eqnarray}
&&\frac{1}{a^2} \equiv \frac{1}{2\sigma_p^2}\;,\;\;\;\frac{1}{b^2} \equiv \frac{(1-l)^2}{2(1+l)^2}\left(\frac{1}{\sigma_s^2}\right)\;,\;\;\\
&&\frac{1}{c^3} \equiv \frac{1}{\sigma_s^2\sqrt{2\pi\sigma_p^2}}\frac{2(1-l)}{(1+l)^3}\;,\;\;
\label{eq;s2Dmatrix11}
\end{eqnarray}
which compactify (\ref{eq;s2Dmatrix10}) into:
\begin{eqnarray}
W(p,l) dp dl
=\exp{\left[-\frac{(p-p_o)^2}{a^2}-\frac{p^2}{b^2}\right]}
\frac{p^2dp}{c^3} dl\;.\;\;\;\;\;\;\;\;\;\;\label{eq;s2Dmatrix12}
\end{eqnarray}
The final answer is:
\begin{eqnarray}
W(l)
=\frac{\sqrt{\pi}\left(2ab^5p_o^2+a^5b^3+a^3b^5\right)}{2c^3(a^2+b^2)^{5/2}}\exp{\left[-\frac{p_o^2}{a^2+b^2}\right]}
 \;.\;\;\;\;\;\;\;\;\;\;\label{eq;s2Dmatrix13}
\end{eqnarray}
Figure (\ref{fig:shear-shear-invariant-AA-1}) shows the probability curves for $l$ for the two cases of 
pressure and shear stress distributions. The curves in the figure obtained using the derived formula (\ref{eq;s2Dmatrix13}) 
coincide with the curves obtained by means of MC simulations.


\end{document}